\documentclass[12pt,a4paper]{article}
\pdfoutput=1
\usepackage{jheppub}

\usepackage{amsfonts}
\usepackage{ccaption}
\usepackage{mathrsfs}

\makeatletter
\@addtoreset{equation}{section}

\makeatletter
\renewcommand\section{\@startsection {section}{1}{\z@}%
                                   {-3.5ex \@plus -1ex \@minus -.2ex}
                                   {2.3ex \@plus.2ex}%
                                   {\normalfont\large\bfseries}}
\renewcommand\subsection{\@startsection{subsection}{2}{\z@}%
                                     {-3.25ex\@plus -1ex \@minus -.2ex}%
                                     {1.5ex \@plus .2ex}%
                                     {\normalfont\bfseries}}
  \captionnamefont{\bfseries}
  \captiontitlefont{\small\sffamily}
  \captiondelim{: }
  \hangcaption

\def\vev#1{\langle\, #1 \, \rangle}

\def\Tr#1{{\rm Tr}\left(#1\right)}

\def\AdS#1{AdS$_{#1}$}

\definecolor{rust}{rgb}{0.8,0.2,0.2}
\definecolor{green}{rgb}{0.1,0.8,0.2}

\title{Holographic thermal field theory on curved spacetimes}

\author{Donald Marolf$^{\;a,b}$,}
\author{ Mukund Rangamani$^{\,c}$,}
\author{  Toby Wiseman$^{\, d}$}

\affiliation[a]{Department of Physics, University of California, Santa Barbara, CA 93106, USA}
\affiliation[b]{Department of Applied Mathematics and Theoretical Physics,\\ University of Cambridge, CB3 0WA  UK}
\affiliation[c]{Center for Particle Theory \& Department of Mathematical Sciences, \\
Science Laboratories, South Road, Durham DH1 3LE, UK}
\affiliation[d]{Theoretical Physics Group, Blackett Laboratory, Imperial College, London SW7 2AZ, UK }

\emailAdd{marolf@physics.ucsb.edu}
\emailAdd{mukund.rangamani@durham.ac.uk}
\emailAdd{t.wiseman@imperial.ac.uk}

\abstract{
The AdS/CFT correspondence relates certain strongly coupled CFTs with large effective central charge $c_\text{eff}$ to semi-classical gravitational theories with AdS asymptotics.
We describe recent progress in understanding gravity duals for CFTs on non-trivial spacetimes at finite temperature, both in and out of equilibrium.  Such gravity methods provide powerful new tools to access the physics of these strongly coupled theories, which often differs qualitatively from that found at weak coupling.

Our discussion begins with basic aspects of AdS/CFT and progresses through thermal CFTs on the Einstein Static Universe and on periodically identified Minkowski spacetime.  In the latter context we focus on states describing so-called plasma balls, which become stable at large $c_\text{eff}$.   We then proceed to out-of-equilibrium situations associated with dynamical bulk black holes. In particular, the non-compact nature of these bulk black holes allows stationary solutions with non-Killing horizons that describe time-independent flows of CFT plasma.   As final a topic we consider CFTs on black hole spacetimes.  This discussion provides insight into how the CFT transports heat between general heat sources and sinks of finite size.  In certain phases the coupling to small sources can be strongly suppressed, resulting in negligible heat transport despite the presence of a deconfined plasma with sizeable thermal conductivity.
We also present a new result, explaining how this so-called droplet behaviour is related to confinement via a change of conformal frame.
}

\begin{document}

\begin{flushright} \small{DCPT-13/51} \end{flushright}

\maketitle

\flushbottom

%
\section{Introduction}
\label{sec:intro}
%

The study of quantum field theories in curved spacetimes has proven to be a fruitful endeavor over many years. The physics of vacuum polarization, particle production, etc., engendered by the presence of background curvature forms the basis for modern discussions of Hawking radiation and cosmological fluctuations.\footnote{ For reviews of these topics we invite the reader to consult the textbooks \cite{Birrell:1982ix, Wald:1994fk,Mukhanov:2007zz} or reviews \cite{Jacobson:2003vx, Ross:2005sc}.} These important insights link the subject naturally to that of thermal field theory, and in particular thermal field theory in curved spacetime.  But what are the properties of the resulting thermal states? Since detailed studies have largely been confined to either free fields or those that interact weakly, quantum fields with strong self-interactions can bring novel effects.  The goal of this review is to describe how the holographic AdS/CFT correspondence can be used to explore such strongly-coupled physics and to survey interesting such results to date.

The AdS/CFT correspondence \cite{Maldacena:1997re, Gubser:1998bc, Witten:1998qj} maps a class of non-gravitational quantum field theories in $d$-dimensions onto certain string theories, which in particular contain dynamical gravity. In general, the class of quantum field theories we discuss will be conformal field theories characterized by two basic parameters: $\lambda$, which measures the strength of the coupling between the microscopic constituents, and an effective central charge $c_\text{eff}$ which is a measure of the number of degrees of freedom.\footnote{ We use this term even when $c_\text{eff}$ is not associated with either anomalies or a central term in any algebra.}

In the limit of large $c_\text{eff}$, the CFT should simplify dramatically.
For a gauge theory this corresponds to the planar limit. More generally, one expects the path integral to localize around a saddle point that defines a new notion of semi-classical behaviour with effective Planck's constant $\hbar \propto \frac{1}{c_\text{eff}}$ and an associated effectively semi-classical description in terms of some so-called master fields.  While an explicit construction of such fields has proven elusive in general, the AdS/CFT correspondence gives a simple description of the result for an infinite class of field theories.  For generic $\lambda$ it is given by semi-classical strings in a non-trivial higher-dimensional spacetime with AdS asymptotics known as ``the bulk.''

One can further simplify the description by taking the limit $\lambda \to \infty$.  In the bulk massive string states then become infinitely heavy and decouple, leaving only the dynamics of semi-classical (super-)gravity.\footnote{ The mass of string states scales as $\lambda^\alpha $ with $\alpha >0$ depending on the spacetime dimension; for $3+1$ dimensional supersymmetric field theories $\alpha = \frac{1}{4}$.}  The result is that complicated questions about the dynamics of quantum fields can be rephrased in terms of semi-classical gravitational physics in higher dimensions precisely in the strongly-coupled limit $\lambda \to \infty$  where the CFT is most difficult to study.

This abstract discussion can be made explicit in many examples. As an illustration, consider the prototypical case where the CFT is the maximally supersymmetric $SU(N)$ gauge theory in $3+1$ spacetime dimensions.  This is in fact a superconformal field theory (SCFT) called ${\cal N}=4$ super Yang-Mills (SYM) whose free parameters are just the positive integer $N$ describing the rank of the gauge group and a dimensionless coupling $g_\text{YM}$ (with vanishing beta function).   The bulk dual turns out to be a 9+1 theory of so-called IIB strings with \AdS{5} $\!\times \,{\bf S}^5$ asymptotics.  This theory is reviewed in standard texts such as \cite{Polchinski:1998rr,Johnson:2003gi,Becker:2007zj,Kiritsis:2007zza}; see also  \cite{Marolf:2011zs} for a review of the supergravity limit designed to be accessible to relativists.
The IIB parameters are the (dimensionless) string coupling $g_s$, the the string length $l_s = \sqrt{\alpha'}$, and the AdS length scale $\ell$ (which is also the radius of the ${\bf S}^5$).  One may also discuss the Planck length $l_p = g_s^{1/4} l_s$ and the associated Newton constant $G_{10} = 8 \pi^6\, l_p^8$.  It is useful to note that the IIB theory contains no bare cosmological constant, but that \AdS{5} $\!\times \,{\bf S}^5$ emerges as a solution supported by the flux of a certain gauge field.  As a result, charge quantization forces $\ell$ to take certain discrete values in a manner determined by $g_s$ and $l_s$. Furthermore, since there is no independent external standard of length, any physical properties of the theory can depend only on the dimensionless ratio $l_s/\ell$ and not on $l_s$, $\ell$ separately.

The two theories above are connected by a classic argument of Maldacena \cite{Maldacena:1997re}. By considering a low energy limit of D3-branes, Maldacena argued  that they are in fact equivalent (`dual') when one identifies
\begin{align}
g_\text{YM}^2 = g_s \; , \qquad \text{and} \qquad \lambda \equiv N g_\text{YM}^2 = \left( \frac{\ell}{l_s} \right)^4.
\label{adsparmap}
\end{align}
In particular, the above-mentioned discreteness of $\ell$ on the IIB side is equivalent to the SYM requirement that $N$ be an integer.  In \eqref{adsparmap} we have introduced the the gauge theory 't Hooft coupling $\lambda$ that controls perturbation theory at large $N$.
Type IIB string theory becomes classical in this planar $N\gg 1$ limit and will be the focus of our discussion.

We hasten to emphasize that the AdS/CFT correspondence is by no means restricted to four-dimensional field theories with maximal supersymmetry. There is by now an infinite class of field theory examples in diverse dimensions $d=(d-1)+1$ with varied amounts of supersymmetry which are described by gravitational dynamics asymptotic to some \AdS{d+1} $\! \times \, X$, where the curvature of $X$ is typically also of order $\ell$ \cite{Morrison:1998cs,Sparks:2010sn}. While the choice of $X$ depends on CFT details, the bulk theory always admits a universal sector describing direct product solutions ${\cal M} \times X$ where ${\cal M}$ is a $(d+1)$-dimensional Einstein space (locally) asymptotic to \AdS{d+1}.  In other words, the dynamics of this universal sector is described just by Einstein-Hilbert gravity with a negative cosmological constant.\footnote{ In standard supergravity parlance, Einstein-Hilbert gravity with negative cosmological constant is a universal consistent truncation of a wide class of supergravity theories arising in string constructions.   In field theory this means that states created from the vacuum by acting with stress tensors have vanishing one-point functions for other local operators, at least in the limit of large $c_\text{eff}$.} Note that the Newton constant for this truncated theory is
$G_{d+1} = \frac{1}{\text{vol}(X)}\, G_D$, where $G_D$ is the Newton constant for the full $D = d + 1 + \rm{dim}(X)$ dimensional theory.  Here $D = 10$ for string theory and $D = 11$ for M-theory. In general, as explained in \S\ref{sec:alads}, $G_{d+1}$, $\ell$ are related to our effective central charge by%
\begin{equation}
c_\text{eff} = \frac{\ell^{d-1}}{16\pi\, G_{d+1}} = \frac{\ell^{d-1}\; \text{vol}(X)}{16\pi\, G_{D}}\,.
\label{ceffdef}
\end{equation}	
For $\mathcal{N} = 4$ SYM with $d=4$ we have $c_\text{eff}= N^2/8\pi^2$.

We will focus largely on this universal sector below, though we also address certain aspects of the physics that depends explicitly on $X$.  An illustrative example is provided
by the physics of black holes in AdS. The well known Hawking-Page transition (see \S \ref{sec:HP}) indicates a phase transition in the canonical ensemble between a thermal gas and black hole in global AdS.  This discussion involves only AdS gravity with negative cosmological constant and is thus universal.  In contrast, the physics of small AdS black holes is not universal since, as reviewed in \S\ref{sec:globalGL},  below some $X$-dependent threshold sufficiently small black holes will tend to develop additional structure on the internal space. Throughout this review we will try to emphasize which aspects of the physics are universal and which are not.

Our primary interest will be situations where the CFT is at finite temperature, or at least where regions of it are thermally excited. Hot regions of CFT containing excited plasmas are associated with black hole horizons in the bulk. One of the most powerful aspects of AdS/CFT is due to the remarkable fact that black holes (at least outside a horizon) become more classical at higher energies, totally at odds with non-gravitational behaviour where higher energies typically probe shorter more quantum scales.  Thus the master field description of the CFT as semi-classical  bulk gravity remains useful even when we consider the CFT in highly energetic thermal states.

We are interested here in using AdS/CFT to study the physics of strongly coupled quantum field theory on some curved spacetime ${\cal B}_d$.  Note that the metric on ${\cal B}_d$ is fixed and non-dynamical.  As a result, we will investigate Asymptotically (locally) AdS${}_{d+1}$ (henceforth AlAdS$_{d+1}$) geometries whose conformal boundaries agree with ${\cal B}_d$;  i.e., the choice of ${\cal B}_d$ provides a boundary condition for the bulk gravitational problem.  We review physical lessons learned from such solutions below, as well as certain details of their construction.

In particular, we will consider static such plasmas in static spacetimes, stationary plasma flows in static spacetimes, and truly dynamical plasmas in time dependent spacetimes. We will be able to study these plasmas in their hydrodynamic regime and also beyond. To appreciate the strength of this statement, recall that for gauge theories the typical approach to studying strongly coupled plasmas involves the Euclidean lattice. While this approach successfully computes both equilibrium thermal behaviour and certain transport properties, Euclidean lattice methods break down once the plasma begins to flow.  Yet AdS/CFT continues to provide a valuable and powerful computational tool in the strongly coupled regime.

We begin our review by explaining further aspects of the correspondence in \S \ref{sec:2} below.  Our treatment will be brief. For further details we direct the interested reader to more complete reviews in e.g. \cite{Aharony:1999ti,Klebanov:2000me,DHoker:2002aw,Polchinski:2010hw}; see also \cite{Fischetti:2012rd} for a review of key features intended to be accessible to relativists.   We then discuss equilibrium configurations on simple static spacetimes ${\cal B}_d$ containing sphere and circle factors in \S \ref{sec:3}.  Such situations are associated in the CFT with various phase transitions (including colour confinement) and thus also with mixed phases describing domain walls and so-called plasma balls.  The bulk physics involves related transitions as well as, in some cases, analogues of the Gregory-Laflamme instability \cite{Gregory:1993vy}.

Moving beyond strict equilibrium, we consider plasma flow in \S \ref{sec:4}.  In particular, we review the fluid/gravity correspondence  \cite{Bhattacharyya:2008jc}, which provides a general tool for situations where gradients are small. More general results from bulk numerics beyond this regime are summarized as well.  Of special interest are stationary flows with entropy production that are described in the bulk by novel stationary black holes with non-Killing future horizons and singular past horizons.

It then remains to discuss CFTs on black hole spacetimes in \S \ref{sec:5}.  As described there, this may be viewed either as the specific study of Hawking radiation in strongly coupled CFTs, or as part of a more general study of how strongly coupled CFTs transport heat between sources and sinks of finite size.  An intriguing aspect is that the coupling to well-localized heat sources can be strongly suppressed, resulting in negligible heat transport despite the presence of a deconfined plasma with sizable thermal conductivity. We discuss aspects of the phase transition associated with this decoupling, and how it is related to colour confinement by a change of conformal frame.  The results reviewed include both equilibrium computations and those describing fully non-linear stationary flows between black holes at different temperatures, where the latter again involve stationary black holes with non-Killing future horizons and singular past horizons.  We close with some final discussion in \S\ref{sec:6}.

%
\section{Overview of AdS/CFT: Boundary conditions \& CFT data}
\label{sec:2}
%

We now review a few key features of the AdS/CFT correspondence, focussing on the universal $(d+1)$ gravity sector described in the introduction. Thus the bulk equation of motion
\begin{align}
\label{bEOM}
R_{AB} +  \frac{d}{\ell^2} \,g_{AB} = 0 \,.
\end{align}
is just that of Einstein-Hilbert gravity with negative cosmological constant $\Lambda = - \frac{d\,(d-1)}{2\,\ell^2}$.  We use the uppercase Latin alphabet for bulk indices and reserve the lowercase Greek alphabet for boundary indices. The bulk metric in $d+1$ dimensions will be denoted by $g_{AB}$ and, when it appears below, the boundary metric on ${\cal B}_d$ will be $h_{\mu\nu}$.  The bulk theory also has an additional parameter $G_{d+1}$ that does not appear in  \eqref{bEOM} and determines the size of bulk quantum effects.

A critical element of the correspondence is the notion of an asymptotically locally AdS (AlAdS) spacetime.  Solutions of \eqref{bEOM} are said to be AlAdS if they have a timelike conformal boundary ${\cal B}_d$ which lies at infinite distance from the bulk.  As an example we may consider pure AdS$_{d+1}$ in so-called global coordinates where the line element takes the form
\begin{align}
ds^2 &= -(1+\frac{\rho^2}{\ell^2})\, dt^2 + \frac{d\rho^2}{1+\frac{\rho^2}{\ell^2}} + \rho^2\, d\Omega_{d-1}^2 \,
\cr
&= \frac{\rho^2}{\ell^2} \left( -(\frac{\ell^2}{\rho^2}+1)\, dt^2 + \frac{\ell^2}{\rho^2} \frac{d\rho^2}{1+\frac{\rho^2}{\ell^2}} + \ell^2\, d\Omega_{d-1}^2 \right).
\label{gadsX}
\end{align}
Here the boundary ${\cal B}_d$ may be said to lie at $\rho = \infty$ and a metric on ${\cal B}_d$ may defined by writing the $d+1$ bulk as a diverging conformal factor times a line element that remains finite near ${\cal B}_d$. Doing so as shown on the second line of \eqref{gadsX}, the metric induced on ${\cal B}_d$ by this finite piece is that of the Einstein Static Universe, ESU$_{d} = {\mathbb R}\times {\bf S}^{d-1}$.  Of course, we could also have obtained any other conformally related metric on ${\cal B}_d$ by extracting a different overall conformal factor in \eqref{gadsX}. But since we wish only to investigate conformal field theories on ${\cal B}_d$, this ambiguity will not affect physical results.\footnote{ This is strictly true only when $d$ is odd.  For even $d$ one should include affects associated with the conformal anomaly; see the end of \S\ref{sec:alads} for a brief discussion.  For simplicity we will generally ignore such effects below.} The decision to extract a particular conformal factor from the bulk metric is said to define a conformal frame on ${\cal B}_d$.  Below we use the symbol ${\cal B}_d$ to mean a manifold together with a metric defined up to changes of conformal frame.

A powerful summary of the AdS/CFT duality is given by equating the bulk path integral over all spacetimes with some fixed conformal boundary ${\cal B}_d$ and the CFT path integral on the background spacetime ${\cal B}_d$ \cite{Witten:1998qj}.  In both cases the result depends on the non-dynamical metric on ${\cal B}_d$, or at least on the associated conformal structure.  This statement is especially useful in Euclidean signature where we may take ${\cal B}_d$ to be compact so that there is no need to specify further boundary conditions.   In particular, dual bulk and CFT thermal states may be computed using corresponding Euclidean path integrals when ${\cal B}_d$ has a $U(1)$ isometry associated with some ${\bf S}^1$ factor in its geometry.  In a given conformal frame, the length $\beta$ of this ${\bf S}^1$ controls the temperature $T = 1/\beta$ of each thermal state (where we have set $\hbar =1 =c$).  Although the numerical value of $T$ depends on the choice of conformal frame, the fact that physics can depend only on dimensionless ratios ensures that physical predictions remain conformally invariant.

Since we focus on the bulk semi-classical limit, the bulk path integral can be evaluated using the steepest descent approximation.  It thus becomes a sum over saddle points satisfying \eqref{bEOM} and having conformal boundary ${\cal B}_d$, where the contribution from each saddle is $\exp(-S_E)$ for an appropriate Euclidean action $S_E$.\footnote{ \label{holreg} The infinite volume of any AlAdS spacetime causes the Einstein-Hilbert action to diverge.  But this naive action does not in fact provide a good variational principle in either Euclidean or Lorentzian signature.  The elegant theory of how to construct a proper variational principle with finite on-shell action is called holographic regularisation; see  \cite{Balasubramanian:1999re,Henningson:1998ey,Henningson:1998gx,deHaro:2000xn,Skenderis:2006uy} for original works, \cite{deHaro:2000xn} and \cite{Skenderis:2002wp} and  comprehensive reviews, and \cite{Fischetti:2012rd} for a review targeted at an audience of relativists.}  In contexts with finite temperature $T$, these bulk saddles will often contain black holes with surface gravity $\kappa = 2 \pi \,T$.  As we review in \S\ref{sec:alads} below, specifying ${\cal B}_d$ is analogous to fixing Dirichlet data for more familiar equations of motion.  In Euclidean signature, where equation \eqref{bEOM} is elliptic (up to gauge issues), one expects at most a discrete set of such solutions.  In our semi-classical limit where each $S_E$ is large, this means that most saddles make negligible contribution when compared with the one minimizing $S_E$.  This least-action saddle is then said to dominate the associated ensemble.  As usual, varying parameters in ${\cal B}_d$  -- such as the temperature -- can cause the actions assigned to various saddles to change in such a way that two distinct saddles exchange dominance, causing a first-order phase transition.

As a special case of the above argument we may take the limit $T \rightarrow 0$ to conclude that the ground state of the CFT maps to the lowest energy state in the bulk.  In the semi-classical limit, this reduces to just the lowest energy classical solution.  By the AdS positive energy theorem \cite{Townsend:1984iu}, when ${\cal B}_d = {\mathbb R}\times {\bf S}^{d-1}$ this is just \eqref{gadsX}.  Since general CFT arguments guarantee that the CFT vacuum is invariant under the full group $SO(d,2)$ of conformal isometries of ${\mathbb R}\times {\bf S}^{d-1}$, it is satisfying to note that this group also gives the isometries of \eqref{gadsX}.

Finding the bulk ground state for general ${\cal B}_d$ may require more work, perhaps involving numerics.  We will discuss some such cases in \S\ref{sec:circleap}.  But another tractable case occurs when ${\cal B}_d$ is the Minkowski spacetime ${\mathbb R}^{d-1,1}$.  Perhaps the simplest argument is that ${\mathbb R}^{d-1,1}$ can be conformally mapped to a region of ${\mathbb R}\times {\bf S}^{d-1}$, and that the CFT vacuum on ${\mathbb R}^{d-1,1}$ agrees with the vacuum on ${\mathbb R}\times {\bf S}^{d-1}$ in this region.  So we obtain the bulk dual from \eqref{gadsX} by restricting attention to a region containing an appropriate piece of the conformal boundary and making an appropriate change of conformal frame.  A natural such piece is the so-called Poincar\'e patch, where the line element takes the form
\begin{align}
ds^2 = \frac{\ell^2}{z^2} \left( dz^2 + \eta_{\mu\nu}\, dx^\mu dx^\nu \right) \
\label{padsX}
\end{align}
in Poincar\'e coordinates.  Here ${\cal B}_d={\mathbb R}^{d-1,1}$ lies at $z=0$ and the Minkowski metric on ${\cal B}_d$ is manifest. The surface $z=\infty$ is a degenerate Killing horizon known as the Poincar\'e horizon.

Although we are physically interested in Lorentz signature, much of the above discussion has focussed on Euclidean solutions.  To the extent that we seek static solutions, one may readily pass back and forth between signatures by changing $t \to it$.  When solving for static Lorentz-signature bulk solutions, we need only bear in mind that we will typically wish to fix the temperatures (surface gravities) on any black hole horizons, perhaps in some cases taking them to vanish.  In particular, we require any such horizons to be smooth.

The merely stationary context turns out to be more subtle.  As we review in sections \S\ref{sec:flows} and \S\ref{sec:Hawking}, stationary bulk solutions can have non-Killing event horizons.  So there is no well-defined notion of temperature to fix, and one must think more carefully about boundary conditions to be imposed.  Two natural options are to use an analogue of excision (i.e., to move the boundary of the computational domain inside the horizon) \cite{Figueras:2011gd} or to simply use smoothness of the future horizon as the defining principle \cite{Fischetti:2012ps}.  As we shall see, in this context it is natural to allow singular past horizons as stationarity in the far past may be unphysical.

In fully dynamical Lorentz-signature contexts, the boundary condition implied by ${\cal B}_d$ suffices to allow the evolution of appropriate initial data for some period of time.  When ${\cal B}_d$ has compact Cauchy surfaces we may expect the result to have qualitative features resembling \eqref{gadsX}, perhaps with the addition of black holes.  But when the Cauchy surfaces of ${\cal B}_d$ are non-compact the maximal development can be more like \eqref{padsX} and end at a Cauchy horizon (which might either be similar to the Poincar\'e horizon of \eqref{padsX} or to a finite-temperature horizon).  Note that time-dependent metrics on ${\cal B}_d$ are allowed in this general scheme.  In particular, choosing a time-dependent ${\cal B}_d$ given by ${\mathbb R}^{d-1,1}$ before some $t=0$ and using an analogue of retarded boundary conditions naturally leads to bulk spacetimes that agree with \eqref{padsX} in the far past (and, in particular, on the past Poincar\'e horizon) but differ in the future.

The discussion above summarizes most of the AdS/CFT features that will be relevant for our review.  The one point that remains is to explain how the CFT stress tensor in some state can be recovered from the dual bulk classical solutions. After briefly describing this so-called boundary stress tensor and some associated technology in \S\ref{sec:alads} we provide some illustrative examples in \S\ref{sec:vacuathermal}.

\subsection{The Fefferman-Graham expansion and the boundary stress tensor}
\label{sec:alads}

It is convenient to describe the boundary stress tensor using Fefferman-Graham coordinates, which are an analog of Gaussian normal coordinates at the conformal boundary ${\cal B}_d$. The `radial' coordinate normal to the boundary is traditionally called $z$ with ${\cal B}_d$ located at $z=0$ and the additional $d$ coordinates $x^\mu$ running along ${\cal B}$.  In a neighborhood of ${\cal B}_d$, any metric can be written in the Fefferman-Graham form
\begin{eqnarray}
\label{FGmet}
g_{AB}\, dx^A dx^B = \frac{\ell^2}{z^2} \left( dz^2 + h(z,x)_{\mu\nu} \,dx^\mu dx^\nu \right).
\end{eqnarray}
But in contrast to actual Gaussian normal coordinates the choice of $z$ is not unique.  To emphasize this point, note that any such coordinates are naturally associated with the conformal frame where the metric on ${\cal B}_d$ is just $h_{\mu \nu}(x) = h_{\mu \nu}(0,x)$.  One may transform to similar Fefferman-Graham coordinates associated with any other conformal frame (i.e., with $h_{\mu\nu}(x) \to h_{\mu\nu}(x)/\phi_0(x)$) by introducing $z' = z \,\phi(x, z)$ for appropriate $\phi(x,z)$ satisfying $\phi(x,0) = \phi_0(x)$ and redefining the $x^\mu$ so as to preserve the form \eqref{FGmet}.

Solving \eqref{bEOM} order by order in $z$ yields an asymptotic series \cite{FG}
\begin{align}
\label{FGseries}
h_{\mu\nu}(z,x) &= {h}_{\mu\nu} + z^2 \,{h_{(2)}}_{\mu\nu}  + z^4\, {h_{(4)}}_{\mu\nu}  + \ldots + z^{2 n}\, {h_{(2 n)}}_{\mu\nu}  + \ldots \nonumber \\
& \qquad \qquad +\; z^d \,t_{\mu\nu}  + z^d\, \log{z} \;{h_{(d)}}_{\mu\nu}+ {\cal O}( z^{d+1} ) \,,
\end{align}
where $h_{(i)}$ (including $h_{(d)}$) vanishes unless $i$ is even.  Here the terms $h_{\mu\nu}(x)$ and $t_{\mu\nu}(x)$ provide sufficient data to determine all other terms in the expansion.   In particular,
the terms ${h_{(2 n)}}_{\mu\nu}$ for $0 < n \le d/2$ are local tensors on ${\cal B}_d$ built entirely from $h_{\mu\nu}(x)$ and its derivatives.   The terms going as $z^p$ for $p>d$ depend on both $h_{\mu\nu}(x)$ and $t_{\mu\nu}(x)$. It is natural to regard $h_{\mu\nu}$ as Dirichlet data for the bulk solution and $t_{\mu\nu}$ as Neumann data, associated with normal derivatives.  One should keep in mind that \eqref{FGseries} is generally only an asymptotic series which need not give the exact solution (or even converge) at any $z > 0$.

The asymptotic equations of motion impose two constraints on  $h_{\mu\nu}$ and $t_{\mu\nu}$, which are just radial analogues of the well-known Hamiltonian and momentum constraints of general relativity.  These take the form
\begin{equation}
\label{htconstr}
h^{\mu \nu} t_{\mu \nu} =0, \ \quad \nabla^\mu t_{\mu \nu} = 0
\end{equation}
where $\nabla_\mu$ is the covariant derivative on ${\cal B}_d$ compatible with $h_{\mu \nu}$ and where we use $h_{\mu \nu}$ to raise and lower Greek indices.  Both constraints will play important roles below.  These are the only constraints within the context of the asymptotic expansion \eqref{FGseries} but, just as providing both Dirichlet and Neumann data on a timelike surface fails to define a good boundary value problem for the wave equation, one expects that no simple relations between $h_{\mu \nu}$ and $t_{\mu \nu}$ can guarantee existence of a smooth bulk.  Indeed, in Euclidean signature with compact ${\cal B}_d$ we have already noted that $h_{\mu \nu}$ alone should suffice to determine the entire bulk solution -- and thus in particular $t_{\mu \nu}$ -- up to a possible discrete choice.

The expansion \eqref{FGseries} allows a convenient presentation of the so-called boundary stress tensor $T_{\mu \nu}$, which is dual to the vacuum expectation value $\vev{T^{CFT}_{\mu\nu}}$ of the stress tensor for the CFT.  In fact, in our bulk classical limit the probability distribution for $T^{CFT}_{\mu\nu}$ is sharply peaked around its mean so that we may treat $T^{CFT}_{\mu\nu}$ as a c-number.  We therefore drop the angle brackets $\vev{}$ and henceforth identify $T_{\mu \nu} = T^{CFT}_{\mu\nu}$.  Using the equality of bulk and boundary path integrals, one may compute $T_{\mu \nu}$ by varying the bulk partition function with respect to $h_{\mu\nu}$.  Evaluating the result using the saddle point method gives just $T_{\mu \nu} = -\frac{2}{\sqrt{|h|}} \frac{\delta S_{bulk}}{\delta h^{\mu \nu}}$ for the appropriate $S_{bulk}$ (see footnote \ref{holreg}).  The result takes the simple but dimension-dependent form\footnote{ One may alternatively express $T_{\mu \nu}$ in terms of the intrinsic geometry and extrinsic curvature $K_{\mu\nu}$ of ${\cal B}_d$ \cite{Balasubramanian:1999re} which gives a more covariant and universal expression
$$
T_{\mu\nu} = \bigg\{ \frac{r^d}{8\pi G_{d+1}}\, \left[-\,K_{\mu\nu} + \, K\, h_{\mu\nu} -\,(d-1)\, h_{\mu\nu} + \frac{1}{d-2}\left( ^{\,h\!}R_{\mu\nu}-\frac{1}{2}\, ^{\,h\!}R\, h_{\mu\nu}\right)+ \cdots \right] \bigg\}_{r\to \infty} \,.
$$
related to the quasi-local stress tensor introduced at finite boundaries by Brown and York \cite{Brown:1992br}.}

\begin{align}
\label{bst}
	T_{\mu\nu} =d\, c_\text{eff}\; \left( t_{\mu\nu} + C_{\mu\nu}[ h ] \right) \,,
\end{align}
where $C_{\mu\nu}$ is a local divergence-free tensor built from $h_{\mu\nu}$  and derivatives with the property that $C_{\mu \nu}$ vanishes for Ricci-flat ${\cal B}_d$. In particular, for fixed boundary metric $h_{\mu \nu}$ the bulk stress tensor depends on the interior solution only through $t_{\mu \nu}$. A key point to note is that the expectation value of this stress tensor must be conserved
\begin{equation}
\nabla_\mu T^{\mu\nu} = 0 \,,
\label{consT}
\end{equation}	
due to \eqref{htconstr}. This provides as an important restriction on solutions of Einstein's equations \eqref{bEOM}.

The details of $C_{\mu \nu}$ are dimension-dependent, though for $d=2,4$ we have
\begin{align}
\label{eq:C}
d=2: && \!\! C_{\mu \nu} &= -\frac{1}{4}\; {}^{\,h\!}R\; h_{\mu \nu} \,,
\cr
d=4: && \!\! C_{\mu \nu} &= - \frac{1}{8} h_{\mu\nu} \left( \left( \mathrm{Tr} h_{(2)} \right)^2 -   \mathrm{Tr} h_{(2)}^2 \right) - \frac{1}{2} \left( h_{(2)}^2 \right)_{\mu\nu} + \frac{1}{4} {h_{(2)}}_{\mu\nu}  \mathrm{Tr} h_{(2)} \,,
\end{align}
where ${h_{(2)}}_{\mu\nu} = \frac{1}{2} \left( ^{\,h\!}R_{\mu\nu}  - \frac{1}{6} ^{\,h\!}R\, {h}_{\mu\nu} \right)$ and ${}^{\,h\!}R_{\mu\nu}$, ${}^{\,h\!}R$ denote the Ricci tensor and scalar on ${\cal B}_d$.

This $C_{\mu\nu}$ term arises from the the log term $h_{(d)}$ in the expansion \eqref{FGseries}; hence it vanishes for odd $d$. It is related to the well known CFT conformal anomaly that may occur in even dimensions.  Although the trace of the stress tensor vanishes at the classical level for any CFT, at the quantum level on a curved spacetime with even $d$ this anomaly gives a non-zero trace related to the local geometry of ${\cal B}_d$.  The bulk dual of this statement is just that by \eqref{htconstr} the trace of $T_{\mu  \nu}$ is given by $h^{\mu \nu} C_{\mu \nu}$. It is this result that motivates the definition \eqref{ceffdef}, where for uniformity of presentation we have extrapolated conventional definitions in terms of the even $d$ conformal anomaly to all integers.  The fact that \eqref{bst} is proportional to $c_\text{eff}$ implies that bulk classical gravity is a good probe of CFT physics which locally involves $\sim {\cal O}(c_\text{eff})$ CFT degrees of freedom. Phenomena that deform the stress tensor by amounts of lesser order in $c_\text{eff}$ cannot be resolved by classical bulk gravity and correspond to quantum or stringy effects in the bulk.

\subsection{Simple bulk geometries: CFT vacuum and thermal state}
\label{sec:vacuathermal}

We now apply this technology to a few simple examples.  For Minkowski ${\cal B}_d$ it is natural to study the zero temperature vacuum.  As described earlier this is dual to \eqref{padsX}, which is already in Fefferman-Graham form and gives $t_{\mu \nu} =0$. Since ${\cal B}_d$ is Ricci flat, we also have $C_{\mu\nu}[h]=0$  and the boundary stress tensor vanishes.

At finite temperature we may consider the the planar Schwarzschild-AdS$_d$ solution
 \begin{align}	
 \label{eq:PoincareSchwarz}
ds^2 = \ell^2 \left( - r^2\,f(r)\, dt^2 + \frac{1}{r^2\, f(r)}\, dr^2 + r^2\, d{\bf x}^2 \right)  , \quad f(r) = 1 - \left( \frac{r_h}{r} \right)^d,
\end{align}
where $r_h$ specifies the position of the bulk horizon and is related to the Hawking temperature via
\begin{equation}
T = \frac{d}{4 \pi}\, r_h.
\label{planarT}
\end{equation}	
To compute the boundary stress tensor we merely write \eqref{eq:PoincareSchwarz} in Fefferman-Graham form by introducing an appropriate coordinate $z = z(r)$.  The result is
 \begin{align}
ds^2 = \frac{\ell^2}{z^2} \left( dz^2 + \left( - 1 + \frac{d-1}{d} \, r_h^d \,z^d + \ldots \right) dt^2 + \left( 1 + \frac{1}{d} \, r_h^d\, z^d+ \ldots \right) d{\bf x}^2  \right) ,
\end{align}
which yields the boundary stress tensor
\begin{align}
\label{eq:stresspoincare}
T_{\mu\nu} \,dx^\mu dx^\nu  =  c_\text{eff}\, \left( \frac{4 \pi\, T }{d} \right)^d \bigg[ (d-1) \,dt^2 + d{\bf x}^2 \bigg]
\end{align}
in terms of the temperature \eqref{planarT} and the effective central charge \eqref{ceffdef}. We note that \eqref{eq:stresspoincare} is the stress tensor for a conformally-invariant perfect fluid.  In particular, the stress tensor scales homogeneously as $T^4$.  This may be seen from the fact that thermal CFT states at any two non-zero temperatures are related by the action of the dilation symmetry.  For the bulk solutions \eqref{eq:PoincareSchwarz}, the analogous statement is that, for any $\lambda$, changing  $r_h \to \lambda^{-1} r_h$ in \eqref{eq:PoincareSchwarz}  is equivalent to the coordinate transformation
\begin{align}
r \to \lambda^{-1} \, r \; , \quad t  \to \lambda \,t \; , \quad x^i \to \lambda \,x^i .
\end{align}

It is instructive to perform the same calculation for ${\cal B}_d = {\mathbb R} \times {\bf S}^{d-1}$.  At finite temperature we may then take the bulk to be the global Schwarzschild-AdS$_{d+1}$ solution,
\begin{align}
\label{globAdSS}
ds^2 = - g(\rho)\, dt^2 + \frac{d\rho^2}{g(\rho)} + \rho^2 \,d\Omega^2_{d-1} \,, \quad g(\rho) = \frac{\rho^2}{\ell^2} + 1 - \left( 1 + \frac{\rho_h^2}{\ell^2} \right) \left( \frac{\rho_h}{\rho} \right)^{d-2},
\end{align}
where $\rho_h$ now gives the radius of the horizon. The Hawking temperature is
\begin{align}
T = \frac{d\, \rho_h^2 + (d-2)\, \ell^2}{4\pi\, \rho_h\, \ell^2} \,.
\label{globalT}
\end{align}
Transforming to Fefferman-Graham coordinates yields,
 \begin{align}
 \label{eq:globalAdScalc}
ds^2 & =  \frac{\ell^2}{z^2} \Big[ dz^2
- \left(  1 + \frac{1}{2} \frac{z^2}{\ell^2} +\frac{1}{16} \frac{z^4}{\ell^4} +
\frac{d-1}{d}  \left( 1 +  \frac{\ell^2}{\rho_h^2} \right) \frac{\rho_h^d\, z^d}{\ell^{2 d}} + \ldots \right) dt^2 \nonumber \\
& \qquad \qquad \quad
 + \left( 1  - \frac{1}{2} \frac{z^2}{\ell^2}  + \frac{1}{16} \frac{z^4}{\ell^4} + \frac{1}{d}  \left( 1 +  \frac{\ell^2}{\rho_h^2} \right) \frac{\rho_h^d \,z^d}{\ell^{2 d}} + \ldots \right) \ell^2 \,d\Omega^2_{d-1}  \Big].
\end{align}
The leading term of each component shows that the boundary metric is $h_{\mu\nu} = -dt^2 + \ell^2\, d\Omega^2_{d-1}$, so the spatial section of the ESU$_d$ has radius $\ell$.

Evaluating the boundary stress tensor requires us to compute the dimension-dependent contribution from $C_{\mu\nu}[ h ]$ \eqref{eq:C}.  For $d=4$ one obtains,
\begin{align}
\label{eq:stressglobal4d}
T_{\mu\nu} \,dx^\mu dx^\nu =  \frac{c_\text{eff}}{ \ell^4} \left( \frac{\rho_h^2}{\ell^2} + \frac{1}{2}  \right)^2 \bigg[ 3\, dt^2 +  \ell^2 \,d\Omega^2_{3} \bigg].
\end{align}
We note that for empty global AdS (i.e., $\rho_h = 0$) the stress tensor does not vanish but instead represents a Casimir energy for the CFT on ESU$_4$.

If we wish to study only the behaviour of the CFT on a fixed ${\cal B}_d$ and we are not interested in comparing results for different metrics, it may be convenient to subtract off the contribution from $C_{\mu \nu}[h]$.  Denoting the result $T^{sub}_{\mu\nu}$, we may compute the answer directly from the $z^d$ terms in Eq.~\eqref{eq:globalAdScalc} to find
\begin{align}
\label{eq:stressglobal}
T^{sub}_{\mu\nu} \,dx^\mu dx^\nu  = \frac{c_\text{eff}}{\ell^{d}} \left( 1 +   \frac{\rho_h^2}{\ell^2} \right) \left( \frac{\rho_h}{\ell} \right)^{d-2} \bigg[ (d-1)\, dt^2 +  \ell^2 d\Omega^2_{(d-1)} \bigg]
\end{align}
We analogously define the subtracted total energy $E^{sub}$ and the subtracted free energy $F^{sub}$ respectively.

%
\section{Thermal CFT on spatially compact geometries}
\label{sec:3}
%

We now discuss the simplest examples of non-trivial (i.e., not Minkowski) spacetimes ${\cal B}_d$.  We begin with ${\cal B}_d = {\mathbb R} \times {\bf S}^{d-1}$ and thus bulks asymptotic to global AdS \eqref{gadsX}.   In this context the CFT exhibits a phase transition associated with colour confinement, and the dual bulk theory provides a beautiful geometrization as the Hawking-Page transition of black holes in global AdS$_{d+1}$. This is a feature of the canonical ensemble of the universal gravity sector and is thus common to all holographic CFTs at finite temperature.

We then move on to discuss a more subtle but equally basic example, where one of the spatial directions is compactified into a circle (with particular boundary conditions imposed on fermions). On such spacetimes the CFT exhibits a more complex behaviour associated with confinement, again elegantly geometrized by various exotic black holes called `plasma balls'. This physics is again universal.

Finally, we discuss two non-universal phenomena which go beyond the $(d+1)$-gravity sector. The first concerns phase transitions in the micro-canonical ensemble due to localization on the internal space $X$. Whilst qualitatively similar behaviour may be expected for all $X$,  the details of the localization will depend on both the full $D = d+1 + \rm{dim}(X)$-dimensional supergravity theory and the properties of $X$.   The second phenomenon concerns  transitions where the spatial Minkowski directions are compactified into a torus (with opposite fermion boundary conditions to the circle case above). In this case a phase transition occurs in both canonical and micro-canonical ensembles, now associated with a stringy instability in the bulk dual and providing an example of non-universal phenomena going beyond the supergravity approximation of the CFT dual.

%
\subsection{Thermal CFT on the Einstein Static Universe}
\label{sec:HP}
%

We noted in \S\ref{sec:vacuathermal} that thermal states of a CFT on Minkowski space at any two non-zero temperatures are related by the action of a dilatation symmetry.  As a result, there can be no phase transitions in the canonical at any $T > 0$, or in the micro-canonical ensemble for $E > 0$.\footnote{\label{planarphases} For $T > 0$ there are in fact two relevant solutions, given by planar  Schwarzschild-AdS \eqref{eq:PoincareSchwarz} and Poincar\'e AdS \eqref{padsX} filled with a thermal gas (which produces no back-reaction in the bulk semi-classical limit).  In the Euclidean formulation, the latter is described by Wick rotating $t \rightarrow i \tau$ in
\eqref{padsX} and then taking $\tau$ to have period $\beta = 1/T$.  One may thus say that the theory has two phases for $T>0$, though the planar Schwarzschild-AdS phase always dominates the ensemble.}  However, the situation is very different if another scale is introduced.  An interesting case occurs when the CFT is placed on ${\mathbb R} \times {\bf S}^{d-1}$.  As we have seen, this corresponds to bulks asymptotic to global AdS${}_{d+1}$.
We take the sphere to have radius $\ell$, whence the boundary and bulk timelike Killing vector fields with standard normalizations coincide.\footnote{ By the dilation action we can rescale the sphere radius at the expense of rescaling the boundary time-translation generator relative to the bulk Killing field.}
Due to scale invariance the actually value of this radius is not of physical interest, though the dimensionless product of temperature and radius is a physical coupling that can be dialed to change the behaviour.  We now consider the Hawking-Page phase transition that occurs in the canonical ensemble for the universal $(d+1)$-dimensional bulk gravity sector as this coupling is varied. We emphasize this is a universal phenomena and thus will occur for all holographic CFTs, the details depending only on the dimension $d$.

\subsubsection{The Hawking-Page transition in gravity}
\label{sec:hpgrav}

We are interested in bulk solutions that might be dual to thermal CFT states on ESU$_d$.  We expect such thermal states to static and spherically symmetric, so it makes sense to seek bulk solutions with the same symmetries.  These are easily classified using the AdS Birkhoff's theorem.  Since we work in vacuum, they are just the global Schwarzschild-AdS$_{d+1}$ \eqref{globAdSS}. The empty global AdS case $\rho_h =0$ is special and we will treat it separately.

We wish to fix the temperature $T$.  We see from  \eqref{globalT} that black holes (with $\rho_h > 0$) have a minimum temperature
\begin{align}
 T_{min} = \frac{ \sqrt{d \,( d - 2)} }{ 2 \pi \,\ell },
\end{align}
corresponding to a horizon size $\rho_{h0} = \ell \sqrt{(d-2)/d}$.
Below this temperature non-rotating bulk black holes do not exist within our universal sector.   Moreover, $\rho_h(T)$ is double-valued.  Thinking in terms of the Euclidean path integral, we have in fact two black hole saddle points at any given temperature $T>T_{min}$. These are the {\em large} and {\em small} Schwarzschild-AdS black holes, with $\rho_h > \rho_{h0}$ and $\rho_h < \rho_{h0}$ respectively. One can observe that the specific heat capacity of the small black holes is always negative, whereas for the large black holes it is always positive.

For $T< T_{min}$ we should look for a new saddle point to the bulk equations of motion. The remaining obvious candidate is just global AdS itself.  Note that this solution may be assigned any temperature $T$, as the corresponding Euclidean solution (known as thermal AdS) remains smooth when the time direction is compactified with any period $\beta$.  A key difference between thermal AdS and our black holes is that thermal AdS  has (subtracted) energy  $E^{sub} = 0$. This statement should be read to mean  $E \sim {\cal O}(1)$ in the large $c_\text{eff}$ limit, since we are working in the semi-classical gravity approximation of the bulk theory.

Armed with the three saddle points of thermal AdS, large, and small Schwarzschild-AdS black holes, we are in a position to discuss the phase structure.  In the canonical ensemble, for all temperatures below $T < T_{min}$ we have only thermal global AdS$_d$, and its (subtracted) free energy vanishes. On the other hand, we may compute the subtracted energy $E^{sub}$ of global Schwarzschild-AdS$_{d+1}$ by integrating the $T_{tt}$ component of \eqref{eq:stressglobal} over the sphere. The entropy of the black hole is of course given by the Bekenstein-Hawking formula:
\begin{align}
E^{sub} =  c_\text{eff} \;\frac{( d-1 ) \,\omega_{d-1}}{\ell} \left( 1 +   \frac{\rho_h^2}{\ell^2} \right) \left( \frac{\rho_h}{\ell} \right)^{d-2}  \,, \qquad S = \frac{ \rho_h^{d-1} \;\omega_{d-1} }{ 4 G_{d+1} }
\end{align}
where $\omega_{d-1}$ is the area of an unit ${\bf S}^{d-1}$. Knowing the Hawking temperature of the black hole \eqref{globalT} we compute the (subtracted) free energy of Schwarzschild-AdS$_{d+1}$
\begin{align}
\label{eq:globalfreeerg}
F^{sub} =  c_\text{eff} \;  \frac{ \omega_{d-1}}{\ell} \left( 1 -   \frac{\rho_h^2}{\ell^2} \right) \left( \frac{\rho_h}{\ell} \right)^{d-2}.
\end{align}
The result \eqref{eq:globalfreeerg} is negative for $\rho_h > \ell$ but becomes  positive for $\rho_h < \ell$.
In particular, the relation $\ell > \rho_{h0}$ implies that
\eqref{eq:globalfreeerg} is positive at $T_{min}$ and that thermal AdS continues to dominate the canonical ensemble even a bit above this threshold.

However, the large Schwarzschild-Ads black hole dominates for $\rho_h > \ell$, indicating a first order phase transition at temperature
\begin{eqnarray}
T_{HP} = \frac{d-1}{2 \pi \ell} \, ;
\end{eqnarray}
this is  the Hawking-Page transition \cite{Hawking:1982dh, Witten:1998zw}.  The
order parameter for this transition can be simply taken to be $F/c_\text{eff}$ (or alternatively the entropy $S/c_\text{eff}$): this quantity vanishes at low temperatures  $T < T_{HP}$ and
is negative definite for $T>T_{HP}$. The small Schwarzschild-AdS black hole is always a subdominant saddle in the canonical ensemble.\footnote{ Note that by taking the limit $\rho_h \gg \ell$ we recover the behaviour of the planar Schwarzschild-AdS black hole. As a result we recover our earlier statement that planar black holes correspondingly always dominate the canonical ensemble over thermal AdS for CFT on ${\mathbb R}^{d-1,1}$. }

The above analysis above applies for $d \ge 4$. In $d = 3$ we must consider the BTZ \cite{Banados:1992wn} black hole for which there is no analog of the small black hole solution. Nevertheless, the qualitative nature of the Hawking-Page phase transition remains unchanged \cite{Birmingham:2002ph}.

The phase structure in the micro-canonical ensemble is a bit more involved and non-universal - it depends on the details of the internal space $X$. We refer the reader to \cite{Horowitz:1999uv,Aharony:2003sx} for discussions and simply note here that now the small Schwarzschild-AdS black holes play a more prominent role since they have positive entropy of order $c_\text{eff}$. As we review in \S\ref{sec:globalGL}, The Schwarzschild-AdS$_{d+1}$ geometries turn out to dominate this ensemble for $\rho_h > \rho_{hX}$  (or equivalently $E> E_X $) \cite{Hubeny:2002xn}.  At lower energies the dominant solutions turn out to be $D$-dimensional black holes localized on the internal space $X$. And at least in string theoretic constructions the dominant solution at  ultra-low energies is always a gas of strings.

%
\subsubsection{Confinement/deconfinement in the CFT}
\label{sec:confdeconf}
%

The Hawking-Page transition described above has a very nice interpretation in gauge field theories as the confinement/deconfinement transition \cite{Witten:1998zw}.

Consider for a moment pure $SU(N)$ Yang-Mills rather than our CFT whence $c_\text{eff} \sim N^2 $.  In pure Yang-Mills precisely such a phase transition occurs as one varies the temperature. At high temperatures the Yang-Mills is weakly coupled, due to asymptotic freedom, and the high energy behaviour is heuristically governed by the ${\cal O}(N^2)$ gluons. Hence the energy density goes as $\sim {\cal O}(N^2)$. However, at low temperature the gluons become strongly coupled, and the low energy behaviour is dominated by colour singlet composites of gluons, the so called glue balls. Being colour singlets these number ${\cal O}(1)$ in $N$ counting. Correspondingly the energy density is of $\sim {\cal O}(1)$.

For conformal gauge theories one does not encounter any phase transition on ${\mathbb R}^{d-1,1}$. However, on a spatial sphere we expect a potential phase transition when the curvature scale $\ell$ is commensurate with the temperature. In the strongly coupled limit, for large $c_\text{eff} \sim N^2 $ the gravitational Hawking-Page transition provides the realization of this phenomenon. In fact, one can ascertain the existence of a similar phase transition in free and weakly coupled large $c_\text{eff}$ CFTs \cite{Sundborg:1999ue,Aharony:2003sx}. One expects that the Hawking-Page transition is a strong coupling continuation of this weak coupling deconfinement phase transition.\footnote{An important distinction from the physics of deconfinement in flat space is that whereas pure Yang-Mills exhibits a phase transition with temperature at finite $N$, here the Hawking-Page transition in the CFT on a sphere only occurs at infinite $N$.
This is necessary to achieve the thermodynamic limit on a compact spatial volume.}

Apart from the large $N$ order parameter $F/N^2$, a more conventional order parameter for confinement in Minkowski spacetime is the energy of a probe quark placed into the vacuum. We expect this is infinite in a confining phase, but finite in a deconfined phase. From the Euclidean perspective this probe quark energy, $E_q$, is measured by considering a Wilson loop wrapping around the Euclidean time circle (the Polyakov loop), through
\begin{align}
 \big\langle |  \Tr W | \big\rangle \equiv
 \bigg\langle  \left| \frac{1}{N} \Tr{ P e^{i \oint \,A_\tau d\tau} } \right| \bigg\rangle
 \sim  e^{- \beta \,E_q}
\end{align}
Thus for Minkowski we expect that in the confining phase $ \langle | \Tr W | \rangle=0$, whereas for the deconfined phase $ \langle | \Tr W | \rangle\ne0$.
In AdS/CFT such a Wilson loop\footnote{ More precisely, in $\mathcal{N} = 4$ this construction computes the `Maldacena loop', which includes contribution from adjoint matter in the field theory.}
 is estimated by considering a string world-sheet wrapping the loop in the boundary (here the Euclidean time circle) and extending smoothly into the bulk \cite{Maldacena:1998im}. Then if a smooth world-sheet satisfying the classical string equations exists for the loop in the boundary then we have,
 \begin{align}
 \label{eq:Polyakov}
 \langle | \Tr W | \rangle \simeq e^{- S_{classical}}
\end{align}
where $S_{string}$ is the classical string action.\footnote{ Note that care must be taken to define this action which naively diverges.} If no classical smooth world-sheet exists, it indicates $ \langle | \Tr W | \rangle \simeq 0$ \cite{Witten:1998zw}.
This discussion makes the physical significance of the Euclidean time circle toplogy in  the bulk clear. If this circle is contractible, the Polyakov loop operator on the boundary can be estimated by considering the string world-sheet which wraps the circle and smoothly closes off in the bulk  (this is a two dimensional minimal surface). Hence for the Schwarzschild-AdS spacetime where the thermal circle is contractible, we expect $ \langle | \Tr W | \rangle\ne0$ corresponding to a deconfined phase. However, for the thermal AdS geometry which has non-contractible time circle, by topology there can not exist such a smooth classical world-sheet, and correspondingly we expect $\langle | \Tr W |   \rangle \simeq 0$; the phase is confining.

%
\subsection{Thermal CFTs on a spatial circle with anti-periodic fermions}
\label{sec:circleap}
%

We now consider CFT on flat spacetime with one or more of the spatial directions  compactified in to a circle. An important decision that we must make for the fermions in the CFT is whether to choose periodic or anti-periodic (`Scherk-Schwarz') boundary conditions. This choice is correlated with that for bulk fermionic fields.
In this section we consider anti-periodic boundary conditions. Later in  \S\ref{sec:periodic} we consider the periodic case. As we shall see, the phenomenology is quite different in these two cases. In particular, the present discussion will focus on the universal $(d+1)$-gravity sector.

Consider for simplicity ${\cal B}_d = {\mathbb R}^{d-2,1} \times {\bf S}^1_{SS}$,
with  circle radius  $R$, and anti-periodic boundary conditions for the fermions (denoted by the $SS$ subscript). At finite temperature a potential bulk solution is still the planar Schwarzschild-AdS solution which for Euclidean time circle $\tau =i\,t$ is
 \begin{align}
 \label{eq:PoincareSchwarz2}
ds^2 = \frac{\ell^2}{z^2} \left( f(z) \,d\tau^2 + \frac{1}{f(z)} \, dz^2 + d{
\bf x}^2 + d\theta^2 \right)  \; , \quad f(z) = 1 - \left( \frac{z}{z_h} \right)^d\,.
\end{align}
The period of Euclidean time  $\tau \sim \tau + \beta$ is given by the inverse temperature $\beta = \frac{4\pi\,z_h}{d}$.  Now ${\bf x}$ denotes non-compact field theory coordinates on ${\mathbb R}^{d-2}$, and $\theta \sim \theta + R$ is the Scherk-Schwarz circle.

Using the Euclidean geometry to study equilibrium finite temperature, we immediately see that the spatial Scherk-Schwarz circle and the Euclidean time circle, which crucially also has anti-periodic boundary conditions for fermions, are almost symmetric. The only difference is the period with which they are identified. For CFTs once again the dimensionless ratio $\beta/R$ will control the physics of this system. This near symmetry indicates a different possible smooth bulk solution with the roles of $\tau$ and $\theta$ reversed;
 \begin{align}
 \label{eq:soliton}
ds^2 = \frac{\ell^2}{z^2} \left( d\tau^2 + \frac{1}{f(z)} dz^2 + dx_a^2  + f(z)
\, d\theta^2 \right)  \; , \quad f(z) = 1 - \left( \frac{z}{z_{IR}} \right)^d
\end{align}
where now $R = \frac{4 \pi\, z_{IR}}{d}$ to ensure the circle closes smoothly. Thus this solution exists for any $\beta$, and with $z_{IR}$ being determined by $R$. This solution is the AdS-soliton, and was proposed by Witten (in one dimension higher) as an early holographic model for confinement \cite{Witten:1998zw}.

Fixing $R$, these two possible bulk metrics both co-exist for any temperature $T$. They are distinguished by the topology of the Euclidean solution. The    black hole bulk has a contractible Euclidean time, but the spatial circle is not contractible. The soliton has no horizon but the spatial circle smoothly contracts, cutting off the geometry in the IR -- hence leading to a mass gap and confining properties \cite{Witten:1998zw}. A very important point is that if we had taken periodic boundary conditions for the fermions about the spatial circle, then the soliton solution would be disallowed as no spin structure could be defined on the manifold. Thus the soliton bulk solution is only applicable to cases where the fermions are anti-periodic.

In addition to these two bulk metrics, we may also consider thermal AdS, i.e., Poincar\'e-AdS with the $\theta$ coordinate compactified.
At finite temperature we must compare the thermodynamics of these three possibilities  to find the one that dominates the bulk partition function. The stress tensor for thermal AdS vanishes, and hence its free energy density vanishes. Using the stress tensor for the planar black hole \eqref{eq:stresspoincare} and its entropy we obtain a free energy density
\begin{align}
f_{bh} = - c_\text{eff}\, \ell \left( \frac{4 \pi\, T}{d} \right)^{d} \,.
\end{align}
For the soliton the stress tensor can be written down immediately by interchanging $\tau \leftrightarrow \theta$ and taking $z_h \to z_{IR}$ (equivalently $\beta \to R$)
\begin{align}
T^{soliton}_{\mu\nu} \,dx^\mu dx^\nu  =   c_\text{eff} \left( \frac{4 \pi  }{d \, R} \right)^d \left( d\tau^2 + dx_a^2 - (d-1) \,d\theta^2 \right)
\end{align}
Noting that the entropy vanishes (to ${\cal O}(c_\text{eff})$) as there is no horizon,\footnote{ The entropy of thermal graviton fluctuations will contribute at ${\cal O}(1)$.} at large $c_\text{eff}$ we have the free energy density
\begin{align}
f_{soliton} = -c_\text{eff}\, \ell \left( \frac{4 \pi }{d \,R} \right)^{d}\,.
\end{align}

We note that thermal AdS never dominates the other two saddle points.
For temperatures $T < 1/R$ it is the soliton that dominates with the lower free energy, whereas for $T > 1/R$ the black hole dominates. Thus in close analogy with the CFT on a sphere, in the presence of a Scherk-Schwarz circle we have only a confining low temperature phase and a deconfined high temperature phase. The phase transition occurs at temperature  $T_c = 1/R$ and is of first order as the saddle points which dominate in either phase are always distinct from each other. The fact that this first order phase transition occurs precisely at the temperature $T_c$ of course follows from the fact that the two Euclidean saddle points are isometric at that point.

The stress tensors above are normalized to vanish for the vacuum state on Minkowski spacetime. For the soliton, the non-vanishing stress tensor can again be interpreted as a (negative) Casimir energy. However, we may think of an effective confining $(d-2)+1$ QFT obtained by reducing the CFT on the $\theta$ circle.
We term this the QFT$_{d-1}$ theory.
This theory will have a mass gap $\Lambda_g \sim 1/R$, whose confining vacuum state is dual to the  soliton.\footnote{ It is important to note that this reduction to a QFT$_{d-1}$ is formal since the mass scale of the Kaluza-Klein modes on the circle is of order the gap scale, so there is no parametric separation of scales. This has important ramifications for using such a setting to model the phenomenology of confining gauge theories (see for example \cite{Witten:1998zw,Mandal:2011ws}).}
This gap is physical, controlling the spectrum of graviton fluctuations about the soliton, and also controlling the corresponding exponential decay of correlation functions for points separated by distances much greater than $R$ in ${\bf x}$.
Making the appropriate stress tensor subtraction then yields a vanishing result for the soliton, and gives
\begin{align}
\label{eq:subT}
T^{sub,bh}_{\mu\nu} \;dx^\mu dx^\nu  &=  c_\text{eff}\, \left(\frac{ 4 \pi }{d}\right)^d
\left[ - \left( (d-1) T^d +  \frac{1 }{ R^d} \right) d\tau^2  + \left(  T^d +  \frac{(d-1) }{ R^d} \right)   d\theta^2 \right.
\nonumber \\
& \left.
\hspace{3cm} + \;  \left(  T^d - \frac{1 }{ R^d} \right) d{\bf x}^2
\right]
\end{align}
for the deconfined state at temperature $T$. Notice that at the transition temperature $T_c$ the pressures in the ${\bf x}$ directions vanishes. We define $\epsilon_c = c_\text{eff} \left( 4 \pi / R \right)^d / d^{d-1}$ to be the energy density in the deconfined state at the transition temperature.

Finally, note that if more than one spatial direction is compactified (e.g., ${\cal B}_d = {\mathbb R}\times T^{d-1}_{SS}$ with anti-periodic boundary conditions on all cycles), then the same behaviour occurs, with the thermodynamically preferred contractible cycle (including the thermal Euclidean circle) being the one with smallest size. One also expects similar behaviour for more general compact factors; see e.g., \cite{Copsey:2006br}.

\subsubsection{A domain wall between confined and deconfined plasmas}

Given the confined and deconfined phases of the plasma on ${\mathbb R}^{d-2,1} \times {\bf S}^1_{SS}$, the geometric exchange symmetry between the thermal ($\tau$) and spatial ($\theta$) circles at $T=T_c$ leads us to suspect that a domain wall solution exists \cite{Aharony:2005bm}.
Let ${\bf x} = \{x, {\bf y}\}$ to single out a spatial direction normal to such a domain wall which we take to be translationally invariant in the ${\mathbb R}^{d-3}$ parameterized by ${\bf y}$.
The solution of interest is one where for $x \to +\infty$ one asymptotes to the black hole at temperature $T = T_c$, and for $x \to - \infty$ the solution asymptotes to the soliton vacuum, again at temperature $T = T_c$.
Such a solution would have the isometries generated by $\partial/\partial \tau$, $\partial/\partial \theta$ and $\partial/\partial {\bf y}$ with the metric nontrivially depending on the direction normal to the domain wall, $x$, and on the bulk radial coordinate.
We also expect the Euclidean solution would then have a discrete $\mathbb{Z}_2$ symmetry under the interchange $x \to -x$ together with $\tau \leftrightarrow \theta$. Without loss of generality we can pick $x = 0$ to be the surface where both the Euclidean time and space circles simultaneously contract.

The domain wall solution described above should have a characteristic thickness $R$.  To see this, note that
in the vacuum asymptotic region the theory is confining with mass gap $\Lambda_g \sim 1/R$. Hence we expect the CFT correlators decay as $\sim e^{ - r/R}$ for a spatial separation $r$ in the non compact $x, {\bf y}$ directions. Likewise in the thermal asymptotic region, the temperature scale $T = 1/R$ yields a thermal mass, and again correlators will decay in a similar manner. Thus a complicated transition region will exist in the CFT and bulk around $x = 0$, but only with a width of order $\sim R$, and then the solution will exponentially quickly return to the two asymptotic behaviours as one moves away from $x = 0$.

A natural concern might be that whilst it is clear that in each asymptotic region one or other circle contracts smoothly in the IR, at $x = 0$ both must do so coincidentally, and can this occur in a smooth manner? In fact this situation is ubiquitous in toric geometry. In the bulk near this point we have two circles $\theta$ and $\tau$ which shrink on co-dimension one surfaces that coincide at $x=0$. We may pick coordinates $a$ and $b$ in the local neighbourhood of this intersection, which we view as being composed of a bulk radial coordinate and the field theory direction $x$. We may ignore the ${\bf y}$ part of the metric here as it will be smoothly fibered over the rest of this geometry. Then locally the relevant part of the smooth geometry is given by,
\begin{align}
ds^2_{local} \simeq da^2 + \frac{4 \pi^2 \,a^2}{\beta^2}\, d\tau^2 + db^2 + \frac{ 4 \pi^2 \,b^2}{R^2}\, d\theta^2
\end{align}
which we see is simply $\mathbb{R}^4$ written in double polar coordinates. On the surface $a = 0$ lies the horizon, and on the surface $b = 0$ is the closure of the Scherk-Schwarz circle. The intersection is simply the origin of this double polar coordinate system.

The conservation  \eqref{consT} of the boundary stress tensor provides an important constraint. The boundary metric $h_{\mu\nu}$ is flat and in particular $\partial/\partial x$ is a boundary isometry. This vector field is however not a bulk isometry and hence $\nabla^\mu T_{\mu\nu} = 0 \Longrightarrow \partial_x T_{x\mu} = 0$  gives a non-trivial constraint on the stress tensor, and hence the bulk geometry. Due to the symmetries the boundary stress tensor will be diagonal, so this condition implies $T_{xx} = \mathrm{constant}$, i.e., the pressure normal to the wall is constant across the domain wall.

Far from the domain wall, this constraint implies that the pressures in the non-compact $x,{\bf y}$ directions must be equal on either side of the wall in both the deconfined phase and the confined vacuum. Indeed using the subtraction natural to the confining vacuum, then this pressure is zero in the confining vacuum, and from \eqref{eq:subT} in the deconfined phase also vanishes at $T = T_c$.
We may simply view this as the expression of the familiar fact that the phase transition temperature for a first order phase transition occurs precisely when the pressures in the two phases are equal (and hence large bubbles may nucleate and the two phases can coexist).

\begin{figure}
\centerline{\includegraphics[width=12cm]{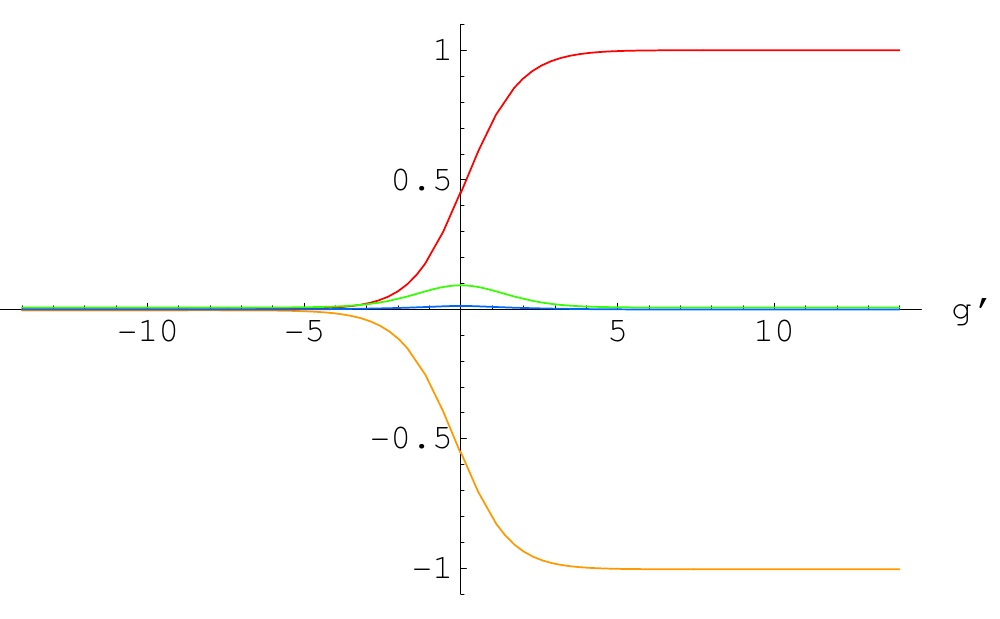}}
\setlength{\unitlength}{0.1\columnwidth}
\begin{picture}(0.3,0.4)(0,0)
\put(8.9,2.8){\makebox(0,0){$2 \pi x/R$}}
\put(5,5.6){\makebox(0,0){$-T_{\mu\nu}(x) / \epsilon_c$}}
\put(7,5.32){\makebox(0,0){$-T_{\tau\tau}$}}
\put(7,1.02){\makebox(0,0){$ -T_{\theta\theta}$}}
\put(7,3.22){\makebox(0,0){$-T_{y^1y^1}$}}
\put(1,4){$\xleftarrow{\text{ Confined vacuum }}$}
\put(7,4){$\xrightarrow{\text{ Deconfined vacuum }}$}
\end{picture}
\caption{
Numerical determination of the components of the (subtracted) stress tensor $T_{\tau\tau}$ (red), $T_{y^1y^1}$ (green), $T_{\theta\theta}$ (yellow), and $T_{xx}$ (blue), as a function of the boundary spatial coordinate $x$ normal to the domain wall for $d = 4$. These are plotted normalized by the asymptotic energy density, $\epsilon_c$, in the deconfined phase. The confining vacuum is  on the left, and the subtracted stress tensor vanishes there. To infinity on the right the theory tends to the deconfined vacuum. We note that $T_{xx}$ vanishes as required by conservation and the pressure tangent to the wall, $T_{y^1y^1}$, is negative everywhere which indicates a positive tension and hence we should expect the wall to be stable. Reproduced from \cite{Aharony:2005bm}. (nb: conventions in \cite{Aharony:2005bm} have opposite sign for stress tensor)
}
\label{fig:stresstensor}
\end{figure}
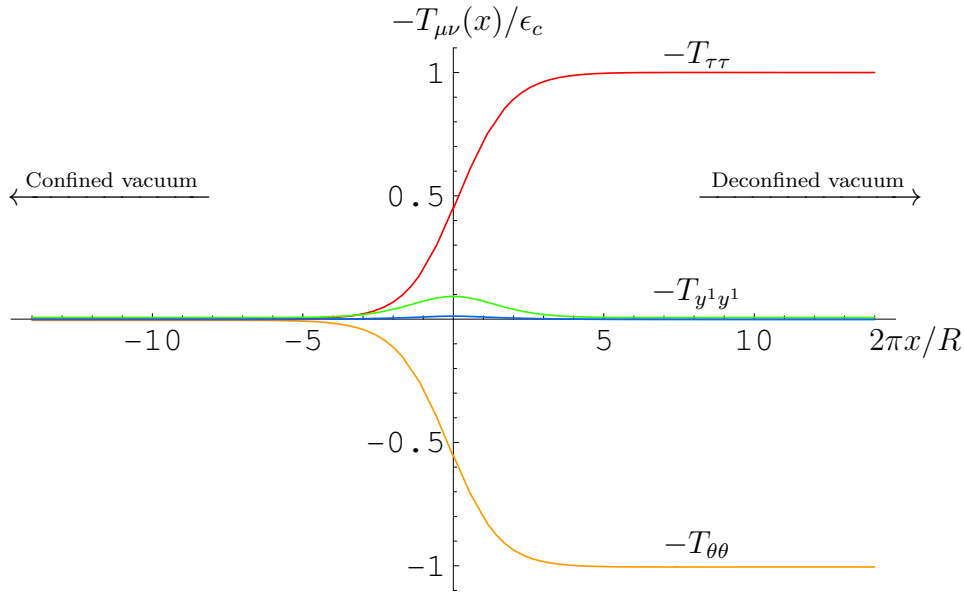

Since the actual bulk metric solution depends non-trivially on both the normal direction to the domain wall, $x$, and also the bulk radial coordinate, it cannot apparently be found analytically. However, thinking of it as an elliptic boundary value problem it has been solved using numerical methods \cite{Aharony:2005bm}. Crucially this allows the boundary stress tensor to be extracted, and this is reproduced in Fig.~\ref{fig:stresstensor} for the case $d=4$. Similar behaviour was also observed for $d=5$. The numerical solution indeed confirms that the domain wall influences the geometry over a distance of order $\sim R$, and then the geometry exponentially returns to the confining and deconfining vacua on either side of the wall.
We may consider the tension $\mu_{d-1}$ of the domain wall, measured in the confining QFT$_{d-1}$ theory.
This tension, $\mu_{d-1}$, may be computed by integrating
the negative of the pressure within the wall over its width and over the circle. This pressure may be determined from
a tangential component of the boundary stress tensor, which we may choose to be in the $y^1$ direction due to the isotropy amongst the ${\bf y}$. Then the tension, 
\begin{eqnarray}
\mu_{d-1} = - \int_{-\infty}^{\infty} dx\, \oint^R d\theta \; T^{sub}_{y^1y^1}
\end{eqnarray}
where $T^{sub}_{\mu\nu}$ is the subtracted stress tensor. This vanishes in the confining vacuum and also for the deconfined phase at $T = T_c$, and hence the integrand only has support near the domain wall.
Parametrically we expect $\mu_{d-1} \sim c_\text{eff}\, R^{2-d} \sim \epsilon_c R^2$, as all scales are controlled by $R$. In fact one finds, 
\begin{align}
\mu_{d-1} =    C_d \,  \epsilon_c \, \left( \frac{R}{2\pi} \right)^2 \,, \quad C_4 = 2.0\,, \quad C_5 = 1.7
\end{align}
with $C_{4,5}$ determined numerically \cite{Aharony:2005bm}.
Importantly, this tension is positive. As Fig.~\ref{fig:stresstensor} illustrates for $d=4$, the tangential pressure  $T^{sub}_{y^1y^1} < 0$ throughout the wall resulting in the positive tension.

\subsubsection{Plasma-balls}
\label{sec:pbs}

The existence of this domain wall solution strongly suggests that there should be black hole configurations where the extent of the horizon in the bulk is finite, rather than infinite as for the domain wall. Such configurations can be viewed as being a bubble of deconfined phase sitting in the confining vacuum. While such solutions are only now under construction in gravity \cite{FT2013} -- again a numerical elliptic problem -- when the bubble of deconfined phase is much larger in extent than the scale $ R$, the properties of these solutions are clear.

\begin{figure}
\centerline{\includegraphics[width=12cm]{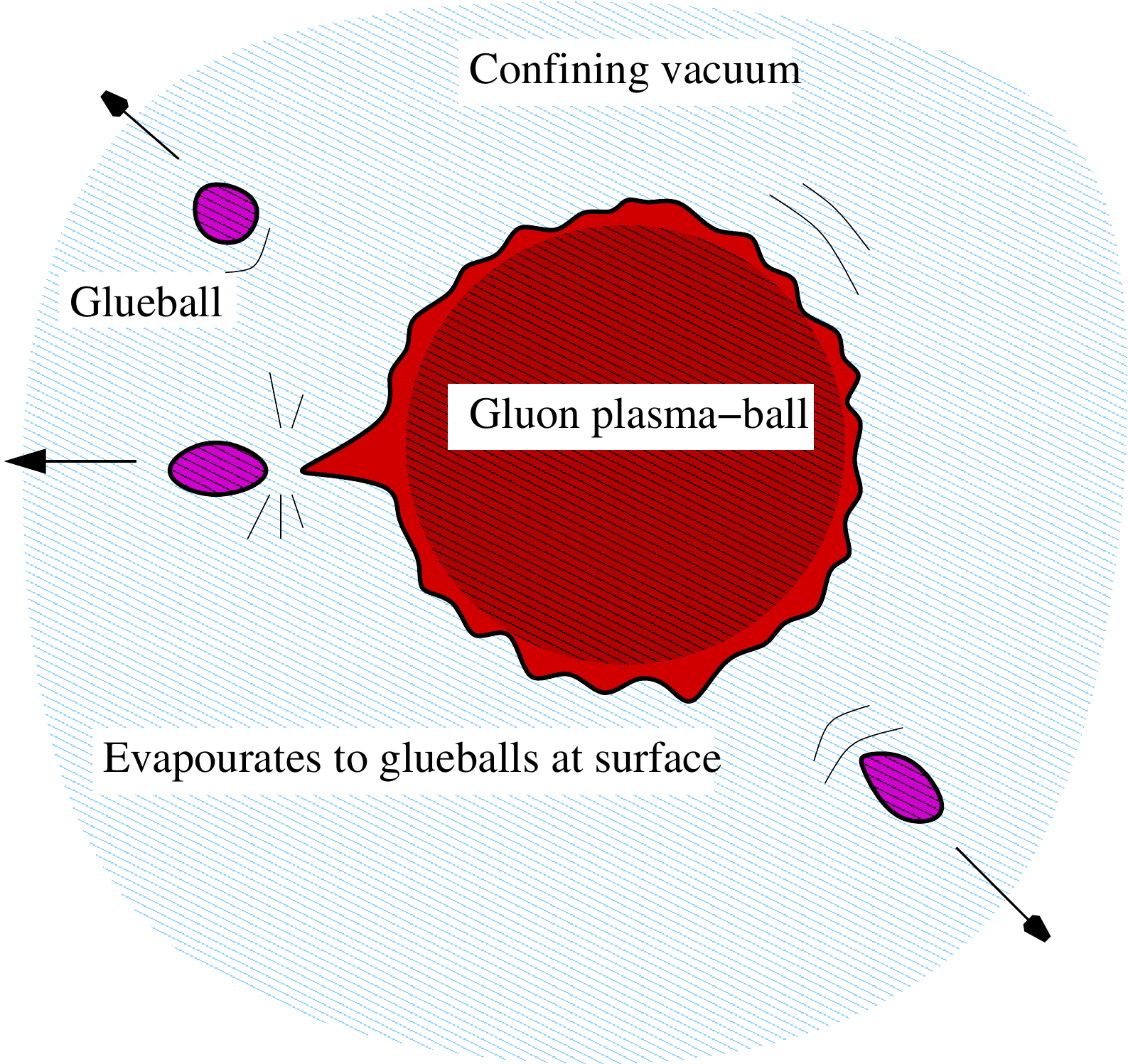}}
\caption{Reproduced from \cite{Aharony:2005bm}: figure depicting a plasma-ball. The size of this object must be much greater than the length scale of the gap in the theory, $1/\Lambda_g \sim L$. Then the plasma ball will appear roughly homogeneous in its interior and exterior, with the surface region being approximately the domain wall solution between the confining and deconfined phases. This object will be long lived in the CFT at large $N$, despite being a hot ball of plasma in zero temperature vacuum, since the radiation rate is slow as it must proceed by emission of colour singlets which are unlikely to be produced in gluon collisions in the plasma region. From a gravitational perspective this evaporation proceeds by Hawking radiation, which is suppressed in the semi-classical limit.
}
\label{fig:plasmaball}
\end{figure}

In either the confining or deconfined phases, static correlations quickly decay on a spatial scale governed by $R$. Hence from the boundary point of view, a bubble of deconfined phase (or `plasma') much larger than $\sim R$ will appear approximately homogeneous in its interior. For a large bubble, the curvature of the surface of this bubble will be small on the scale $R$, and the tension of the bubble wall will be approximately that of the domain wall. Since we know the properties of the homogeneous plasma phase, we may then compute the behaviour of such a bubble solution by simply balancing the internal deconfined phase pressure and that due to the curvature of the bubble wall. Consider a spherical bubble of radius $\rho$ in the $x,{\bf y}$ directions in the QFT$_{d-1}$ theory. Then the energy of the bubble wall  is,
\begin{align}
E_{wall} = \omega_{d-3} \,\rho^{d-3}  \,\mu
\end{align}
In the exterior of the bubble the pressure vanishes, as we are in the confining vacuum. In the interior, the pressure is $P$. Using $d E = P d V$ with the volume $V = \frac{1}{d-2}\, \omega_{d-3} \,\rho^{d-2}$ we obtain,
\begin{align}
P = \left( d - 3 \right) \frac{\mu}{\rho}
\end{align}
where we recall that for the tension to be $\mu$ we have required $\rho \gg R$. From the subtracted stress tensor of the deconfined homogeneous phase we have that the interior pressure of the QFT$_{d-1}$ theory is (recalling we must integrate $T^{sub}_{\mu\nu}$ over the $\theta$ circle),
\begin{align}
P = c_\text{eff} \left( \frac{4 \pi }{ d } \right)^d \left( T^d - \frac{1}{R^d} \right)  R  \; \implies \;  T = \frac{1}{R} \left( 1 + \frac{( d - 3 ) \, C_d}{(2 \pi)^2} \, \frac{R}{\rho} + {\cal O}\left( \frac{R^2}{\rho^2} \right) \right)
\nonumber
\end{align}
As $\rho \to \infty$ then $T \to T_c = 1/R$, the phase transition temperature. For a finite radius bubble we see that the temperature of the interior is $T > T_c$. Note the relation between the wall tension being positive, and the temperature increasing with decreasing radius.
The bulk solution corresponding to such a large bubble will presumably be very similar to the domain wall bulk solution, but the domain wall will be `bent' round into a sphere in the $x, {\bf y}$ directions. An illustration of such an object is given in Fig.~\ref{fig:plasmaball}.

We term such a configuration -- a static finite radius bubble of deconfined phase in the confining vacuum - a `plasma-ball'. We emphasize that these plasma-balls will have  universal behaviour for a holographic CFT on a compact circle length $R$, provided the radius $\rho$ is much greater than the length scale associated to the mass gap, i.e., $\rho \gg R$. As the plasma-ball radius decreases, it is likely the temperature continues to increase until the radius $\rho$ becomes of order $R$. At this point the behaviour of the plasma-ball will no longer be universal, with the dual black hole localizing on the internal space $X$ via the Gregory-Laflamme instability which we discuss in \S\ref{sec:globalGL} in  more detail.

Recall that for asymptotically flat neutral black holes the Hawking temperature decreases as a function of the black hole horizon radius. While the plasma-ball solutions at large radius also have their temperature decreasing, in this case one asymptotically reaches the phase transition temperature $T_c$ (as opposed to zero).

In the Euclidean context, the confining exterior to the plasma ball is imagined to be at finite temperature (with periodic $\tau$). However, in Lorentzian signature we may choose to interpret the solution as comprising of a droplet of  finite temperature plasma ($T \simeq T_c$ for large plasma-balls) sitting in the zero temperature confining vacuum. This is because gravity does not discriminate between the confining vacuum geometry at zero or finite temperature (as in global AdS, cf., \S\ref{sec:HP}).

The gauge theory interpretation of these solutions is perhaps their most interesting feature. The existence of large plasma balls indicates that in the large $N$ (or $c_\text{eff}$) confining QFT created by taking the CFT on a Scherk-Schwarz circle, there exist large \emph{static localized} balls of deconfined plasma. These are regions of hot plasma with temperature $T \simeq T_c$ and energy $\gg N^2 \Lambda_g$, sitting in the zero temperature confining vacuum. In fact in \cite{Aharony:2005bm} it was argued that plasma-ball states should exist in any large $N$ confining gauge theory with a first order confinement/deconfinement transition.

This is to be contrasted with the case of the CFT on flat space (which has no mass gap), where any ball of plasma will necessarily expand indefinitely driven by its internal pressure. Likewise, for the CFT on ${\bf S}^{d-1}$ discussed in \S\ref{sec:HP} either the plasma is homogeneously distributed on the sphere (so not localized) or it is dynamical (so not static).

One might wonder how such a static localized region at finite temperature could exist as a bubble in the zero temperature vacuum. In fact we must be more careful in recalling what the gravity is telling us. Really gravity tells us that a \emph{long lived} rather than actually static plasma-ball exists. In gravity the plasma-ball in a zero temperature vacuum will eventually Hawking radiate away. It is simply that the time taken for this to happen will go roughly as $\sim {\cal O}(N^2)$. From a field theory perspective this is easy to understand given that there is a mass gap and the vacuum is confining. A hot ball of deconfined plasma will certainly want to radiate away its energy. However, since it will be radiating into the confining vacuum, the degrees of freedom available to radiate into will be colour singlets numbering ${\cal O}(1)$, rather than the ${\cal O}(N^2)$ coloured degrees of freedom.
A simple picture is as follows (a more formal discussion in terms of planar graphs is given in \cite{Aharony:2005bm}): The density of the plasma goes as ${\cal O}(N^2)$ and the number of collisions of the gluonic or coloured degrees of freedom per unit time goes as ${\cal O}(N^2)$. However, it is unlikely that upon a collision of two coloured degrees of freedom a colour singlet results. This occurs only 1 in ${\cal O}(N^2)$ collisions. Hence the energy radiated into singlets per unit time will only be ${\cal O}(1)$, whereas there is ${\cal O}(N^2)$ energy to evaporate, hence the evaporation time ${\cal O}(N^2)$.
\\

\subsubsection{Other plasma solutions}

We see from the discussion above that once the domain wall solution and its (positive) tension is known, then if we consider static spherical plasma-balls with masses $M \gg N^2 \Lambda_g$ we can argue their existence simply from a fluid analogy, taking their interior to be the homogeneous deconfined plasma fluid, and the surface to have the domain wall tension. More generally then any stationary solution of this fluid dynamics with surfaces where the surface curvature is weak (i.e., the radius of curvature $\gg R$) and where the interior plasma fluid is approximately homogeneous on scales of order $\sim R$ would then be expected to have a dual gravity stationary black hole solution. An example is slowly rotating plasma balls and plasma rings as discussed in \cite{Lahiri:2007ae,Bhardwaj:2008if,Bhattacharya:2009gm}, or a plasma tube in the $x, {\bf y}$ plane. One can discuss the dynamics of the Gregory-Laflamme instability for these plasma tubes by dynamical simulation of the boundary fluid interpretation (up until curvature scales $\sim \ell$ are reached for the plasma tube walls) \cite{Maeda:2008kj,Caldarelli:2008mv}. Such a setting provides a precise map between stationary solutions of fluids with surfaces and black holes in these confining backgrounds. This is related to the  fluid/gravity correspondence \cite{Bhattacharyya:2007vs}  which we discuss later in \S\ref{sec:flgra}. See also further discussions of the nature and dynamics of plasma configurations in \cite{Cardoso:2009nz} and \cite{Emparan:2009dj} for some exact analytic results.

%
\subsection{Non-universal behaviour}
\label{sec:globalGL}
%

In general for a holographic CFT on ESU$_d$ we have argued there is a universal confinement/deconfinement phase transition in the canonical ensemble dual to the Hawking-Page phase transition \S\ref{sec:hpgrav}. However if we are to consider a string theoretic embedding then we should remember the internal space $X$ in addition to the $(d+1)$ universal gravity sector (e.g., $X = {\bf S}^5$ for ${\cal N}=4$ SYM).
So far we have considered only gravitational dynamics in the universal sector. Whilst it is consistent to do this, there may be dynamics associated to non-trivial behaviour on the internal space or due to stringy physics going beyond bulk gravity. In the example of a holographic CFT on a sphere we will observe a transition in the micro-canonical ensemble associated to dynamics on $X$  \cite{Horowitz:1999uv,Horowitz:2000kx,Hubeny:2002xn}. In order to discuss this we will first quickly review the relevant features of the Gregory-Laflamme gravitational phase transition. We then briefly discuss the example of the CFT on Minkowski with a compact circle with periodic boundary conditions, where phase transitions due to stringy physics occur.

\subsubsection{Review of Gregory-Laflamme transition in pure gravity}
\label{sec:GLreview}

We now briefly review the behaviour of black holes when one space dimension is compact by considering the example of Kaluza-Klein theory, namely pure gravity in 5 dimensions, with one space direction compactified to a circle of length $R$.
The subject has been extensively reviewed elsewhere, and for more detail the reader is referred to the recent reviews \cite{Horowitz:2011cq,Gregory:2011kh,Lehner:2011wc} and the references therein.

Let us consider black hole solutions which asymptote to ${\mathbb R}^{3,1} \times {\bf S}^1_{(R)}$. Note that the geometry may in principle have a complicated topology in the interior, and is only constrained asymptotically to have this product form. If we consider static black holes then there are two distinct behaviours that are immediately evident. Firstly if the black hole is very small then it will not `see' the extra dimension, and hence will behave as a 5d Schwarzschild solution. Secondly, for circle size, we may write an analytic solution which is the product of a 4d Schwarzschild metric with horizon size $r_h$ and the circle length $R$. In the former case the horizon has spherical topology ${\bf S}^3$, but in the latter has topology ${\bf S}^2 \times {\bf S}^1$. We say the first is a \emph{localized} horizon, whereas the latter has a \emph{string}-like horizon as it wraps over the ${\bf S}^1$ of the extra dimension - we call it a homogeneous black string. The term homogeneous refers to the solution preserving the translation symmetry of the ${\bf S}^1$, and we note that these solutions are also known as uniform black strings.

At first one might think it odd that for low mass solutions there are two possible behaviours, as both the localized and string solutions above will exist. Furthermore the string solution appears rather unnatural as the radius of the string will be much narrower than its extent in the circle direction. Indeed Gregory and Laflamme showed that while for $r_h \gg R$ the black string is stable to small perturbations, at a critical radius parametrically given by $R$, so $r_h \sim R$, then the strings develop a dynamical instability that acts to break the translation invariance in the circle direction of the horizon \cite{Gregory:1993vy}. Much as when a tube of fluid becomes too thin, the surface tension acts to break it into droplets, it is believed that a black string that becomes too thin will break into localized black holes. Thus at low masses, the only stable solution is thought to be the localized one, resembling a small 5d black hole.
At high masses, the only solution is the homogeneous string solution. There is an interesting story as to the nature of the localized branch of solution as their mass is increased to scales $G_5 M\sim R^2$ which we return to shortly. The important point here, however, is that this branch of solutions is believed to have an upper mass of $\sim R^2/G_5$.

A nice way to understand this instability is to ask whether the thin homogeneous string with radius $r_h \ll R$ or a localized black hole of the same mass has greater horizon area. The mass and area of a localized 5d Schwarzschild solution with horizon radius $r_l$ go as,
\begin{eqnarray}
M_{loc} \sim \frac{1}{G_5} r_l^2 \; , \quad A_{loc} \sim r_l^3 \quad \implies \quad A_{loc} \sim \left( G_5 M_{loc} \right)^{\frac{3}{2}}
\end{eqnarray}
whereas the mass and area for the homogeneous black string radius $r_h$, goes as,
\begin{eqnarray}
M_{string} \sim \frac{R}{G_5} r_h \; , \quad A_{string} \sim r_h^2 R  \quad \implies \quad A_{string} \sim \frac{1}{R} \left( G_5 M_{string} \right)^{2}
\end{eqnarray}
So we consider taking $M_{loc} = M_{string} = M$ with $G_5 M \ll R^2$ so that the radius of the localized solution is $\ll R$ so that the 5d Schwarzschild approximation is good. Then we find
\begin{eqnarray}
  \frac{ A_{string} }{R^3}  \sim \left( \frac{ G_5 M }{R^2} \right)^{2} \ll  \frac{ A_{loc} }{R^3} \sim  \left( \frac{ G_5 M }{R^2} \right)^{\frac{3}{2}} .
\end{eqnarray}
Hence the localized solution has a greater area, and thus entropy, than the homogeneous string solution of the same mass. This shows in the micro-canonical ensemble that at high masses the homogeneous string dominates, being the only solution, but at low masses the localized solution dominates, with a phase transition occurring at intermediate masses $G_5 M \sim R^2$.  Whilst this does in no way prove that there is a dynamical instability for a low mass homogeneous string solution, it is suggestive of this, and certainly is compatible with the localized black hole being the end state of this instability. A detailed dynamical linear perturbation theory calculation confirms the existence of this instability.
An analogous calculation fixing temperature rather than mass, and considering free energy rather than entropy shows that the same first order phase transition between localized and homogeneous solutions occurs in the canonical ensemble.

The natural question to ask is what happens to the localized solutions as their mass increases. At some point they will `see' their images in the circle direction and become strongly deformed from a spherical shaped horizon. An elegant picture has emerged. The localized solution is thought to grow in size as its mass is increased. Moving further along this branch of solutions, however, the horizon eventually touches itself and then merges into a horizon that wraps around the circle, now having topology ${\bf S}^2 \times {\bf S}^1$. However the solution is not translationally invariant. This type of solution is said to be an `inhomogeneous black string' \cite{Horowitz:2001cz,Gubser:2001ac,Wiseman:2002zc} (also referred to as a `non-uniform black string').
Such a topology change was postulated in \cite{Kol:2002xz} (see also \cite{Harmark:2003eg}) and there is now numerical evidence from construction of the localized and inhomogeneous solutions \cite{Kudoh:2004hs,Headrick:2009pv}.
Further moving along this branch of static solutions the inhomogeneous strings become increasingly homogeneous with radius $\sim R$, and eventually meet the homogeneous string branch precisely at the point when the homogeneous strings are marginally stable. The marginally stable deformation is a precisely static linear perturbation that is tangent to the inhomogeneous black string branch of solutions.

In summary, a horizon homogeneously `smeared' over a compact circle direction may become unstable to `clumping' into a localized solution when the horizon radius becomes much smaller than the size of this circle. In the pure gravity context this transition is of discontinuous or first-order type. As one decreases the mass the dynamically preferred solution jumps from being a homogeneous string to a localized black hole.\footnote{ We note that in fact in sufficient dimensions the transition can become a continuous one or of second order type in pure gravity \cite{Sorkin:2004qq}.}

\subsubsection{The Einstein Static Universe and localization on the internal space}

We have discussed the Hawking-Page transition that occurs in the canonical ensemble on ESU$_d$ due to the negative cosmological constant scale $\ell$. Let us consider the full bulk geometry with an internal space $X$. Since it will be compact, we may associate a natural length scale to it, let us say $R_X$. In most supergravity constructions $R_X \sim {\cal O}(\ell)$, as for example for the ${\bf S}^5$ of the $\mathcal{N}=4$ SYM case.
Then as for Kaluza-Klein theory discussed above, when the radius of the black hole horizon, $\rho_h$, becomes much smaller than this compact space scale, so $\rho_h \ll R_X$ we expect that the black hole will want to localize itself in the compact space, rather than being smeared over it.

Now since $R_X \sim \ell$ this shows that generally the small  Schwarzschild-AdS black hole solution is in fact not the dominant bulk geometry in the micro-canonical ensemble.  One must instead consider the potential for this  solution  $\rho_h < \ell$ to be unstable to localization in all the $D$ spacetime dimensions.\footnote{ The situation in the canonical ensemble as discussed in \S\ref{sec:HP} is different since the small black holes are never thermodynamically favoured. Large Schwarzchild-AdS black holes are stable to localization on $X$ \cite{Hubeny:2002xn}.} At sufficiently low $\rho_h$, such localized $D$ dimensional black holes will completely evaporate due to the Hawking effect. This observation justifies our comments at the end of \S\ref{sec:hpgrav}about the micro-canonical phase structure of holographic field theories.

We emphasize that the precise nature of the transition from the smeared global AdS black holes to the fully localized ones depends on the details of the internal space, and thus is not a `universal' phenomena for CFTs with a holographic dual.\footnote{ We however expect that the qualitative phenomenon  of the localization to be generic.}  In particular, the energy (equivalently horizon radius $\rho_h$) at which the small black hole wants to localize depends on the spectrum of gravitational fluctuations and hence on the detailed geometry of $X$ \cite{Hubeny:2002xn}. A simple way to understand this is to estimate the analog of the Gregory-Laflamme zero mode cf., \cite{Reall:2001ag}. The spectrum of fluctuations on compact $X$ being discrete, we will encounter an instability only when the Euclidean negative mode of the small Schwarzschild-AdS$_{d+1}$ black hole \cite{Prestidge:1999uq} is balanced by the eigenmodes of $X$, i.e., the fluctuations `fit in on $X$', to allow for an incipient tachyonic (or marginally stable) mode in the physical spectrum \cite{Hubeny:2002xn}.

From a field theoretic perspective, the localization instability should be interpreted as spontaneous symmetry breaking in the micro-canonical ensemble.
Recall that isometries of $X$ are related to global symmetries of the CFT and the localization instability indicates that as the energy is lowered past a critical point, the preferred state of maximal entropy breaks these symmetries. The Goldstone mode associated with the symmetry breaking is of course the zero mode described above.

As noted in our discussion of the Hawking-Page transition, focussing on scales much smaller than the sphere size we recover the behaviour of planar black holes in AdS. The correspondence is between the `large' global black holes and the planar ones, and hence we see that we do not expect localization on the internal space to occur in the case of the CFT on $ {\mathbb R}^{d-1,1}$. Heuristically the planar black hole horizon is always infinite in size, being non-compact, and so is much bigger than the internal space, and hence would not wish to localize.

In fact, a variant of the above argument says that BTZ black holes in $d=2$ are stable to localization on any compact space, i.e.,  BTZ $\times \,X$ is stable. Intuitively one can see this  by noting that BTZ black holes are always large in AdS$_3$. Alternately, one can argue that spectrum of fluctuations about the  BTZ geometry should have no static Euclidean negative mode (preserving the boundary rotational symmetry) by noting that the Euclidean manifold is identical to that of Euclidean AdS$_3$. This follows because the BTZ geometry unlike higher dimensional AdS black holes is constrained to be locally AdS$_3$.

We expect the same phenomenology to occur in plasma balls associated with CFTs on a Scherk-Schwarz circle, length $R$. For large masses the black hole will be a large plasma ball, as discussed in \S\ref{sec:pbs}. But at small enough masses the bulk black hole will develop structure on the scale $\ell$, and it will want to localize on the internal space $X$.

The upshot of the above discussion is that  one should consider the issue of instabilities on the internal space whenever the universal gravity sector has deformations in the bulk of order the scale $\ell$. Outside of the hydrodynamic regime \S\ref{sec:flgra}, such deformations are ubiquitous when the CFT is put on curved spacetimes, and we will be careful in the remainder of this review to note when the physics may be subject to non-universal instabilities on the internal space.

\subsubsection{Circles with periodic fermions}
\label{sec:periodic}

Suppose we consider our thermal CFT on flat space, and compactify a spatial circle with periodic fermion boundary conditions, length $R$. One might naively expect that this would lead to precisely the same phenomenology as anti-periodic boundary conditions. The subtlety is that one cannot define a spin structure in the bulk on a manifold where a cycle shrinks and has periodic fermion boundary conditions.\footnote{ This is familiar in the case of Kaluza-Klein theory where we know the Witten bubble of nothing naively signals that the vacuum is disastrously unstable, but this solution is disallowed in a theory with fermions for the same reason \cite{Witten:1981gj}.} Hence the AdS-soliton vacuum solution is not allowed. At any finite temperature, the only bulk dual is the planar Schwarzschild solution with a space direction compactified.

Since there is no instability to localization on $X$ for the planar black hole, neither is there when one compactifies a circle. Thus one might naively conclude that as for the CFT on Minkowski, this planar black hole with circle compactification describes the CFT on a periodic circle at all temperatures. In fact the story is more subtle, as the size of the compact circle at the horizon is now a physical length scale in the bulk that becomes very small at low energy densities. For the case of $\mathcal{N} = 4$ SYM one may deduce this circle size becomes comparable to the string scale when $T R \lesssim 1/\lambda^{1/4}$. In fact \cite{Susskind:1997dr, Barbon:1998cr,Li:1998jy,Martinec:1998ja, Aharony:2004ig,Harmark:2004ws} argue that in both the canonical and micro-canonical ensembles bulk strings winding around this circle in the bulk become unstable, although not until the somewhat lower scale $T R \lesssim 1/\lambda^{1/2}$ is reached, resulting in phase transitions.\footnote{
Such phase transitions have been argued to be analogous to confinement/deconfinement transitions and like confinement have order parameters given in terms of a Wilson loop \cite{Aharony:2004ig}. Rather than in the case of usual confinement where the loop is about Euclidean time, here it is about the spatial circle. Evidence for these transitions has been seen in numerical lattice simulation \cite{Catterall:2010fx}, are they appear to continue to the weakly coupled theory \cite{Aharony:2004ig,Kawahara:2007fn,Mandal:2009vz}, and also more generally in non-holographic large $N$ gauge theories compactified on circles \cite{Kiskis:2003rd,Aharony:2005ew}.
}

To summarize, whilst the thermal CFT on Minkowski has no scale other than $T$, on other spaces typically a length scale -- say $L$ -- is introduced. At low energies or temperatures, $T L \sim {\cal O}(1)$, phase transitions associated to the non-universal gravity sector may occur,  such as localization on $X$. At very low temperature scales, $T L \ll {\cal O}(1)$, phase transitions  associated to bulk stringy physics may occur.

%
\section{Out-of-equilibrium CFT on curved spacetimes}
\label{sec:4}
%

Our discussion thus far has focussed on the physics of thermal field theories in global equilibrium. In particular, we saw in \S\ref{sec:3} the holographic dual spacetimes for the thermal states of field theories on ${\cal B}_d = {\mathbb R}^{d-1,1}$ and ${\cal B}_d = {\mathbb R}\times {\bf S}^{d-1}$ map respectively to planar and global Schwarzschild black holes. The former of course can be obtained from the latter by zooming in on a small region of the boundary while passing to the limit of high temperature, and thus the limit of large size for the bulk black hole.

However, these backgrounds are clearly a limited set of examples. Global thermal equilibrium can be attained in much more general settings with a time-translation symmetry. For any stationary ${\cal B}_d$, it is thus natural to seek stationary bulk solutions with a Killing horizon and to interpret the corresponding surface gravity as the temperature of a dual thermal state in the CFT. Here we require the stationary Killing field $\partial_t$ to be globally timelike on ${\cal B}_d$; the case where ${\cal B}_d$ contains black holes will be discussed in \S \ref{sec:5}.

There is a simple intuitive argument for existence of stationary black hole solutions with a Killing horizon, which for simplicity we present when $\partial_t$ is a static Killing field.\footnote{ Similar statements can be made in the stationary case; we refer the reader to  \cite{Banerjee:2012iz,Jensen:2012jh} for arguments from a field theoretic perspective and construction of an equilibrium partition function as a functional of the background metric in these cases.}   Geometries ${\cal B}_d$ with the time translational isometry can always be brought to a canonical form:
 \begin{align}
ds^2  = -e^{2\,\varphi(x)}\,dt ^2 + \gamma_{ij}(x)\, dx^i\, dx^j \,,
\label{stybdy}
\end{align}
 where the metric functions depend on the spatial coordinates alone. Placing a field theory in these backgrounds and considering the thermal density matrix is equivalent to demanding that the Euclidean thermal circle parameterized by $\tau = i \,t$ be of fixed period and performing the associated Euclidean path integral. From the bulk perspective, corresponding to each such choice of ${\cal B}_d$ one may expect to find a stationary black hole solution ${\cal M}_{d+1}$ characterized by the Euclidean thermal circle shrinking in the interior of the spacetime.\footnote{As we will see later it is possible that there are multiple such solutions. However at any given value of temperature generically
 at most one will dominate and provide the gravitational description of the thermal density matrix.} For future reference we note that by a conformal transformation the geometry \eqref{stybdy} can be brought to the ultra-static form:
 \begin{equation}
{\widetilde ds^2 } = -dt ^2 + {\tilde \gamma}_{ij}(x)\, dx^i\, dx^j \,,
 \label{ultrastat}
 \end{equation}	
 This form of  the metric will be useful in our discussion of heat flow in strongly coupled theories in \S\ref{sec:5}.

For generic static boundary geometries there is no known  explicit construction of the corresponding bulk black hole spacetimes. But one may construct approximate solutions when ${\cal B}_d$ is weakly curved. In fact, one can do a lot better: one can construct solutions slowly varying both in space and time, and thereby  deviate from global thermal equilibrium. In field theory, it is easy to envisage a circumstance wherein these deviations occur at length scales which are large compared to the characteristic mean free path of the system, effectively ensuring that we have local equilibration. For curved ${\cal B}_d$ consistency requires that the curvature scales are also large compared to the mean free path.
In these situations the hierarchy between the length scales of variation $L$ and the mean free path $\ell_\text{mfp}$ gives a natural small parameter $\varepsilon = {\ell_\text{mfp}}/L \ll 1$. The effective field theory obtained by studying the system perturbatively in $\varepsilon$ is just hydrodynamics.

As shown in \cite{Bhattacharyya:2008jc} (see \cite{Rangamani:2009xk, Hubeny:2011hd} for reviews of this subject), the corresponding bulk geometries may also be studied perturbatively in $\varepsilon$, defining what is known as the fluid/gravity correspondence.  Here we emphasize that $\varepsilon$-expansion is not that of familiar perturbative gravity.  The solutions we will describe below are fully non-linear, though they remain approximate in the sense that they solve the equations only to some order in $\varepsilon$. We will describe this construction briefly in \S\ref{sec:flgra}.

The long-wavelength perturbation expansion is another universal feature of holography. Since we start with large black holes and perturb around thermal equilibrium these solutions can be constructed within the AdS geometry by solving Einstein's equations for AlAdS geometries with no matter sources. As a result they describe the dynamics of fluids arising from an infinite class of conformal field theories (obtained by changing the internal space $X$). It is rather remarkable that holography allows us to explicitly derive the constitutive relations for a strongly coupled plasma in a simple and straightforward manner. The corresponding computation in perturbative gauge field theories is in fact much more challenging \cite{Arnold:2000dr,Arnold:2003zc,York:2008rr}.

 It is also interesting to go beyond this slow variation approximation to consider large departures from equilibrium. To do so one can consider either linear response to some applied perturbation or the full-fledged non-linear computation. The former amounts to working in the linearized regime in the dual gravity, while the latter require a complete solution of Einstein's equations -- typically using numerical techniques. Impressive advances in numerical general relativity have led to interesting new results in the non-linear problem and we will describe the salient features of some of these in \S\ref{sec:numericalGR}.  As we will emphasize there while some the features here are universal, there is potential for non-universal behaviour once we depart from the long-wavelength regime.

%
\subsection{The fluid/gravity correspondence}
\label{sec:flgra}
%

 Fluid dynamics is best thought of
 as an effective field theory valid for interacting quantum systems in local thermal equilibrium, valid when the length scales of departures from equilibrium are large in the units of local energy density (or temperature). As described above one has in this limit a natural small parameter $\varepsilon$ which can be used to set up a perturbation theory. In field theory this effective description is conventionally given in terms of equations of motion; one typically writes down the expressions for the conserved charges as functionals of thermodynamic parameters and a flux vector (the co-moving fluid velocity). As in any other effective field theory we allow the conserved charges/currents to be linear combination of the associated operators with arbitrary coefficients. The latter are called transport coefficients in the fluid dynamics context and are fixed by the underlying microscopic theory.

 In the gauge/gravity context, configurations of locally equilibrated field theory quanta are described by inhomogeneous, dynamical black hole asymptotically locally AdS spacetimes. These can be constructed explicitly in a gravitational analog of the $\varepsilon$-expansion as was first explained in \cite{Bhattacharyya:2008jc}.
The logic is rather simple and is easiest to explain when we consider departures from equilibrium on ${\cal B}_d = {\mathbb R}^{d-1,1}$; we can then generalize to curved boundary manifolds with ease.

 Recall that global equilibrium is dual to the planar Schwarzschild-\AdS{d+1} black hole spacetime \eqref{eq:PoincareSchwarz}, which is naively parameterized by one parameter $r_h$. However, since the field theory in question in relativistic, and the vacuum is Poincar\'e invariant, it follows that when we turn on a temperature we get also to make a choice of inertial frame. In particular, by an appropriate boost parameterized by a normalized four-velocity $u^\mu$ we can bring the energy momentum tensor in \eqref{eq:stresspoincare} to a familiar perfect fluid form:
 \begin{equation}
T_{\mu \nu } = c_\text{eff}\, \left( \frac{4 \pi T}{d} \right)^d\, \left(\eta_{\mu\nu} + d\, u_\mu u_\nu \right) \,, \qquad u^\mu\,u_\nu = -1 .
 \label{zeroTmn}
 \end{equation}	
 The choice of inertial frame  in the boundary can  be identified with a boost of the corresponding black hole solution in the bulk spacetime; we can by an appropriate coordinate transformation bring \eqref{eq:PoincareSchwarz} to a form that will prove useful below:
\begin{equation}
ds^2 = 2\, u_\mu \, dx^\mu\, dr - f(r,r_+)\, u_\mu\,u_\nu\, dx^\mu \, dx^\nu + r^2\, P_{\mu\nu} \,dx^\mu \, dx^\nu
\label{fg0}
\end{equation}	
In writing this metric we have relabeled $r_h \to r_+$ and changed the black hole metric into ingoing Eddington-Finkelstein coordinates $(t,x^i) \to (v,x^i)$. $P_{\mu\nu}$ is the spatial projector onto directions orthogonal to $u^\mu$, $P_{\mu\nu} \equiv \eta_{\mu\nu }+ u_\mu \, u_\nu$. The reason for switching to these coordinates is to ensure that the starting point of the perturbation theory possess a regular coordinate chart in the vicinity of the future horizon.

While the bulk solution has undergone a simple diffeomorphism to convert it into a $d$-parameter solution  labeled by $(r_+, u^\mu)$, on the boundary we still have a Minkowski metric with the identification $t \to v$. This solution captures global equilibrium in an inertial frame not aligned with the timelike Killing field. If we want to describe deviations away from equilibrium then all we need to do on the boundary is to allow the  temperature $T$ and boost velocity $u^\mu$, to be slowly varying functions of the coordinates $(v,x^i)$.

We can implement this slowly varying ansatz directly into the metric \eqref{fg0} leading to a geometry:
\begin{equation}
ds^2 = 2\, u_\mu(x) \, dx^\mu\, dr - f(r, r_+(x))\, u_\mu(x)\,u_\nu(x)\, dx^\mu \, dx^\nu + r^2\, P_{\mu\nu}(x) \,dx^\mu \, dx^\nu \,,
\label{fg0x}
\end{equation}	
which does not solve the Einstein's equations \eqref{bEOM} anymore, since we promoted parameters to functions. However, the parameters $r_+,\; u^\mu$ are moduli in the black hole solution -- all we have done is make them dependent on boundary coordinates. In an appropriate slow variation limit, we can correct for augmenting them to functions perturbatively order by order in gradients $\nabla_\mu T \simeq \nabla_\mu r_+$  and $\nabla_\mu u_\nu$ (thus constructing the collective coordinate field theory for the moduli). More specifically one considers an expansion
of the bulk metric data
\begin{align}
& g_{AB} = \sum_{k=0}^\infty \; \varepsilon^k \, g_{AB}^{(k)}\left(r_+(\varepsilon\, x),u^A(\epsilon\,x)\right)  \,, \;\;  \nonumber \\
&r_+(x) = \sum_{k=0}^\infty \; \varepsilon^k \, r_+^{(k)}(\varepsilon\, x) \,, \;\;
u^A = \sum_{k=0}^\infty \; \varepsilon^k \, u^{A\,(k)}(\varepsilon \,x) \,,
\label{}
\end{align}
with $g^{(0)}_{AB}$, etc., given by the solution \eqref{fg0}. Writing down the Einstein's equations for a spacetime with negative cosmological constant \eqref{bEOM} and expanding in the perturbation parameter $\varepsilon$ we then obtain equations for $g^{(k)}_{AB}$ etc., in terms of data involving $g^{(j)}_{AB}$ with $j < k$.

The details of how this works has been extensively reviewed in \cite{Rangamani:2009xk, Hubeny:2011hd}, so for now we will simply quote the main results of the analysis. Before doing so we note that while we motivated the discussion in terms of the field theory on ${\mathbb R}^{d-1,1}$ the generalization to slowly varying curved ${\cal B}_d$ is immediate. Consistency of the $\varepsilon$-expansion requires that we also expand curvatures of ${\cal B}_d$ in powers of $\varepsilon$ (equivalently we work in local patches of Gaussian normal coordinates with the patches being large in units of the local temperature), which means that we can perturbatively include the effects of curved sources. To first order in gradients this makes no difference (since we don't see the curvature at this order), but there will be explicit couplings of the fluid to curvature at higher orders. In any event the upshot of the perturbative construction to first order can be summarized by the metric:
\begin{align}
ds^2 &= -2\, u^\mu\, dx^\mu \, dr- r^2\, f(r, r_+(x)) \, u_\mu(x)\, u_\nu(x)\, dx^\mu\, dx^\nu + r^2\, P_{\mu\nu}(x)\, dx^\mu\, dx^\nu
\nonumber \\
& + \; \left[2\, \frac{r^2}{r_+(x)}\,F(r, r_+(x))\, \sigma_{\mu\nu}\,+ \frac{2}{d-1}\, r\, u_\mu(x)\, u^\nu(x)\, \nabla_\lambda u^\lambda -r\, u^\rho\, \nabla_\rho(u_\mu\, u_\nu) \right] dx^\mu \,dx^\nu
\nonumber \\
& \qquad \text{with}\;\; F(x) = \, \int_x^\infty\, dy \, \frac{y^d-1-1}{y\,(y^d-1)}
 \label{fg1}
\end{align}
Here $\sigma_{\mu\nu} =
P^{\mu\alpha}\, P^{\nu \beta} \left(\nabla_{(\alpha} u_{\beta)} - \frac{1}{d-1} P_{\alpha\beta}\, \nabla_\gamma u^\gamma\right)$ is the shear tensor of the velocity vector $u^\mu$.

The slow variation in the moduli space approximation of the gravitational solution corresponds in the boundary field theoretic language to constructing the effective equations of motion for a relativistic fluid order by order in gradients of temperature and velocity. To see the link, we recall that given a regular AlAdS solution\footnote{ By a regular solution we refer to a Lorentzian spacetime free of naked singularities in the bulk.} we may extract the boundary CFT stress tensor using the methods discussed in \S\ref{sec:2}, cf., Eq.~\eqref{bst}. Indeed at leading order (zeroth order in perturbation theory) we saw that the stress tensor took the perfect (conformal) fluid form \eqref{zeroTmn}. Once we solve for $g^{(k)}_{AB}$ we can construct the boundary stress tensor at order $\varepsilon^k$ by using the same procedure. For example at first order in the gradient expansion we obtain from \eqref{fg1} the relativistic viscous fluid stress tensor
\begin{equation}
T_{\mu\nu} = c_\text{eff}\, \left( \frac{4 \pi T}{d} \right)^d\,  \left(\eta_{\mu\nu} + d\, u_\mu u_\nu \right) - 2\, \eta\, \sigma_{\mu\nu} \,, \qquad \eta = \frac{s}{4\pi} = \frac{c_\text{eff}}{4\pi}\, T^{d-1}
\label{}
\end{equation}	
The particular value of the shear viscosity coefficient $\eta$ is a consequence of having a regular solution in the interior of the spacetime. The ratio of $\eta/s$ is the lowest known value amongst all known fluids in nature \cite{Kovtun:2004de} and hence the holographic computations are believed to be relevant for nearly ideal fluids such as the quark-gluon plasma or cold atoms at unitarity, see \cite{Schafer:2009dj}.

Not only do we recover the transport coefficients  from this construction, but the structure of Einstein's equations ensures that we obtain the correct dynamics (Ward identities). Indeed, the boundary stress tensor obtained from the solution is required to be conserved. More importantly however, the way the perturbation expansion has been implemented in the fluid/gravity context is such that the bulk metric $g^{(k)}_{AB}$ solves all the Einstein's equations iff the fluid dynamical data $\{u^\mu(x), T(x)\}$ solve the energy-momentum conservation equations at order $\varepsilon^{k-1}$.\footnote{ The offset of one in the powers of $\varepsilon$ has to do with the fact that the boundary conservation equations have an explicit derivative.} In effect, the correspondence sets up an isomorphism between the equations of gravitation and those of viscous relativistic fluid dynamics.

The explicit construction has been carried out to second order\footnote{ A particular class of third order terms were extracted in \cite{Brattan:2011my}.} in various dimensions \cite{Haack:2008cp, Bhattacharyya:2008mz} in pure gravity, as well as in the presence of additional fields \cite{Bhattacharyya:2008ji, Erdmenger:2008rm, Banerjee:2008th}. In all cases one can go on to show that the solution to Einstein's equation thus obtained in regular and one can in fact determine the location of the event horizon explicitly order by order in $\varepsilon$ \cite{Bhattacharyya:2008xc}. The construction has paved the way for exploring new phenomena in fluid dynamics (notably the imprint of quantum anomalies \cite{Son:2009tf}) and we refer the reader to \cite{Landsteiner:2012kd, Jensen:2013kka} for further details.\footnote{ We  note in passing that the blackfold approach to black branes  employs a similar gradient expansion to construct approximate gravitational solutions. These ideas have not yet been applied to general curved world-volumes (see however footnote \ref{f:probestring}). We refer the reader to \cite{Emparan:2009cs,Emparan:2009at, Camps:2012hw} for a review of these concepts.
}

Now that we have the general hydrodynamic solution, we can specialize to static boundary geometries of the form \eqref{stybdy} and look for time-independent
solutions. Clearly, plugging in the solutions to the fluid dynamical conservation equations on \eqref{stybdy} with no time dependence for the fields into the bulk metric $g^{(k)}_{AB}\left(u^\mu(x), T(x)\right)$ one obtains immediately a stationary black hole AlAdS spacetime. In some specific cases one can check the consistency of this approximation against explicitly known solutions; for e.g., one can take the exact form of the (neutral) rotating AdS black hole  \cite{Gibbons:2004ai} and cast it into the form predicted by the fluid/gravity construction. This was carried out explicitly in \cite{Bhattacharyya:2008mz}. One
novel feature they found was that not only did the solution match with predicted result (up to second order), but it also hints that certain higher order transport coefficients\footnote{ Since we are talking about
stationary states on compact spaces these transport data are more accurately described as thermodynamic response parameters (they are adiabatic in that they cause no entropy production). } are  suitably related to allow for a neat resummation of the fluid perturbation expansion into the exact solution of Einstein's equations. Whether this phenomena persists for more general stationary configurations is unclear, but construction of bulk stationary black holes with more general boundary metrics could help shed light on this issue.

To forestall confusion as to whether such a resummation would be possible in the full hydrodynamic expansion, we hasten to add that hydrodynamics as any effective field theory is an asymptotic expansion with zero radius of convergence. From the bulk perspective this is immediately clear since the moduli which we treated as collective coordinates correspond to the lowest tier of the quasinormal mode spectrum (the so called hydrodynamic quasinormal modes). One expects the perturbation theory to break down when we encounter the higher quasinormal modes \cite{Bhattacharyya:2008jc} and this has been spectacularly verified by the recent numerical work \cite{Heller:2011ju}.

%
\subsection{Plasma flows in stationary spacetimes}
\label{sec:flows}
%

Let us now return to the static
boundary spacetime \eqref{stybdy}. While we argued above that there is a preferred equilibrium configuration which is stationary in these geometries, it is not the only such configuration that preserves the Killing field $\partial_t$. To illustrate the basic point consider a simple static boundary metric in 2+1 dimensions \cite{Figueras:2012rb} (see \cite{Khlebnikov:2010yt, Khlebnikov:2011ka} for some interesting earlier work on this topic)
\begin{equation}
\varphi(x) = a_i(x) =  0 \,, \qquad \gamma_{ij}(x) \, dx^i\,dx^j = dx^2 + \gamma(x) \, dy^2
\label{sty1}
\end{equation}	
which has a spatial Killing field $\partial_y$.  Suppose that we wish to study configurations of our field theory on these backgrounds with energy-momentum flux in the non-Killing $\partial_x$ direction. If we use our intuition from the fluid description above, we could for example set up a flow of the field theory plasma with velocity $u^t = \frac{1}{\sqrt{1-v(x)^2}}, \ u^x = \frac{v(x)}{\sqrt{1-v(x)^2}}, \ u^y =0$ with the velocity profile and the local temperature obtained by solving the hydrodynamic conservation equations in terms of $\gamma(x)$ and its derivatives. Since the flow is time-independent it follows that the configuration is stationary. Nevertheless, it is easy to check that such a flow has a non-trivial shear and leads to local entropy production.

What is the holographic dual of such flows? The dual solution must respect the symmetries which the boundary configuration possesses: in the present case these are the Killing symmetries $\partial_t$ and $\partial_y$. For slowly varying functions $\gamma(x)$ one can plug  the solution to the fluid dynamical conservation equation as before into the fluid/gravity solution to learn immediately that the bulk solution is a stationary black hole geometry. However, this black hole horizon is not a Killing horizon since the horizon generators have a non-vanishing component along the non-Killing $\partial_x$ direction. One can explicitly check this using the horizon generators for the fluid/gravity solutions \cite{Bhattacharyya:2008xc}. The stationary but non-Killing nature of the horizon is consistent with the rigidity theorems \cite{R1,R2,Hollands:2006rj} since the horizon is not compactly generated.

In fact, one can go well beyond the hydrodynamic regime and construct numerical solutions  \cite{Figueras:2012rb,Fischetti:2012vt} which allow one to examine the situation when $\gamma(x)$ has rapid variations on the boundary.\footnote{ We refer the reader to these reference for discussions of methods, though we mention that \cite{Figueras:2012rb,Fischetti:2012vt} take rather different approaches to boundary conditions.  In particular,
\cite{Figueras:2012rb} uses a form of excision that requires finding the solution somewhat inside the horizon, while
\cite{Fischetti:2012vt} uses smoothness of the horizon as a guide and solves only for the exterior.} The numerical simulations in fact reveal interesting phenomena in this regime
\cite{Figueras:2012rb}. The flow velocity identified by writing the boundary stress tensor in the form
\begin{align}
\frac{v}{1+v^2} = \frac{T^{tx}}{T^{tt}+T^{xx}}
\label{flowvel}
\end{align}
can be superluminal ($v>1$)! Note that the  stress tensor itself remains sensible and satisfies the appropriate Ward identities expected from field theory.
What is curious is the interpretation of the energy-momentum flow in terms of a velocity is breaking
down -- though such a break-down is natural when the gradients become large and we exit the hydrodynamic regime.
What has yet to be examined is whether such flows are stable.  However, some intuition for this question may be developed by considering rotating black holes in \AdS{d+1} \cite{Gibbons:2004ai} which, as remarked earlier, correspond to rigidly rotating plasma flows in ESU$_{d}$ \cite{Bhattacharyya:2007vs,Bhattacharyya:2008mz}. For large black holes, these flows are hydrodynamic, but for small black holes these flows no longer look like a conformal fluid. When a small black hole rotates sufficiently rapidly the dual flow velocity defined by
\eqref{flowvel} becomes superluminal. In fact, this occurs precisely when there is no longer a globally timelike Killing vector in the bulk, and hence a classical superradiant instability develops \cite{Hawking:1999dp}. It is thus natural to expect superradiant instabilities of any flow with $v>1$ as defined by \eqref{flowvel}, though the precise extent to which this is so remains to be investigated.

It is also interesting to examine these flows from a different viewpoint. By an appropriate boost one can convert the stationary flow to an explicitly time dependent one on the boundary. In this presentation the time translation is not manifest, so that we may interpret the flow in terms of dynamically perturbing some otherwise stationary plasma.  It is thus related to the physics of quenches in strongly coupled systems.

More generally, similar flows are quite interesting in the related context of heat transport in strongly coupled field theories. In this setting one naturally considers flows driven by temperature differences \cite{Fischetti:2012vt} instead of by the initial velocity of \cite{Figueras:2012rb}.  As particular examples,
by a suitable conformal transformation we can convert a wide class of spacetimes of the form \eqref{sty1} to a black hole metric so that the temperature difference is provided by the Hawking effect.  We will  therefore return to these examples  in \S\ref{sec:5} when we discuss field theories on black hole backgrounds.

%
\subsection{Quenches \& equilibration on curved	 spacetimes}
\label{sec:numericalGR}
%

 While the fluid/gravity correspondence allows us to explore the space of slowing varying solutions as described above, it fails to capture the full physics in the presence of more rapid spatio-temporal variations. While the details of any internal space $X$ may also become relevant in this regime, studies to date have focussed on the universal physics of the AlAdS factor and we shall do so as well below. Such examples are not only interesting in their own right, but also serve to illustrate how far one can push the fluid/gravity approximation discussed above. Impressive strides in numerical general relativity in recent years, mainly motivated by the desire to use AdS/CFT to understand dynamics of strongly coupled systems encountered in particle physics (heavy-ion collisions) or condensed matter applications, have resulted in an improved understanding of dynamical phenomena in asymptotically locally AdS spacetimes (cf., \cite{Hubeny:2010ry} for a review).

As a simple example where one needs to go beyond hydrodynamics, consider the process of equilibration following a rapid disturbance of an interacting quantum system. If the external perturbation, which can be induced by varying background sources such as the metric of ${\cal B}_d$, is carried out over a short time-interval then one is said to be discussing a quench of the system.\footnote{ In condensed matter physics a quench refers to a situation wherein one prepares an energy eigenstate of some Hamiltonian and then changes the evolution operator by some external control parameters. This has the effect of making the initial state behave typically as an excited state of the new Hamiltonian and the interesting questions in this context relate to the process of attaining equilibrium in the subsequent evolution.}

 In \cite{Chesler:2010bi} the authors analyzed the effect of changing the boundary metric over a period of time on the CFT dynamics; in their model the metric on ${\cal B}_d$ was taken from Minkowski in the far past to the Minkowski metric in the future, albeit with a change in the spatial length scales. To wit, the geometry on ${\cal B}_4$ was taken to be
 \begin{equation}
ds^2 = -dt^2 + e^{B_\partial(t)}\, dx_\perp^2 + e^{-2\,B_\partial(t)}\, dx_\parallel^2 \,, \qquad B_\partial(t) = \frac{\alpha}{2}\, \left(1 -\tanh(t/\tau)\right)
 \label{}
 \end{equation}	
$\alpha$ gives control over the amplitude of the perturbation and $\tau$ controls the temporal rate of change in the source. This is clearly a special case of \eqref{stybdy}.

This temporal change in the metric of the background does {\em work} on the quantum degrees of freedom. The system which starts out at early time in a state with spatial anisotropy equilibrates at late time to a hydrodynamic state. The dual geometry which must now be constructed numerically is an explicit dynamical black hole solution \cite{Chesler:2008hg}, but one that at late times is very well approximated by the fluid/gravity metric.

One  can in fact understand the process of black hole formation by a different kind of perturbation expansion as described in \cite{Bhattacharyya:2009uu}. Let us take the amplitude of the perturbation $\alpha$ to be small
(together with $\tau \to 0$ for simplicity).
We now have a small parameter $\alpha$ and can attempt to understand its effect in a perturbation expansion -- except that now we can take recourse to the familiar ground of linearized gravity.
One can explicitly work out the effect of the change in the boundary condition on the bulk metric analytically, order by order in the $\alpha$-expansion.

However, this perturbation theory is a bit more subtle and one needs to resum the perturbative series to see the correct physics. The authors of \cite{Bhattacharyya:2009uu} argue that the process of resummation shifts the starting point of the perturbation expansion from pure AdS to a Vaidya-AdS geometry in the causal future of the boundary perturbation. In the limit $\tau \to 0$ this change is instantaneous for the boundary observer -- more precisely, the boundary stress tensor changes from the vacuum value to the thermal stress tensor \eqref{zeroTmn} discontinuously at $\tau =0$. While we phrased the results for a homogenous change in the metric in the spatial directions, introducing inhomogeneities which are slowly varying spatio-temporally will only result in a fluid-dynamical stress tensor. One can in effect combine the perturbation expansion of linearized gravity with the long-wavelength fluid approximation to construct the bulk spacetime to any desired order approximately. This was already presaged in the original work and was recently verified in \cite{Balasubramanian:2013oga} (see also \cite{Buchel:2013gba} for universal physics in abrupt quenches).

The numerical methods developed to tackle the problem of strong coupling isotropization described above, have been applied to a wide variety of problems in recent years. In particular, much effort has gone into understanding:
\begin{itemize}
\item
The  boost invariant Bjorken flow \cite{Bjorken:1982qr} solution of relativistic hydrodynamics relevant for heavy-ion collisions \cite{Chesler:2009cy, Beuf:2009cx,Beuf:2009zz, Heller:2011ju, Heller:2012je}.
\item Detailed studies of black hole formation leading to isotropization and thermalization in strongly coupled systems \cite{Bantilan:2012vu, Heller:2012km, Heller:2013oxa}; in particular focussing on the onset of validity of hydrodynamics.
\item Numerical analysis of colliding shock wave solutions which again result in black hole formation \cite{Chesler:2010bi, Casalderrey-Solana:2013aba, Chesler:2013lia}.
\item Thermalization following a quench in a strongly coupled system and its interplay with phase boundaries \cite{Bhaseen:2012gg, Buchel:2012gw, Buchel:2013gba, Buchel:2013lla}.
\item Numerical construction of gravitational solutions corresponding to turbulent flows in relativistic hydrodynamics \cite{Adams:2013vsa, Green:2013zba} (cf., \cite{Carrasco:2012nf} for studies of relativistic turbulence and \cite{Adams:2012pj} for holographic superfluid turbulence).
\item  The nature of thermalization in global AdS which may be non-linearly unstable to formation of black holes \cite{Bizon:2011gg,Dias:2012tq,Buchel:2013uba}.
\end{itemize}
For a survey of these ideas and numerical techniques employed we refer the reader to \cite{Chesler:2013lia}. In general while the numerical methods provide complete characterization of  the bulk spacetime, the general picture that emerges is that a  judicious use of perturbative methods often allows one to understand the physics at a semi-quantitative level. In the numerical simulations the fluid/gravity solutions appear to capture the geometry (and thus boundary stress tensor dynamics) better than at the 10\% level even when the pressure anisotropies are of the order 50\%. In general however one should note that hydrodynamization is not equivalent to thermalization -- see \cite{Chesler:2009cy, Chesler:2010bi, Heller:2011ju,Bantilan:2012vu} for studies of the validity of the hydrodynamic approximation.

An open question in this area is understanding which of these results apply universally to holographic field theories independently of the choice of internal space $X$. In particular, since many of the above bulk solutions contain structures small compared to the
AdS scale $\ell$, one should consider whether there is a proclivity to localize on $X$. From a field theory perspective, such localization would indicate that rapid changes in the system induce spontaneous breaking of global symmetries.

%
\section{CFTs on black hole backgrounds}
\label{sec:5}
%

Perhaps the deepest and most interesting aspects of curved spacetime QFT arise for field theories on black hole backgrounds.  The production of Hawking radiation, with its close connections to black hole thermodynamics and the information problem, continues to be a source of inspiration, frustration, and fascination in modern theoretical physics.  From the QFT in curved space point of view it is thus natural to ask what effects strong coupling might have on Hawking radiation, and in particular what might be learned from AdS/CFT.

Now, at the beginning we should state clearly that one does not expect to learn fundamental lessons about either black hole thermodynamics or the information problem in this way.  Those subjects belong to the realm of dynamical gravity where in particular the Hawking radiation back-reacts on the geometry.  While AdS/CFT does indeed have implications for (at least stringy) quantum gravity, see e.g., \cite{Almheiri:2013hfa,Marolf:2013dba}, one sees this by using the dual QFT to describe dynamical quantum gravity in the bulk.  In contrast, we focus here on placing the dual CFT on a non-dynamical black hole background ${\cal B}_d$. This means that our CFT Hawking radiation cannot back-react on ${\cal B}_d$; one may say that the dual CFT has vanishing gravitational constant: $G_{CFT} = 0$. Thus we speak of a `boundary' black hole that does not evaporate, and which has effectively infinite entropy and heat capacity. In other words, it plays the role of a heat bath.\footnote{\label{brane} It is possible to make this $G_{CFT}$ non-zero by altering the AdS/CFT set up.  One may either move the boundary in to finite distance, as in Randall-Sundrum brane-world constructions \cite{Randall:1999ee,Randall:1999vf}, or simply alter the AdS boundary conditions \cite{Compere:2008us} if one is willing to accept the resulting ghosts \cite{Andrade:2011dg}. However, both result in so called weak/weak dualities which provide no insight into non-perturbative quantum gravity.  In particular,
following \cite{deHaro:2000wj}, and
as explained in detail in \cite{Figueras:2011gd}, in the Randall-Sundrum context the case of small $G_{CFT} > 0$ is perturbatively connected to our $G_{CFT}=0$ setting and the results may be extrapolated accordingly.  From the brane-world point of view, the limit of small $G_{CFT}$ is precisely the limit of large brane-world black holes.}

It is this observation which provides the real motivation for studying CFTs on black hole backgrounds.  Heat transport is a classic problem of field-theoretic interest, and we will see that AdS/CFT provides interesting insights into how this process proceeds.   The most novel phenomena are associated with transport between finite-sized heat sources at finite separations.  Although the idea may seem surprising to the uninitiated, the universal coupling of all fields to gravity means that black holes are in fact an excellent way to introduce model heat baths that treat all CFT degrees of freedom on an equal footing.  In particular, there is no need to choose separate coupling constants for each field.  While one must still specify the details of the black hole metric to be chosen, one may hope for robust results that depend only on basic scales associated with the black hole such as its horizon size $R_{\partial BH}$ and Hawking temperature $T_{\partial BH}$. Here the symbol $\partial$ indicates that we discuss a black hole in what will be the boundary manifold ${\cal B}_d$ of our bulk AlAdS spacetimes.  We discuss below the extent to which this hope has been verified, as well as what remains to be explored.

Below we discuss only aspects of the problem that arise in the universal bulk gravity sector in the sense of \S\ref{sec:2}.
While the bulk solutions and associated CFT phases below thus exist for any choice of internal space $X$, many interesting bulk solutions will have structure on the AdS scale $\ell$, which is generally comparable to scales in $X$.  As a result, the details of $X$ may affect the interpretation of such solutions.  For example, the stability of such solutions to localization on the internal space (in a manner analogous to that described for Schwarzschild-AdS in \S\ref{sec:GLreview}) may well depend on the choice of $X$.  Such effects have not yet been investigated in detail for the solutions below, and we encourage the reader to bear this in mind.  We will return briefly to this point in \S\ref{sec:DFTstory}, but for now we simply remind the reader that the issue cannot arise in the hydrodynamic regime of \S\ref{sec:flgra}. It is thus of concern only when this approximation fails or at its margins.

%
\subsection{Droplets, Funnels, and CFTs on asymptotically flat black holes}
\label{sec:funnels}
%

For concreteness, let us first place our CFT on an asymptotically flat spacetime containing a single spherically symmetric static black hole.  For $d \ge 4$ we may take this to be the familiar Schwarzschild solution or a higher dimensional generalization, though the fact that the metric on ${\cal B}_d$ need not satisfy any field equations allows us far greater freedom if desired.

\subsubsection{Droplets and funnels in thermodynamic equilibrium}
\label{sec:eqmfunnels}

General path integral arguments tell us that even strongly interacting QFTs admit a Hartle-Hawking state $|HH\rangle$ which describes a thermal state at temperature $T_{CFT} = T_{\partial BH}$ in such black hole backgrounds, is regular at the event horizons \cite{Kay:1988mu}, and is invariant under all spacetime symmetries; see e.g., \cite{Jacobson:2003vx,Ross:2005sc} for reviews.  In free QFTs, the Hartle-Hawking state far from the black hole is characterized by a manifestly thermal stress tensor of magnitude dictated by the QFT density of states, associated with a thermal density  of  particles at infinity.

\begin{figure}[h]
\begin{center}
\includegraphics[scale=0.3]{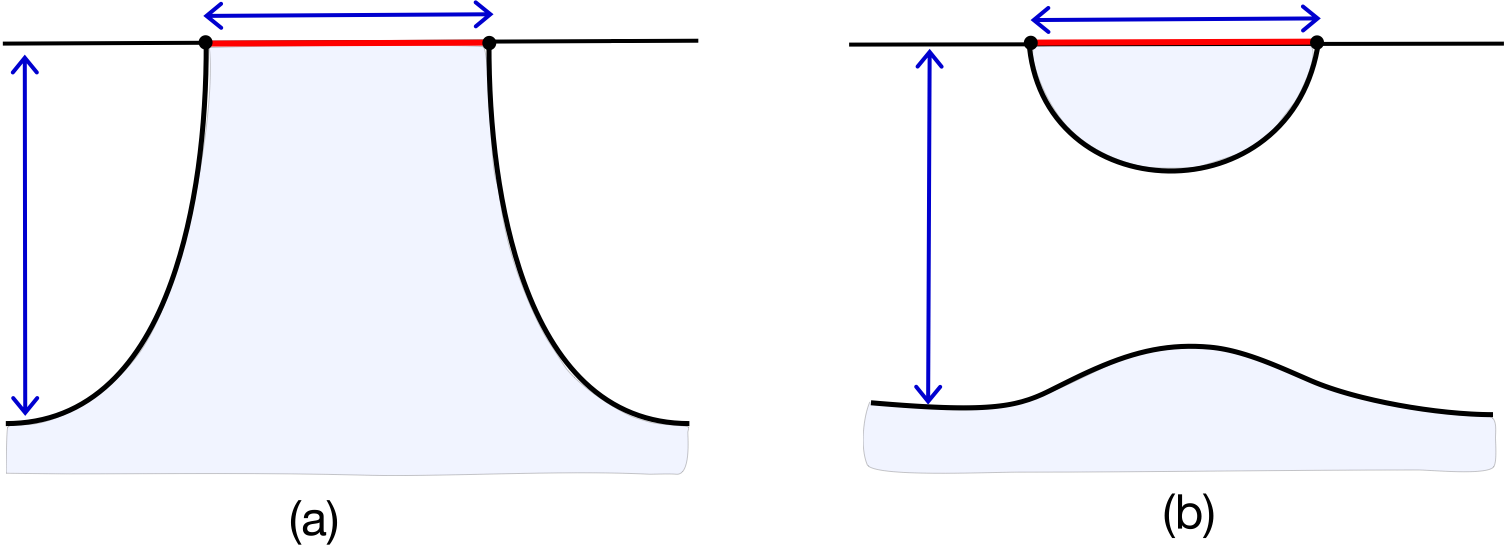}
\setlength{\unitlength}{0.1\columnwidth}
\begin{picture}(0.3,0.4)(0,0)
\put(-2.5,4.4){\makebox(0,0){$R_{\partial BH}$}}
\put(3.3,4.36){\makebox(0,0){$R_{\partial BH}$}}
\put(1.25,2.6){\makebox(0,0){$T_{\partial BH}^{-1}$}}
\put(-4.25,2.6){\makebox(0,0){$T_{\partial BH}^{-1}$}}
\put(-3,2.5){black funnel}
\put(2.8,2.8){$\nearrow$}
\put(2,2.5){black droplet}
\put(1.82,1.1){(deformed) planar black hole}
\end{picture}
 \caption{{\bf (a):} Black funnel and  {\bf (b):} black droplet above a deformed planar black hole.  The top line corresponds to the boundary with the dots marking the horizon of the boundary black hole.  The shading indicates the region behind the bulk horizons.}		
\label{f:fundrop}
\end{center}
\end{figure}
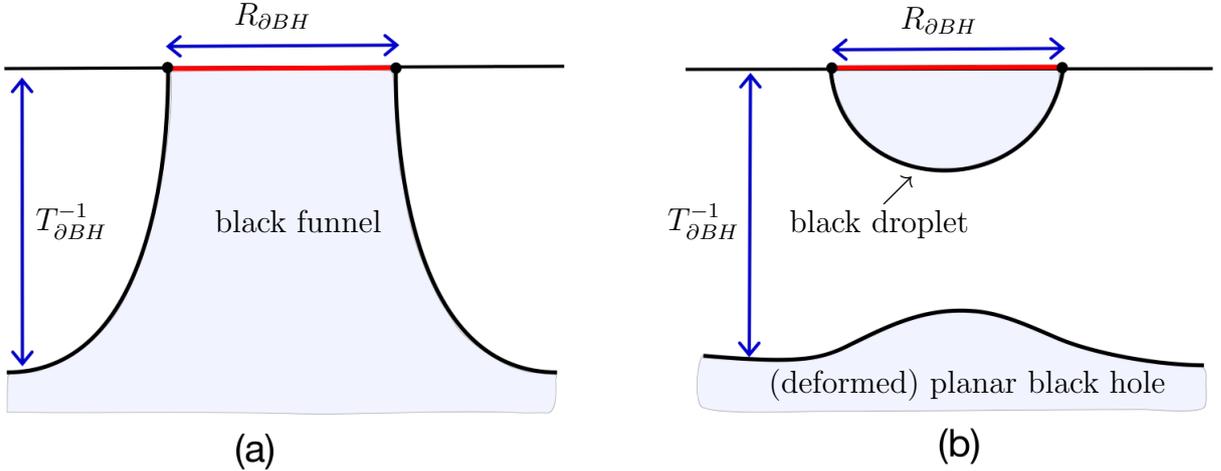

Furthermore, in free QFTs $|HH \rangle$ can in fact be {\it defined} by regularity at both past and future horizons.  It is therefore natural to attempt to construct bulk duals to $|HH\rangle$ by looking for smooth static AlAdS spacetimes with bifurcate Killing horizons at temperature $T_{bulk-BH} = T_{\partial BH}$ and having conformal boundary ${\cal B}_d$.  This in particular requires the bulk horizon to connect to that of the boundary black hole; see Fig.~\ref{f:fundrop}.  We may say that the bulk horizon ends on the boundary horizon, though the fact that this `end' is infinitely far away means that the bulk horizon is not compactly generated.

While there is much interesting physics to discuss here, a useful starting point is the observation that theories at large $c_\text{eff}$ can admit multiple phases, and that these are associated in AdS/CFT with multiple (say, static) bulk solutions satisfying identical boundary conditions, and in particular with the same conformal boundary ${\cal B}_d$. The importance of phase transitions for the current problem was recognized in \cite{Fitzpatrick:2006cd}, and the particular bulk phases relevant to the current problem were described in \cite{Hubeny:2009ru}, whose treatment we largely follow below.

Let us begin by recalling from \S\ref{sec:3} (see footnote \ref{planarphases}) that the dual CFT has two phases at finite temperature, even when ${\cal B}_d$ is just flat Minkowski space.  One phase corresponds to empty planar AdS space in the bulk, but this phase is sub-dominant for all $T > 0$.  In the dominant phase the bulk solution is a planar Schwarzschild-AdS black hole.  Each solution corresponds to a saddle point of the bulk Euclidean thermal path integral.  Since $|HH\rangle$ can be defined by a similar Euclidean path integral, we expect the dominant bulk solutions to approximate planar Schwarzschild-AdS near the asymptotically flat region of ${\cal B}_d$ and thus to have a finite temperature horizon far from the boundary black hole. On the other hand, the horizon of the boundary black hole must also extend into the bulk.  There are then two natural classes of possible bulk solutions:  If the above two horizons connect to form a single smooth horizon, the solution is said to describe a black funnel.    If they are instead disconnected, the solution describes a black droplet, suspended above a (deformed) planar black hole.  These two situations are sketched in Fig.~\ref{f:fundrop}.

The above argument may also be read in reverse: In a droplet solution small changes to the boundary black hole, including changes to its temperature, need not affect the distant parts of the planar horizon.  In contrast, since the surface gravity must be constant along any connected Killing horizon, changing $T_{\partial BH}$ in an equilibrium funnel solution necessarily requires a corresponding change in the part of the bulk solution that describes the CFT plasma in the asymptotic region of ${\cal B}_d$.  In other words, funnel solutions efficiently transport heat between the boundary black hole and the distant regions of the CFT while droplet solutions do not.\footnote{ However, as discussed further in \S\ref{sec:DFTstory}, the finite temperature of the droplet horizon does indicate that a region of the CFT near the boundary black hole has been filled with what one may call a thermal plasma.  So even in the droplet phase heat is efficiently conducted some finite distance from the horizon of the boundary black hole.}
This argument, presented in \cite{Hubeny:2009ru}, has been verified by direct studies of connected \cite{Fischetti:2012ps,Figueras:2012rb,Fischetti:2012vt} and isolated \cite{Haehl:2012tw} boundary black holes. Some of the salient results will be reviewed in \S\ref{sec:Hawking}. See also \cite{Emparan:2013fha} for a simple solvable example describing heat transport along an analogous horizon.
 It is very interesting to ask how this fundamental difference arises from the field theory point of view and, while we will provide a useful perspective on this issue in \S\ref{sec:DFTstory} below, much remains to be understood.
We emphasize that a weakly coupled field theory would have no transitions of this sort, so that holography is giving us an important insight into new phenomena that can occur at strong coupling.

Since only the conformal structure of ${\cal B}_d$ can be relevant, and since the natural parameters $R_{\partial BH}$, $T_{\partial BH}$ admit a single scale-invariant combination $R_{\partial BH}T_{\partial BH}$, the phase structure should be determined by this product.  Using the intuition that any structure of size $R$ on the boundary naturally produces a bulk image extending to a depth $z \sim R$ in terms of the natural Fefferman-Graham coordinate $z$ and the fact that the planar black hole horizon \eqref{eq:PoincareSchwarz} resides at $z_h = \frac{4\pi}{d\,T}$, \cite{Hubeny:2009ru} argued that droplets should dominate at small $R_{\partial  BH}T_{\partial BH}$ while funnels should dominate when this product is large. For a Schwarzschild boundary black hole, $R_{\partial  BH}T_{\partial BH} = {\cal O}(1)$, and hence our argument says little about whether the preferred equilibrium bulk solution is a funnel or droplet.

Verifying the above phase transition requires constructing both funnels and droplets for a family of boundary black holes with varying $R_{\partial  BH}T_{\partial BH}$.
Unfortunately finding the equilibrium bulk funnel and droplet solutions for $d>2$ is very challenging, even for spherically symmetric boundary black holes, due to the geometries depending not only on the bulk radial coordinate, but also on the boundary black hole radial direction. The desired funnel solutions have been found \cite{Santos:2012he} numerically using methods introduced in \cite{Headrick:2009pv} (see \cite{Wiseman:2011by} for a review of these methods). Constructing droplet solutions in thermal equilibrium with planar horizons turns out to be more challenging, though as we discuss in \S\ref{sec:dropsub} below, static droplets have been constructed in the presence of lower-temperature horizons. In all cases we caution the reader that the droplet solutions have structure on the AdS scale $\ell$ and that possible instabilities to localization on some internal space $X$ remain to be investigated.

One may also consider the low-dimensional case $d=2$, which turns out to be under full analytic control \cite{Hubeny:2009ru}. We note that all $1+1$ boundary black holes effectively have $R_{\partial BH} = \infty$, since they are best viewed as dimensional reductions of higher-dimensional planar black holes.  So the above discussion predicts that funnels always dominate when ${\cal B}_2$ contains a black hole.  From the bulk point of view, this turns out to be a consequence of the lack of local degrees of freedom in pure 2+1 gravity; all 2+1 Einstein metrics with negative cosmological constant are locally equivalent to pure AdS${}_3$. Requiring the $1+1$ boundary to have the topology of the plane and taking the solution to be static then ensures that it is globally diffeomorphic to a subset of empty AdS${}_3$.  But all Killing horizons in  AdS${}_3$ are connected, so droplet solutions cannot exist. On the other hand, the funnel solution can be readily constructed \cite{Hubeny:2009ru}.  Furthermore, one can use similar arguments to show that it transports heat as expected in a 1+1 CFT \cite{Fischetti:2012ps}.

\subsubsection{Droplets: dynamic verses thermodynamic equilibrium}
\label{sec:dropsub}

The droplet solutions of Fig.~\ref{f:fundrop}(a) admit a natural static deformation.  If we assume that this solution is stable, the disconnected nature of the horizons should allow us to dial the temperature $T_{planar}$ of the deformed planar horizon independently of the $T_{droplet}$ characterizing the droplet horizon.  The bulk solution, and thus the CFT stress tensor, should remain smooth and static.\footnote{ Smooth spacetimes with multiple horizons at different temperatures by themselves are not exotic. Schwarzschild-de Sitter is a familiar example and many more can be found using the C-metric of \cite{Plebanski:1976gy}; see e.g. the explicit discussions in \cite{Hubeny:2009kz,Caldarelli:2011wa}.}  Since it is the droplet horizon that connects to the boundary black hole, we set $T_{droplet} = T_{\partial BH}$.  But it is the planar horizon that describes the thermal CFT plasma in the asymptotic region of ${\cal B}_d$, so this means that the CFT temperature at infinity ($T_{planar}$) need not match the boundary black hole temperature $T_{\partial BH}$.
While the classical bulk gravity solution is in dynamic equilibrium, it is clearly no longer in thermodynamic equilibrium.

That a static bulk solution, and hence a static CFT stress tensor can occur when $T_{\partial BH}$ is not equal to $T_{planar}$ is very surprising from the point of view of weakly coupled QFT, and thus appears to crucially depend on the  holographic CFT being strongly coupled and having large $c_\text{eff}$.\footnote{
See \cite{Kaloper:2012hu} for another perspective.}
At weak coupling, fixing boundary conditions that set different regions of the CFT to different temperatures would instead lead to a stationary flow of heat with the CFT stress tensor exhibiting a corresponding flux of energy.

Nonetheless, the above droplet picture is verified by explicit construction of the gravitational bulks. Indeed, droplet solutions were first constructed in the extreme limit $T_{planar} \to 0$, so that the CFT stress tensor vanishes at leading order in $c_\text{eff}$ in the asymptotic region of ${\cal B}_d$. In fact, an analytic solution is available for $d=3$ \cite{AM}.  For $d=4$ and Schwarzschild boundaries the solution was found numerically in \cite{Figueras:2011gd}. In these cases, droplets at temperature $T_{\partial BH}$ sit in a zero temperature vacuum but continue to remain static.
As discussed in \S\ref{sec:DFTstory} below, since these solutions have $T_{planar} = 0$ and appear to be stable at least to perturbations on the AlAdS$_d$ factor \cite{FiguerasWiseman}, it is natural to think of them as describing Unruh states of our CFTs on black hole backgrounds (as opposed to the Hartle-Hawking states considered in \S\ref{sec:eqmfunnels}).

It is important that classical bulk gravity is not the full story, but as we have seen in \S\ref{sec:alads} only describes the contribution to the stress tensor of order $\sim {\cal O}(c_\text{eff})$. Indeed, bulk Hawking radiation also contributes to the CFT stress tensor, but is only seen at subleading order $\sim {\cal O}(1)$. In a case where $T_{\partial BH} \ne T_{planar}$ there will certainly be a flux of bulk Hawking radiation between the localised droplet horizon and infinity. Thus the surprise for field theory is that for $T_{\partial BH} \ne T_{planar}$ the CFT stress tensor is static at order ${\cal O}(c_\text{eff})$, although it does contain a stationary heat flux at subleading order in $c_\text{eff}$.

In the CFT at leading order in $c_\text{eff}$ we may picture droplet solutions as representing a static halo of hot plasma surrounding the boundary black hole horizon. The halo is attracted to the horizon gravitationally, and is also strongly interacting with itself in an attractive manner (recall that bulk gravity tells us the CFT interaction between regions of positive mass is attractive). This balances the radiation pressure within the halo due to it being in contact with the boundary horizon (heat conduction within the halo is efficient). At subleading order in $c_\text{eff}$ this hot halo gently exchanges radiation with infinity and the horizon leading to a small energy flux.

Such behaviour is very different to the expectation from weakly coupled field theory, where naive Hawking radiation arguments would indicate a heat flux at order ${\cal O}(c_\text{eff})$. Indeed related weak coupling arguments in the closely related context of large braneworld black holes (see footnote \ref{brane}) led \cite{Tanaka:2002rb,Emparan:2002px} to claim that no large static black holes could exist in the Randall-Sundrum braneworld model and thus placing significant phenomenological constraints on such theories.  This created considerable controversy despite the existence of a static analytic solution for $d=3$ \cite{Emparan:1999wa}. The controversy was exacerbated by the fact that early numerical solutions for $d=4$ braneworld black holes could not reach large sizes \cite{Wiseman:2001xt,Casadio:2002uv,Karasik:2003tx,Kudoh:2003xz,Kudoh:2003vg,Kudoh:2004kf,Karasik:2004wk,Yoshino:2008rx}. However, subsequently using the relation between large braneworld black holes and droplets, large static braneworld black holes have finally been numerically constructed \cite{Figueras:2011gd,Figueras:2011va,Abdolrahimi:2012qi,Abdolrahimi:2012pb,FiguerasWiseman}.

To make our droplet discussion more concrete, we now take a moment to briefly review the $d=3$ analytic solution of \cite{AM} with $T_{planar} = 0$.  This solution is closely related to the $d=3$ brane-world black holes of \cite{Emparan:1999wa} but has thus far remained unpublished\footnote{ In fact, due to an oversight whereby \cite{Hubeny:2009ru,Hubeny:2009kz} did not examine double roots of the function $F$ below, these works implied that no such solution exists.} -- though it will be further explored in \cite{RMtocome}.  This solution is just the AdS C-metric of \cite{Plebanski:1976gy} in the form
\begin{equation}
\label{Cmetric}
ds^2_{d+1} = \frac{\ell^2}{(x-y)^2} \left(-F(y)\, dt^2 + \frac{dy^2}{F(y)} + \frac{dx^2}{G(x)} + G(x)\, d \phi^2    \right),
\end{equation}
with $F(y) =  y^2 + 2 \mu \, y^3$ and $G(x) = 1 - F(x)$.  We are interested in the regime $\mu > \frac{1}{3\sqrt{3}}$ where $G$ has a single real zero at some $x_0$.  The bulk metric is regular at $x_0$ when $\phi$ has period $\Delta \phi = 4 \pi/|G'(x_0)|$.  For $\mu < \frac{1}{3\sqrt{3}}$ there is at least one exposed conical singularity in the bulk, while the case $\mu = \frac{1}{3\sqrt{3}}$ can be interpreted as an example of the detuned droplets described in \S\ref{sec:DFTstory}.

The metric (\ref{Cmetric}) solves the bulk Einstein equations with the usual negative cosmological constant and has an AlAdS boundary ${\cal B}_d$ at $z =  x-y =0$.  Introducing $\chi = - 1/x$ and extracting a factor of $y^2$ from the factor in parenthesis in \eqref{Cmetric}, we may write the line element on ${\cal B}_d$ as conformal to
\begin{equation}
\label{Cboundary}
ds^2_{d} = - \left(1 - \frac{2\,\mu}{\chi}\right) dt^2 + \frac{d\chi^2}{\left(1 - \frac{2\mu}{\chi} \right) G(-1/\chi) }  + \chi^2 \,G(1/\chi)\, d \phi^2  .
\end{equation}
Since $G(0) =1$, the metric \eqref{Cboundary} is asymptotically flat at $\chi =\infty$.    There is also a Killing horizon of temperature $T_{\partial BH} = 1/2\mu$ and size $R_{\partial BH} =  2\mu$ at $\chi = 2\mu$; i.e., with $T_{\partial BH} R_{\partial BH} = 1$ for all $\mu$.  This Killing horizon extends into the bulk along the surface $y = -1/2\mu$ and there is a further Poincar\'e-like zero-temperature horizon at $y=0$.  Defining the coordinates to satisfy $y \le 0$, $y \le x$ (so that $z > 0$), and $x \le x_0$ one finds\footnote{ Other related solutions are defined by taking either $y \ge 0$ (and thus removing the droplet horizon) or $z \le 0$ (which leads to a naked singularity).} that the solution outside these horizons is smooth. One may also identify the horizon at $y = -1/2\mu$ as a droplet horizon since it ends at the rotation axis $x= x_0$.  One may check that the boundary stress tensor vanishes at large $\chi$ like $1/\chi^3$.

The numerical construction of droplets with $T_{planar} = 0$ has recently been extended to droplets with $T_{planar} \ne 0$ \cite{SW}, and also to $d=5$ rotating droplets -- in which ${\cal B}_5$ contains a rotating black hole metric \cite{Figueras:2013jja,Fischetti:2013hja} -- with $T_{planar} = 0$.  We also mention that the results of \cite{Kaus:2009cg} describe the region near any droplet horizon with $T_{droplet} = 0$.\footnote{\label{f:probestring} We note that droplet like solutions were examined in an approximation scheme (motivated by the blackfold approach) by considering a probe black string in the Schwarzschild-AdS$_5$ geometry in \cite{Haddad:2012ss, Haddad:2013tha}; see also \cite{Grignani:2012iw}.}

Before we leave this section, we mention two more exact bulk solutions describing CFTs on black hole backgrounds which are static, but not in thermal equilibrium.  Their construction rests on the observation that the Poincar\'e AdS solution generalizes readily to any Ricci-flat ${\cal B}_d$, and likewise the AdS-soliton solution generalizes to any  ${\cal B}_d$ which is a product of a Ricci-flat metric with a Scherk-Schwarz circle.
The corresponding line elements \eqref{padsX} and \eqref{eq:soliton} continue to solve the bulk equations of motion when the flat metric on constant $z$ surfaces is replaced by any Ricci-flat geometry.   Taking this to be Schwarzschild (for $d \ge 4$) then yields solutions that may be called respectively the Schwarzschild-AdS black string (see \cite{Chamblin:1999by}) and the Schwarzschild-soliton-string \cite{RMprivate}.  In both cases one finds that the boundary stress tensor is essentially unchanged from that of Poincar\'e-AdS and the AdS-soliton
 due to what may be interpreted as a precise cancelation between negative energy from vacuum polarization and positive energy from a thermal plasma near the boundary black hole; see \S\ref{sec:DFTstory} for further details.

The Schwarzschild-AdS black string is both unstable \cite{Gregory:2000gf} and singular \cite{Chamblin:1999by} at the Poincar\'e horizon.
As emphasized in \cite{Figueras:2013jja} the energy density derived from the boundary stress tensor for the droplet solutions with $T_{planar} = 0$ is actually negative outside the horizon. Thus not only is the Schwarzschild-AdS black string pathological, as mentioned above, but we see that since it has vanishing stress tensor it also has a higher energy than the droplet to which it presumably dynamically decays. Physically it appears that the halo of CFT plasma for the Schwarzschild-AdS black string has higher energy, and some of the plasma must either disperse during the instability or fall into the boundary black hole, leaving the smaller halo of the droplet.

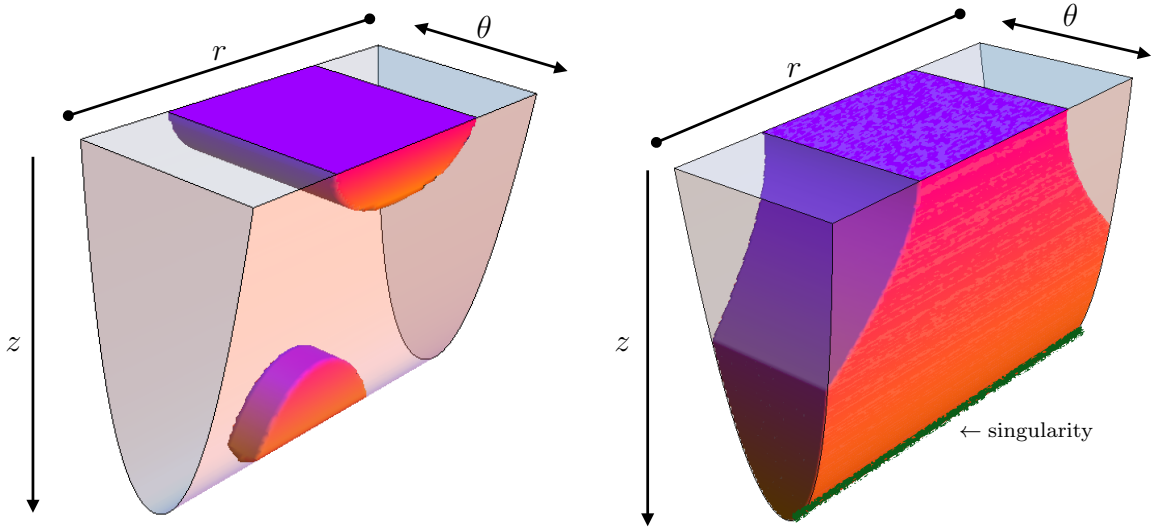
\begin{figure}[t]
\centerline{
\includegraphics[width=1\textwidth]{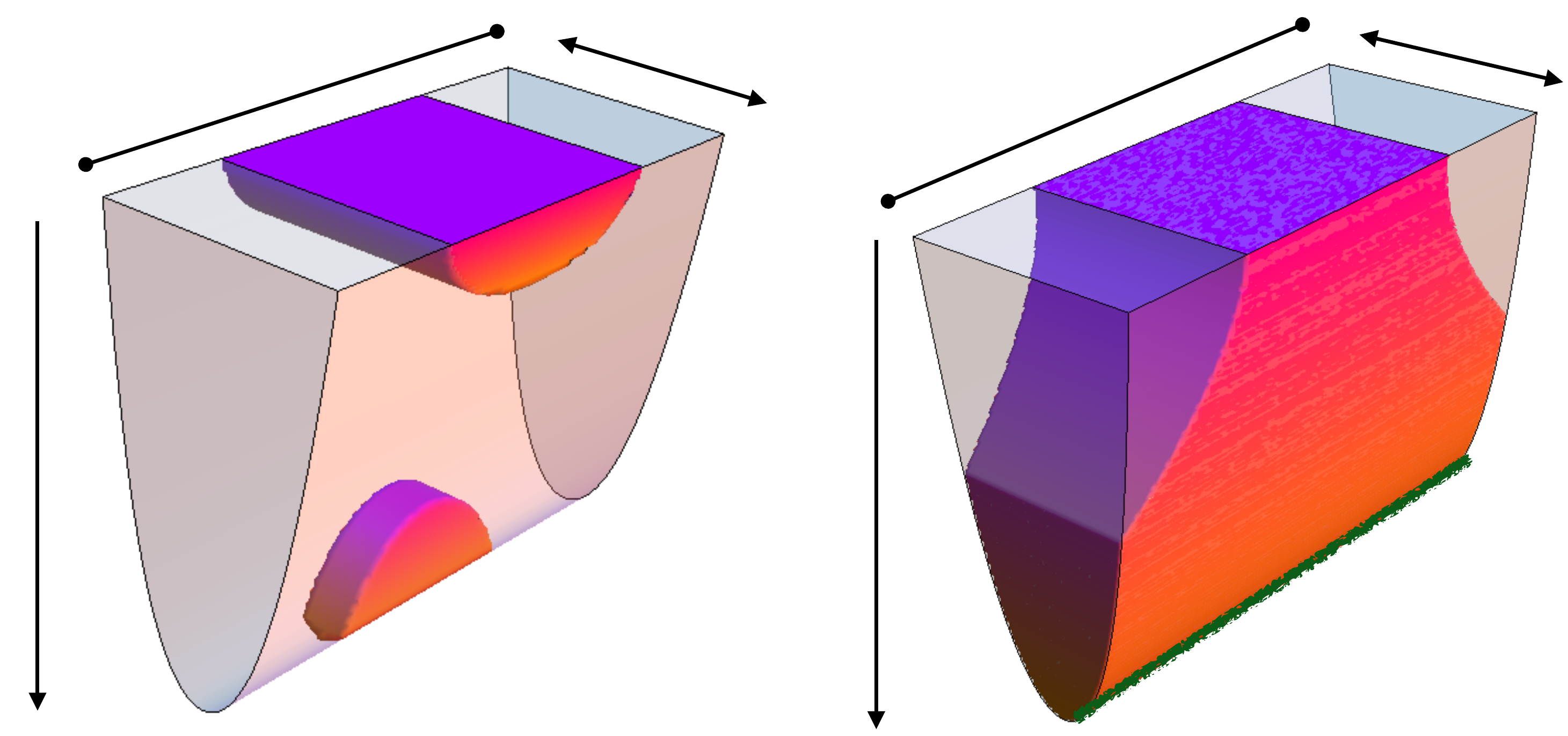}
}
\setlength{\unitlength}{0.1\columnwidth}
\begin{picture}(0.3,0.4)(0,0)
\put(5.3,2){$z$}
\put(0.,2){$z$}
\put(1.8,4.55){$r$}
\put(6.8,4.4){$r$}
\put(4.1,4.75){$\theta$}
\put(9.2,4.85){$\theta$}
\put(8.3,1.25){{\scriptsize $\leftarrow \text{singularity}$}}
\end{picture}
\caption{Illustration of the potential phases when we have a boundary black hole smeared over a Scherk-Schwarz circle. The Scherk-Schwarz circle $\theta$ has been depicted as an interval for ease of visualization and this direction closes off in the bulk smoothly. As indicated in the text, we anticipate having either a droplet phase with a plasma ball at the bottom (left) or a funnel phase (right). The thick green line indicates the singularity hidden behind  the  funnel horizon.}
\label{fig:solphases}
\end{figure}

The Schwarzschild-soliton-string is manifestly everywhere regular, and it is unstable on the AlAdS factor only when the boundary black hole is small compared to the boundary size of the Scherk-Schwarz circle \cite{Haehl:2012tw}. It would be interesting to understand the fate of the instability in this latter case and whether it necessarily results in pinching off the string into a droplet\footnote{ True droplet solutions where the droplet horizon does not reach the tip of the soliton (where the spatial circle contracts) have only recently been constructed numerically \cite{SW2}.} (perhaps with an additional plasma ball sitting at the bottom of the soliton), or whether it can sometimes result in what one might describe as a hot expanding plasma ball that remains connected to the boundary black hole and which grows to approximate a black funnel at large times; see Fig.~\ref{fig:solphases} for an illustration.

%
\subsection{Global Droplets and Funnels}
\label{sec:globalFunDrop}
%

\begin{figure}[t]
\centerline{
\includegraphics[width=0.6\textwidth]{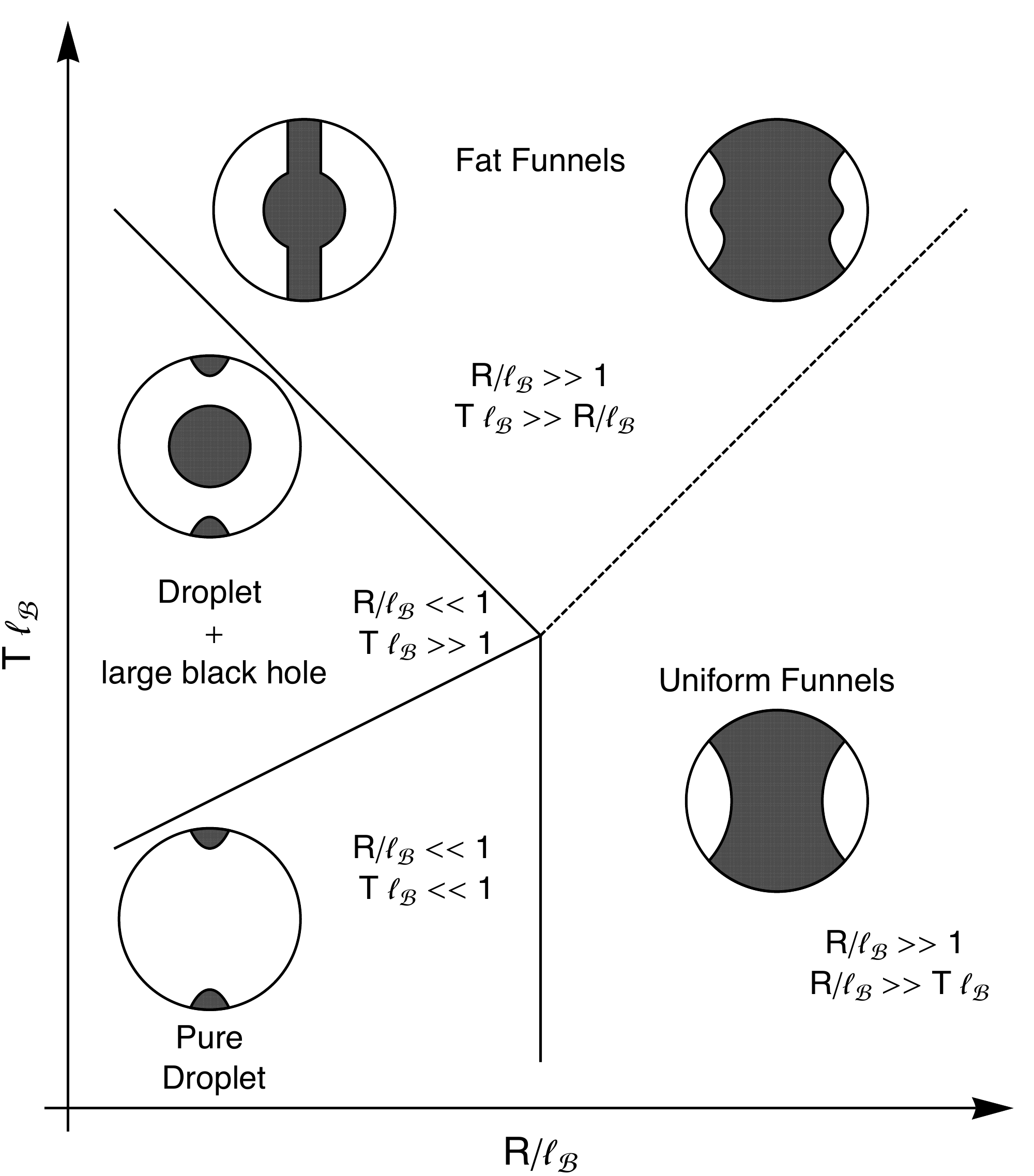}
}
\caption{Phases of droplets and funnels expected to dominate at extreme values of $R_{\partial BH}/\ell_{\cal B}$ and  $T \ell_{\cal B}$ from \cite{MS}. Phase boundaries (solid lines) are rough and approximate, and additional phases could potentially hide near such boundaries.  It is unknown whether the dashed line running up and to the left is a phase boundary or a crossover.  The pictures indicate the rough shape of the horizon in each phase.  On the left side of the fat funnel phase the horizon approximates that of the global Schwarzschild-AdS${}_{d+1}$ black hole \eqref{globAdSS} over much of the spacetime, while on the right side the fat funnel is not parametrically different from the uniform funnel.   See \cite{MS} for details.  We note that when $d=3$ and ${\cal B}_3$
is built from BTZ as in \S\ref{sec:BTZ}, such solutions have $T \propto R_{\partial BH}$ and so run from the pure droplet phase up to the right in the vicinity of the dashed line separating fat and uniform funnels.  A similar $d \ge 4$ construction using Schwarzschild-AdS${}_d$ would run from the upper left corner toward the middle along a curve somewhere near the the solid line between fat funnels and droplets with central black holes.  It would then again run up to the right in the vicinity of the dashed line.}
\label{fig:Full}
\end{figure}

While astrophysical concerns might motivate a focus on asymptotically flat ${\cal B}_d$, it turns out that one can gain further control by generalizing the discussion to static but spatially-compact ${\cal B}_d$ -- spacetimes that one may think of as versions of the Einstein static universe (ESU) that also contain black holes.    Furthermore, this context is at least as interesting as asymptotically flat ${\cal B}_d$ from the perspective of studying heat flow in the CFT between finite-sized heat sources at finite separations.  Below, we first discuss the $d=3$ case for ${\cal B}_3$ built from BTZ black hole solutions (\S\ref{sec:BTZ}).  In this context, \cite{Hubeny:2009rc} found that one may analytically construct all funnel and droplet solutions that do not spontaneously break a certain symmetry. As a result, the expected funnel/droplet phase transition may be explicitly analyzed and confirmed.

The higher dimensional case is also quite interesting and introduces qualitatively new features relative associated with the Gregory-Laflamme instability \cite{Gregory:1993vy} that lead to multiple funnel phases known as uniform, fat, and thin.  While this case is quite interesting, we refer the reader to \cite{MS} for detailed study of the case where ${\cal B}_d$ is built from two copies of Schwarzschild-AdS${}_d$, and also for discussion of expectations when ${\cal B}_d$ describes more general black holes in the ESU.  Here we confine ourselves to simply presenting Fig.~\ref{fig:Full}, which summarizes the phases expected to dominate in various regions of parameter space. In addition to the above-mentioned uniform and fat funnels, these phases contain droplets that may be accompanied by additional central black holes playing a role analogous to that of the planar black hole shown in Fig.~\ref{f:fundrop}(a) in the context of asymptotically flat ${\cal B}_d$.  We also mention that other solutions describing $d=3$ droplets and funnels on spatially compact (though not ESU) ${\cal B}_3$  can be constructed from the AdS C-metric \cite{Hubeny:2009kz,Caldarelli:2011wa}.

\subsubsection{A simple example with $d=3$ and BTZ ${\cal B}_3$}
\label{sec:BTZ}

The starting point for our BTZ discussion is the observation \cite{Hubeny:2009rc} that certain ${\cal B}_3$ built from a pair of BTZ black holes have an $SO(2,1) \times U(1)$ conformal symmetry. From the perspective of the bulk theory, this will be a symmetry of the boundary conditions and corresponds to a bulk isometry of any phase in which it is not spontaneously broken.    The phases preserving this symmetry can then be mapped via double Wick rotation ($t \rightarrow i\tilde \phi$ and $\phi \rightarrow i\tilde t$ for an azimuthal angle $\phi$) to static spherically-symmetric solutions classified by Birkhoff's theorem.  In other words, any transition between phases preserving the above symmetry can be mapped to the Hawking-Page transition described in \S\ref{sec:HP} between thermal AdS${}_4$ and Schwarzschild-AdS${}_4$.

The desired SO(2,1) symmetry may be seen by considering the non-rotating\footnote{ The rotating case was also discussed in \cite{Hubeny:2009rc}.} BTZ metric \cite{Banados:1992wn,Banados:1992gq} in the form
\begin{equation}
ds^ 2_{BTZ} =  \frac{r^2\ell_{\cal B}^2}{R_{\partial BH}^4} \left[ - \left( 1 - \frac{R_{\partial BH}^2}{r^2}\right) \frac{R_{\partial BH}^4\, dt^2}{\ell_{\cal B}^4} + \frac{R_{\partial BH}^4}{r^4} \frac{dr^2}{1 - R_{\partial BH}^2/r^2} +  \frac{R_{\partial BH}^4}{\ell_{\cal B}^2} d\phi^2 \right],
\end{equation}
in which an overall factor of $\frac{r^2}{\ell_{\cal B}^2}$ has been pulled out relative to the usual presentation.  Introducing $\eta = R_{\partial BH}^2 t/\ell_{\cal B}^2$, $\tau = R^2_{\partial BH} \phi/\ell_{\cal B}^2$, and  $\sin \theta = R_{\partial BH}/r$ for $\theta \in [0,\pi]$ this becomes
\begin{equation}
\label{dS2}
ds^2_{BTZ} = \frac{\ell_{\cal B}^2}{R_{\partial BH}^2 \sin^2 \theta} \left[ -\cos^2 \theta \, d\eta^2 + d \theta^2 +  d\tau^2 \right].
\end{equation}
The factor in square brackets is just the $\theta \ge 0$ half the static patch of two-dimensional de Sitter space times an ${\bf S}^1$ associated with the $\tau$ (n\'ee $\phi$)  circle.  The full static patch is then obtained by gluing together two copies of \eqref{dS2} along the surface $\theta = 0$, or equivalently just extending the range of $\theta$ to negative values so that it runs over $[-\frac{\pi}{2}, \frac{\pi}{2}]$.   Including the region behind the BTZ horizons leads to global $\text{dS}_2 \times {\bf S}^1$, as is clear from the fact that this is the maximal analytic continuation preserving periodicity of $\tau.$

\begin{figure}[t]
\centerline{
\includegraphics[width=0.4\textwidth]{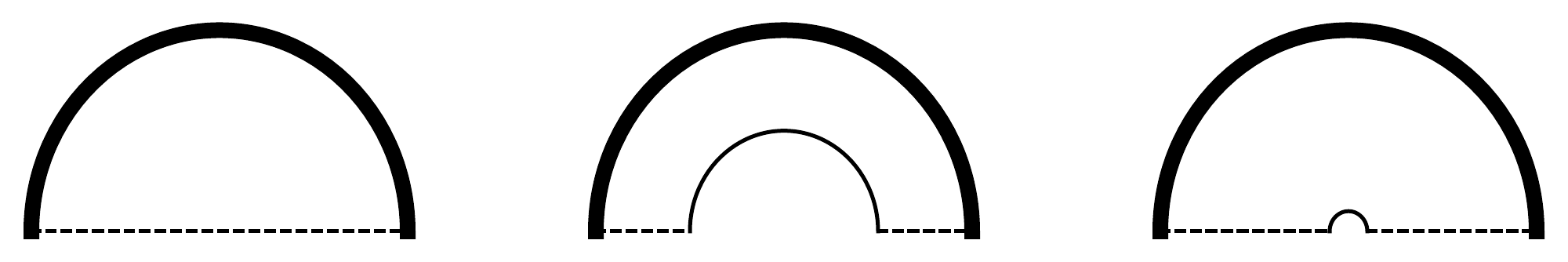}
}
\caption{Surfaces of constant Killing time and constant azimuthal angle for global AdS${}_4$ (left) and the large (center) and small (right) Schwarzschild-AdS${}_4$ black holes.  The upper solid semi-circle is the AdS boundary.  When it exists, the smaller solid semi-circle is the horizon.  Dashed lines are the azimuthal rotation axes.  Horizons and rotation axes are exchanged under double Wick rotations. So interpreting solid lines as rotation axes and dashed lines as horizons, these figures thus also depict static slices of spacetimes with BTZ boundary black holes corresponding to  the funnel (right) and short (center) and long droplet phases discussed in the text.  Figure from \cite{MS}.
}
\label{fig:HP}
\end{figure}
The key point is the $SO(2,1)$ isometry group of $\text{dS}_2$, though there is also another $U(1)$ symmetry associated with the ${\bf S}^1$ factor.  After analytically continuing $\text{dS}_2$ to ${\bf S}^2$ the boundary becomes ${\bf S}^2 \times {\bf S}^1$, which is just thermal (Euclidean) ESU${}_3$.   Applying the same operations to any bulk solution with $SO(2,1) \times U(1)$ isometries gives a spherically-symmetric static solution asymptotic to empty (global) AdS${}_4$.  So in the absence of spontaneous symmetry breaking the phases we seek can be obtained by double Wick rotation of empty (thermal) AdS${}_4$ together with the large and small Schwarzschild-AdS${}_4$ black holes.

The relation to funnels and droplets is best understood by considering a co-dimension 2 surface in each spacetime having constant Killing time $t$ and constant azimuthal angle $\phi$.  As shown in Fig.~\ref{fig:HP}, in the bulk AdS${}_4$ and Schwarzschild-AdS${}_4$ solutions such surfaces end on some combination of horizons (fixed points of $\partial_t$, indicated by solid lines) and the rotation axis for $\phi$ (fixed by $\partial_\phi$ and indicated by dashed lines).  Since double Wick rotation exchanges horizons with rotation axes, we must now reinterpret Fig.~\ref{fig:HP} by taking the solid lines to  describe  rotation axes and the dotted lines to describe horizons.  This makes the solution obtained from AdS${}_4$ a black funnel; it has no rotation axis and a horizon that runs from one side of the ESU${}_3$ boundary to the other.  Explicit calculation shows that it is in fact the BTZ black string of \cite{Emparan:1999fd}.  In contrast, the solutions that come from double Wick rotating the AdS${}_4$ black holes are droplets.  They have a disconnected horizon with the two components separated by a rotation axes.   Before \cite{Hubeny:2009rc}, these droplet solutions had been previously discussed from a different perspective by \cite{Balasubramanian:2002am}.  Following \cite{MS}, we call them short and long droplets based on the depth that their horizons penetrate into the bulk.

The expected phase transition is now verified by studying the Euclidean action of the above solutions.  Since this action is manifestly invariant under double Wick rotation the calculation is equivalent to that performed by Hawking and Page \cite{Hawking:1982dh}, provided of course that one properly understands the mapping of parameters between the two pictures.  The exchange of the time and space circles $(t, \phi) \rightarrow (\tilde \phi, \tilde t)$ implies that the high-temperature behaviour of the Hawking-Page transition maps to that of low-temperature BTZ boundary black holes.  Thus \cite{Hubeny:2009rc} found that the droplet phases exist only for $T_{\partial BH} = T_{BTZ} \le \frac{1}{2 \pi \sqrt{3}\, \ell_{\cal B}}$, and that the short droplet phase dominates the canonical ensemble below $T_{BTZ} = \frac{1}{4 \pi\, \ell_{\cal B}}$. At higher $T_{BTZ}$ the ensemble is dominated by the funnel phase, though this phase exists at all $T_{BTZ}$.   The long droplet phase locally maximizes the Euclidean action and can be interpreted as mediating the transition between the other two phases.

%
\subsection{The ultra-static frame: Field theory interpretations of the \\
droplet/funnel transition}
\label{sec:DFTstory}
%

We have thus far used holography to deduce the existence of both funnel and droplet phases for the relevant strongly coupled large $c_\text{eff}$ CFTs, and to see that the transition is controlled by the parameter $R_{\partial BH} T_{\partial BH}$.  But it is natural to ask how this physics is to be explained, or at least interpreted, on the field theory side of the correspondence.    The funnel phase is of course qualitatively just a large $c_\text{eff}$ generalization of familiar free field Hawking radiation, and it is also what one would expect from the hydrodynamic description of the CFT plasma (see \S\ref{sec:flgra}).  But how can we understand the novel behaviour of the droplet phase shown in Fig.~\ref{f:fundrop}(b) in which the boundary black hole completely decouples from an ambient CFT plasma?

We begin by noting that, even from the field theory side, it is natural for new effects to arise at small $R_{\partial BH} T_{\partial BH}$. This is because the hydrodynamic expansion of \S\ref{sec:flgra} is a gradient expansion with the relevant scale set by the fluid's temperature.  But any metric with a boundary black hole will clearly provide some kind of structure on the scale $R_{\partial BH}$ (and any finite sized heat source will lead to structure on an analogous scale set by its size), so hydrodynamics will fail badly\footnote{ As we will shortly see, the hydrodynamic expansion breaks down for Hartle-Hawking states at black hole horizons in the sense that gradients are not parametrically small.  The same is true a bit farther from the black hole for $R_{\partial BH} T_{\partial BH} \sim 1$. In these contexts hydrodynamics will not be a precision tool, and qualitatively new behaviours are possible though not guaranteed.  On the other hand, we would be surprised if any new behaviours outside the hydrodynamic regime fail to manifest themselves for $R_{\partial BH} T_{\partial BH} \ll 1$.} at small $R_{\partial BH} T_{\partial BH}$. What is interesting about the droplet phase is that it indicates significant qualitative,  and not merely quantitative, differences between the hydro and non-hydro regimes.

A possible explanation for the droplet behaviour, or a least what one might call an interpretation of such behaviour, was suggested in \cite{Hubeny:2009rc}.  The idea is to suppose that a field theory plasma at temperature $T$ far from the black holes admits a useful quasi-particle description.  Suppose also that the quasi-particles, being composite glueball-like objects, have a preferred finite size $R_{quasi}$ of order $1/T$. If the spring constant controlling the energy cost to change the quasi-particle size becomes large in the limit of large $c_\text{eff}$ and $\lambda$, then it will be difficult for black holes with $R_{\partial BH} < R_{quasi}$ to emit and absorb such quasi-particles.  This may be described by an effective grey body factor that vanishes in the limit of large $c_\text{eff}$ and $\lambda$ and thus at leading order by droplet-like behaviour.  One would very much like to derive this picture from the CFT, or at least to use the picture to make an in-principle independent prediction which might be tested, though this hope has yet not been realized.

One may pursue a different approach to explaining droplet physics by changing conformal frames and using the known confining properties of the CFT \cite{DM}.  In particular, given any static boundary metric $ds^2_d$ we may construct the corresponding ultra-static metric $\widetilde{ds}_d^2 = \Omega^2(x) ds_d^2$, where $\Omega(x)$ is independent of time and is chosen so that $|\partial_t|$ has unit norm at all $x$ with respect to the new metric $\widetilde{ds}_d^2$, cf., \eqref{ultrastat}.  In particular, for metrics of the form  \eqref{globAdSS} (for general $g(\rho)$) we find
\begin{equation}
\label{wt}
\widetilde{ds}_d^2 = \frac{ds_d^2}{g(\rho)} = -dt^2 + \frac{d\rho^2}{g^2(\rho)} + \frac{\rho^2}{g(\rho)} \,d\Omega_{d-2}^2.
\end{equation}
So when \eqref{globAdSS} represents a finite temperature black hole, and thus when $g(\rho)$ has a simple zero at some $\rho_0$ corresponding to its horizon, we may define $f' = df/d\rho$ as evaluated at $\rho=\rho_0$ and set $z = \frac{\rho_0}{2} \sqrt{|f'|(\rho -\rho_0)}$ to write \eqref{wt} near $\rho=\rho_0$ in the form
\begin{equation}
\label{wt2}
\widetilde{ds}^2 = ds_d^2 = -dt^2 + \frac{4}{|f'|^2} \left[ \frac{dz^2}{z^2} +\frac{1}{z^2}\,  d\Omega_{d-2}^2 \right] + \, \cdots.
\end{equation}
Here the ellipses represent terms that are subleading at small $z$.  The leading (explicit) term in \eqref{wt} agrees with the leading order term in the metric on $H^{d-1} \times {\mathbb R}$, where $H^{d-1}$ is the Euclidean-signature maximally-symmetric hyperbolic space with length scale $\ell_{hyp} = 2/|f'| = \frac{1}{2\pi \, T_{\partial BH}}$ set by the boundary black hole temperature in the original conformal frame.   The conformal transformation above is essentially the same as that used to discuss Rindler and de Sitter horizons in \cite{Headrick:2010zt,Marolf:2010tg,Hung:2011nu}. For more general static black holes with bifurcate horizons without spherical symmetry the result remains of the form \eqref{wt2} with $|f'|$ replaced by $\pi/T_{\partial BH}$ and with more general metric $d\Sigma^2$; i.e., it remains asymptotic to $\tilde H^{d-1} \times {\mathbb R}$ where $\tilde H^{d-1}$ is an asymptotically locally hyperbolic space analogous to the AlAdS spacetimes of \S\ref{sec:2}.    However, for simplicity we continue to assume spherical symmetry below.

The resulting $H^{d-1} \times {\mathbb R}$ will be filled with a thermal fluid at the original temperature $T_{\partial BH}$.  This may be seen from the fact that temperatures measured with respect to a fixed Killing field $\partial_t$ are invariant under the conformal transformation \eqref{wt} and that we know the bulk to contain a horizon with Euclidean period $\beta = 1/T_{\partial BH}$.   The general static spacetimes with the corresponding symmetries are the hyperbolic black hole (sometimes called topological black hole) solutions of \cite{Emparan:1998he,Birmingham:1998nr,Emparan:1999gf}.
Introducing $d\Sigma_{d-1}^2 = d\xi^2 + \sinh^2\xi\, d\Omega_{d-2}$ as the metric on the unit $H^{d-1}$, the metric for such a black hole with temperature
\begin{equation}
T_{hyp  BH}  = \frac{r_0^2\, \ell^{-2} \, d - (d-2)}{4\pi \, r_0}
\end{equation}
takes the form
\begin{equation}
\label{hypBH}
ds^2_{d+1} = - F(r) \frac{\ell^2\, dt^2 }{\ell_{hyp}^2} + \frac{dr^2}{F(r)} + r^2 d \Sigma^2_{d-1} \,,\quad F(r) = \frac{r^2}{\ell^2} -1 - \frac{r_0^{d-2}}{r^{d-2}} \left(\frac{r_0^2}{\ell^2} -1 \right).
\end{equation}

The particular case $T_{hyp  BH} = T_{\partial  BH} = \frac{1}{2\pi\,\ell_{hyp}}$
turns out to be just empty AdS${}_{d+1}$ in the Rindler-like coordinates associated with some uniformly accelerated observer, expressed in a conformal frame that has sent the boundary Rindler horizon to $\xi = \infty$.  So this statement parallels the usual one that the near-horizon behaviour of a finite-temperature black hole (in this case, on in the boundary spacetime) is well-approximated by Rindler space.  The vanishing of the boundary stress tensor for pure (say, Poincar\'e) AdS may then be said to explain the perhaps otherwise surprising result \cite{Emparan:1998he,Birmingham:1998nr,Emparan:1999gf} that the boundary stress tensor of \eqref{hypBH} vanishes (up to terms associated with the conformal anomaly for $d$ even) for the tuned case $T_{hyp BH} = T_{\partial  BH} = \frac{1}{2\pi\, \ell_{hyp}}$.   In the hyperbolic frame, the fact that the energy density becomes negative at lower temperatures \cite{Emparan:1998he,Birmingham:1998nr,Emparan:1999gf} may be interpreted as a negative energy from vacuum polarization so that the vanishing of $T_{tt}$  for the tuned case represents a fine cancelation between this negative vacuum energy and the positive energy density associated with the thermal plasma (and likewise for the other components of $T_{\mu \nu}$).  The corresponding cancelation near the boundary black hole horizon for a general tuned droplet or funnel is precisely what guarantees smoothness of the boundary stress tensor across the boundary black hole horizon, though for some special solutions (including the Schwarzschild-AdS black string and Schwarzschild-soliton-string mentioned at the end of \S\ref{sec:funnels} as well as their global AdS analogues -- see discussion in \cite{Gregory:2008br,Hubeny:2009rc}) the cancelation may persist much farther from the boundary black hole horizon.

However, it is also interesting that $T_{hyp BH} \neq T_{\partial BH}$ continues to give a smooth bulk geometry.  Following \cite{Headrick:2010zt,Marolf:2010tg,Hung:2011nu} we may invert the above steps to see that it corresponds to a CFT state defined by a Euclidean path integral with period $\beta \neq 1/T_{\partial BH}$, and thus on a Euclidean spacetime with a conical singularity.  In other words, as noted in \cite{Fischetti:2012ps}, it describes a CFT state at temperature $T_{CFT}  = T_{droplet} = T_{hypBH} = 1/\beta$ which is explicitly not in equilibrium with the boundary black hole.  Physically we may think that, while our CFT lives on a black hole background, some other infinitesimally thin heat bath at the new temperature $T_{CFT} = T_{droplet}$ has been placed just outside the black hole horizon.

A full definition of such a state requires some prescription for how to deal with the conical singularity. But for a CFT it is natural to proceed as above and to simply work in the ultra-static conformal frame where no further information is required.  The result \cite{Marolf:2010tg} is that the CFT stress tensor diverges at the boundary black hole horizon, associated with the fact that there \eqref{hypBH} is not AlAdS in the sense of \S\ref{sec:2} (though it again becomes AlAdS to the future of this horizon).  However,  this need not be of concern if our goal is to model general CFT heat sources instead of natural black hole states per se.  Solutions describing such `detuned' spacetimes may be found exactly in 2+1 dimensions \cite{Marolf:2010tg}, where they describe a one parameter generalization of the 2+1 black funnel of \cite{Hubeny:2009ru} to arbitrary $T_{CFT}/T_{\partial BH}$.  Solutions describing similarly `detuned' black holes have also been found numerically in higher dimensions \cite{Fischetti:2012vt}. However, it remains to investigate their stability in detail and indeed the results of \cite{Dias:2010ma,Belin:2013dva,moreHinstab} suggest that for $T_{CFT} < T_{\partial BH}$ there can be non-universal instabilities that depend on the choice of internal space $X$ through the spectrum of matter fields induced by Kaluza-Klein reduction to AdS${}_{d+1}$.

One feature that makes the above construction useful is that the fluid approximation of \S\ref{sec:flgra} becomes highly accurate for  $T_{CFT}/T_{\partial BH} \gg 1$ \cite{Fischetti:2012vt}. In particular, the limit $\ell_{hyp} \rightarrow 0$ is obtained by taking $T_{CFT}/T_{\partial BH} \rightarrow \infty$.
So solutions with general $T_{CFT}/T_{\partial BH}$ can be used to test numerical codes while gradually tuning $T_{CFT}/T_{\partial BH}$ back down to $1$ allows one to systematically study the validity of hydrodynamics. In the other direction, recalling \cite{Birrell:1982ix} that the Boulware vacuum of free theories on black hole backgrounds is defined to be the ground state in the static region,
taking $T_{CFT}=0$ naturally defines a notion of a Boulware state for our CFT \cite{Fischetti:2012ps}.  In contrast, one might define an Unruh state to be a stationary state that agrees with the large time behaviour obtained from placing the CFT on a collapsing black hole geometry subject to the initial condition that it begins in the Minkowski ground state. Assuming that droplet solutions above zero-temperature (Poincar\'e) horizons are stable, in the strict large $N$ (large $c_\text{eff}$) limit\footnote{ At least in certain settings one may expect tunneling to significantly modify this picture at finite $N$.  It would be interesting to understand this in more detail.
}
it is thus natural to expect that they describe such Unruh states \cite{Figueras:2011va}.\footnote{ Note that \cite{Figueras:2011va} incorrectly also assigned these solutions to the Boulware state. We expect the dual to the Boulware state would be a droplet whose bulk horizon was detuned to zero temperature.}

\begin{figure}[t]
\centerline{
\includegraphics[width=0.65\textwidth]{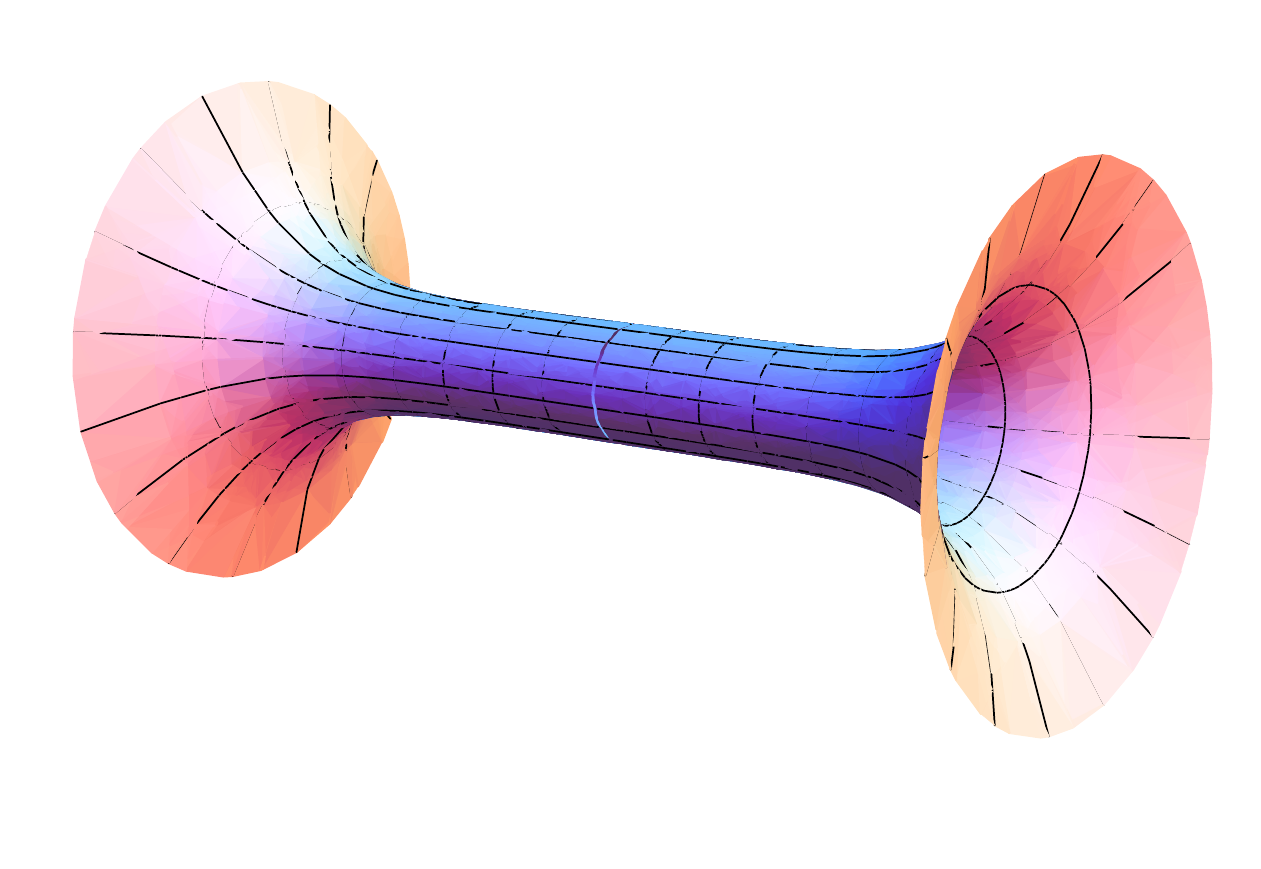}
}
\setlength{\unitlength}{0.1\columnwidth}
\begin{picture}(0.3,0.4)(0,0)
\put(5,3.5){${\bf S}^{d-2}$}
\put(2,4.6){$\xleftarrow{\text{ towards droplet horizon }}$}
\put(7.4,2.6){$\xrightarrow{\text{ towards bulk horizon }}$}
\end{picture}
\caption{A qualitative sketch of the spatial geometry between the droplet and bulk black hole horizon emphasizing the region where the transverse ${\bf S}^{d-2}$ to the black hole horizons is of nearly constant size set by $R_{\partial BH}$.}
\label{fig:longnecks}
\end{figure}
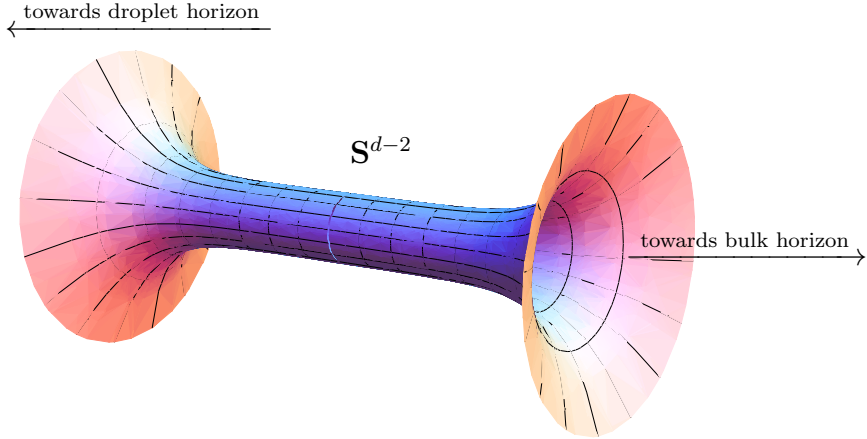

We now have in hand the tools to connect the droplet physics to that of colour confinement. Consider any droplet solution with two finite-temperature bulk horizons, such as the asymptotically flat droplets depicted in Fig.~\ref{f:fundrop}(a) or the global droplet solutions of \S\ref{sec:globalFunDrop}.  In the ultra-static frame there are in fact two asymptotic regions corresponding in the global case to the near-horizon regions of the two boundary black holes and in the asymptotically flat context to both the region near the boundary black hole and to the asymptotic region of the field theory.  In between there are one or more neck-like regions where the spheres ${\bf S}^{d-2}$ are small, with radii of order $R_{\partial BH}$; see Fig.~\ref{fig:longnecks}.

Let us now choose the metric on ${\cal B}_d$ to make $R_{\partial BH} T_{\partial BH}$ very small and also to make the neck-like region sufficiently long and uniform in the ultra-static frame.  This will certainly not be the case when the original metric on ${\cal B}_d$ is Schwarzschild or similar, but one may carefully construct boundary metrics to achieve this end.  For $d=3$ the neck of ${\cal B}_d$ approximates the Scherk-Schwarz setting\footnote{ The anti-periodic Scherk-Schwarz boundary conditions for fermions is natural here as it would be required for any black hole that forms by collapse from spacetimes that are asymptotically flat in the far past (so that the angular ${\bf S}^1$ is contractible).  However, for eternal black holes one could also choose to impose periodic fermion boundary conditions which would then forbid droplet solutions in the bulk.} discussed in \S\ref{sec:circleap}. We thus expect the CFT to confine and the bulk to be given by a horizon-free AdS${}_4$ soliton.  The case $d > 3$ is analogous with the Scherk-Schwarz ${\bf S}^1$ now replaced by ${\bf S}^{d-2}$.
While an analytic analogue of the AdS-soliton is then unavailable, it remains natural to expect confining behaviour. For $d=4$ such a solution is that obtained numerically in \cite{Copsey:2006br} for a boundary metric $\mathbb{R} \times {\bf S}^1 \times {\bf S}^2$,  where the ${\bf S}^1$ is taken to be large, and the ${\bf S}^2$ contracts smoothly in the bulk.
.

Thinking of the CFT as confined in the neck then makes it natural to have droplet-like bulk solutions with horizons on either side, describing plasmas in each asymptotic region that are unable to connect or even to significantly interact \cite{DM}.  In other words, in this conformal frame our droplet solutions are essentially infinite-sized versions of the plasma balls described in \S\ref{sec:pbs}.
The reason that the droplets appear static in the bulk classical gravity approximation is precisely the same as for plasma ball black holes.
In the CFT the hot plasma in one asymptotic region cannot radiate into the confining region at a rate visible in classical gravity. However, the process is seen if one includes Hawking radiation in the bulk.

This picture also leads to the expectation that, when the two horizon temperatures differ, the transition from droplet to funnel behaviour is controlled by the larger of the two temperatures.  Although we have focussed on CFTs, this argument can also be useful for non-conformal theories where the above conformal transformation merely introduces position-dependence into any non-marginal couplings.  Nevertheless, it remains very interesting to better understand the droplet/funnel transition from the field-theoretic point of view in the original black hole conformal frame.

%
\subsection{Flowing funnels and Hawking radiation}
\label{sec:Hawking}
%

We now turn to more dynamical results concerning holographic CFTs on black hole spacetimes. In principle, one might like to perform truly time-dependent calculations in this context analogous to those discussed in \S\ref{sec:numericalGR}, perhaps with ${\cal B}_d$ describing a time-dependent black hole that forms by collapse from initially flat space.  One would then like to observe dynamical formation of funnels and/or droplets as well as the ensuing conduction of heat along what are then called `flowing' funnels or the lack of conduction in droplet spacetimes..  However, such studies have not yet been performed.  Instead, the available results consist of:
\begin{enumerate}
\item[(i).] A perturbative study \cite{Haehl:2012tw} showing that, as expected, droplets do not readily transport heat to infinity.
\item[(ii).] A simple analytic but non-stationary model \cite{Emparan:2013fha} of heat flow along horizons constructed by allowing a black string to fall through a Rindler horizon, as well as older models \cite{FPV,Amsel:2007cw,Friess:2006kw,Figueras:2009iu} of localized particles and black holes falling across Rindler or Poincar\'e  horizons.
\item[(iii).] Analytic solutions describing stationary flowing funnels in AdS${}_3$ \cite{Fischetti:2012ps}.  This $d=2$ case is special because conformal symmetry in 1+1 dimensions guarantees that right- and left-moving heat flows do not interact and simply travel at the speed of light without changing shape.  As a result, the dual bulk AdS${}_3$ stationary flowing funnels have independent left- and right-moving temperatures $T_L,T_R$, each of which is constant along the horizon.
\item[(iv).]  Numerical results \cite{Figueras:2012rb,Fischetti:2012vt} describing heat flow along bulk horizons.   In particular, \cite{Fischetti:2012vt} constructed stationary AdS${}_4$ flowing funnels that describe dual CFT states on ESU${}_3$-like spacetimes (see \S\ref{sec:globalFunDrop}) containing two BTZ-like black holes at different temperatures.   While the flow in \cite{Figueras:2012rb} (recall the discussion in \S\ref{sec:flows}) was due to a non-zero fluid velocity in the asymptotic regions of ${\cal B}_d$ rather than to the temperature difference that drives flowing funnels, the bulk solution should nevertheless have similar properties.
\end{enumerate}

We will focus here on the results of \cite{Fischetti:2012vt}, which should  be indicative of general $d> 2$ stationary flowing funnels. We first discuss some features of the bulk horizon and then mention some salient results for the dual CFT.

While stationary, the bulk event horizon is not a Killing horizon. This is to be expected on general grounds \cite{Hubeny:2009rc} since Killing horizons necessarily have constant temperature while any notion of temperature along the horizon of a flowing funnel must vary continuously between the two boundary black holes.   As for the plasma flows of \S\ref{sec:flows}, since the horizon is not compactly generated, it may be stationary but non-Killing without being in conflict with the rigidity theorems \cite{R1,R2,Hollands:2006rj}.  In contrast, due to the special properties of AdS${}_3$ the $d=2$ flowing funnels of \cite{Fischetti:2012ps} do correspond to Killing horizons related to a certain conformal symmetry of ${\cal B}_2$.

In addition, there are several reasons to expect the past horizon to be singular \cite{Hubeny:2009rc}.  The first is that the presence of a stationary heat flux requires the CFT stress tensor to be singular on the past horizon of the boundary black hole. The second is that the temperature gradient\footnote{ For $d >2$.  As noted above, $T_L,T_R$ are constants for $d=2$.  And indeed the solutions of \cite{Fischetti:2012ps} have smooth past horizons.} suggests a continuous creation of entropy along the horizon, so that the horizon generators have positive expansion $\Theta$.  Thus by the Raychaudhuri equation $\Theta$ must have been infinite at some finite affine parameter in the past, and it is natural for this to occur on a singular past horizon.  This expectation was verified in \cite{Fischetti:2012vt} which found numerically that certain components of the Weyl tensor diverge at the past horizon in any orthonormal frame like $\lambda^{-11/6}$, where $\lambda$ is an affine parameter along each generator.  It would be interesting to understand this power law analytically.

\begin{figure}[t]
\centerline{
\includegraphics[width=0.8\textwidth]{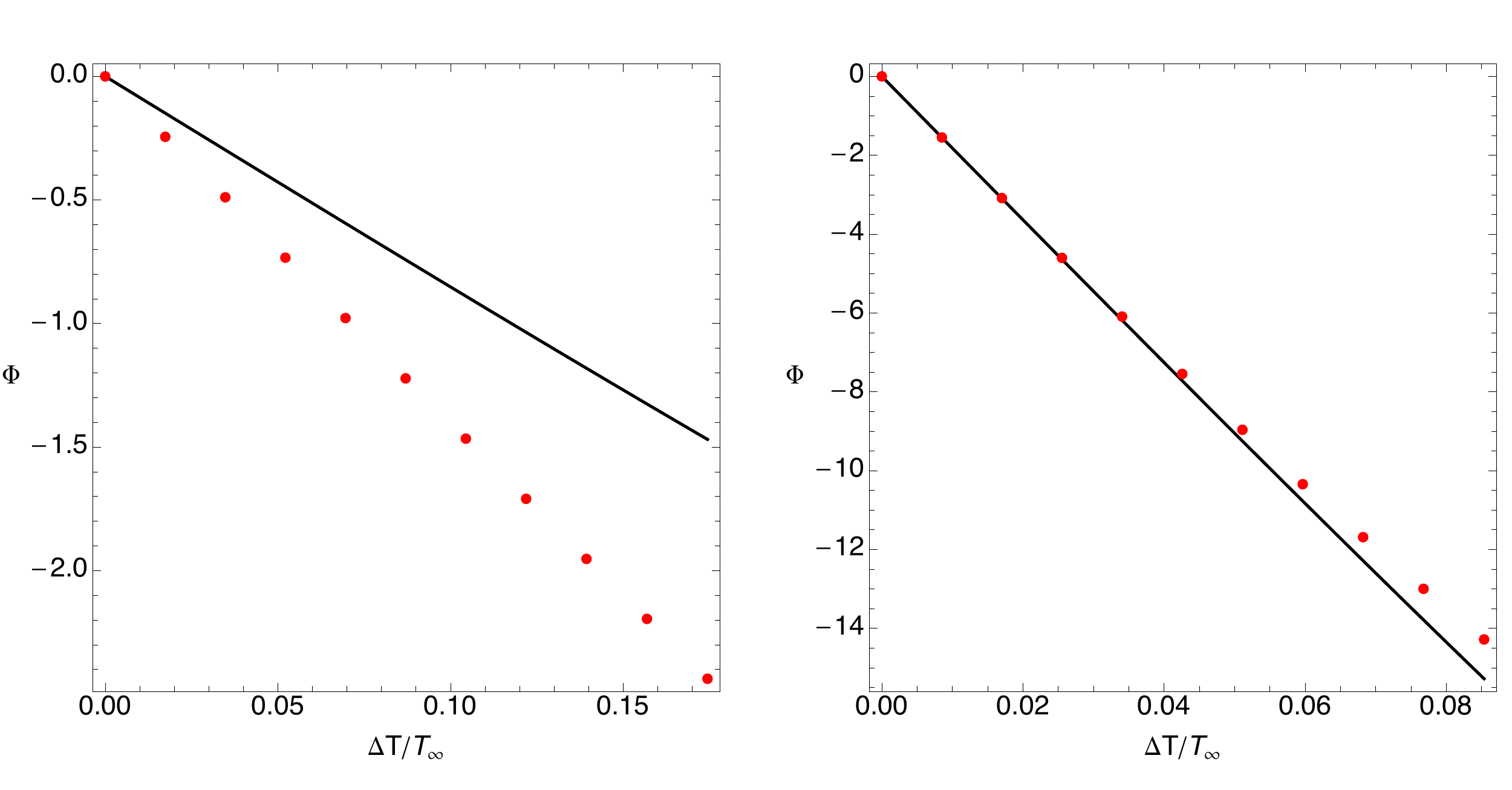}
}
\caption{Results from \cite{Fischetti:2012vt} for the total heat flux $\Phi$ as a function of the temperature difference $\Delta T$ and a parameter $T_\infty$ (roughly the average temperature at small $\Delta T$, see eqn. (5.1) of \cite{Fischetti:2012vt}) when both boundary black holes have $T_{CFT} \approx 1.1 T_{\partial BH}$ (left) and $T_{CFT} \approx 1.43 T_{\partial BH}$ (right).  The negative sign of $\Phi$ indicates a heat flow to the left due to the hotter black hole being placed on the right. The solid lines show predictions from linearized first order hydrodynamics while the points are measured from bulk solutions constructed numerically.  Non-linear and higher order hydrodynamic corrections make negligible difference at these small values of $\Delta T/T_\infty$.  At $T_{CFT} = T_{\partial BH}$ the discrepancy between bulk solutions and hydrodynamic predictions is even larger.
}
\label{fig:FFFlux}
\end{figure}

\begin{figure}[t]
\centerline{
\includegraphics[width=0.8\textwidth]{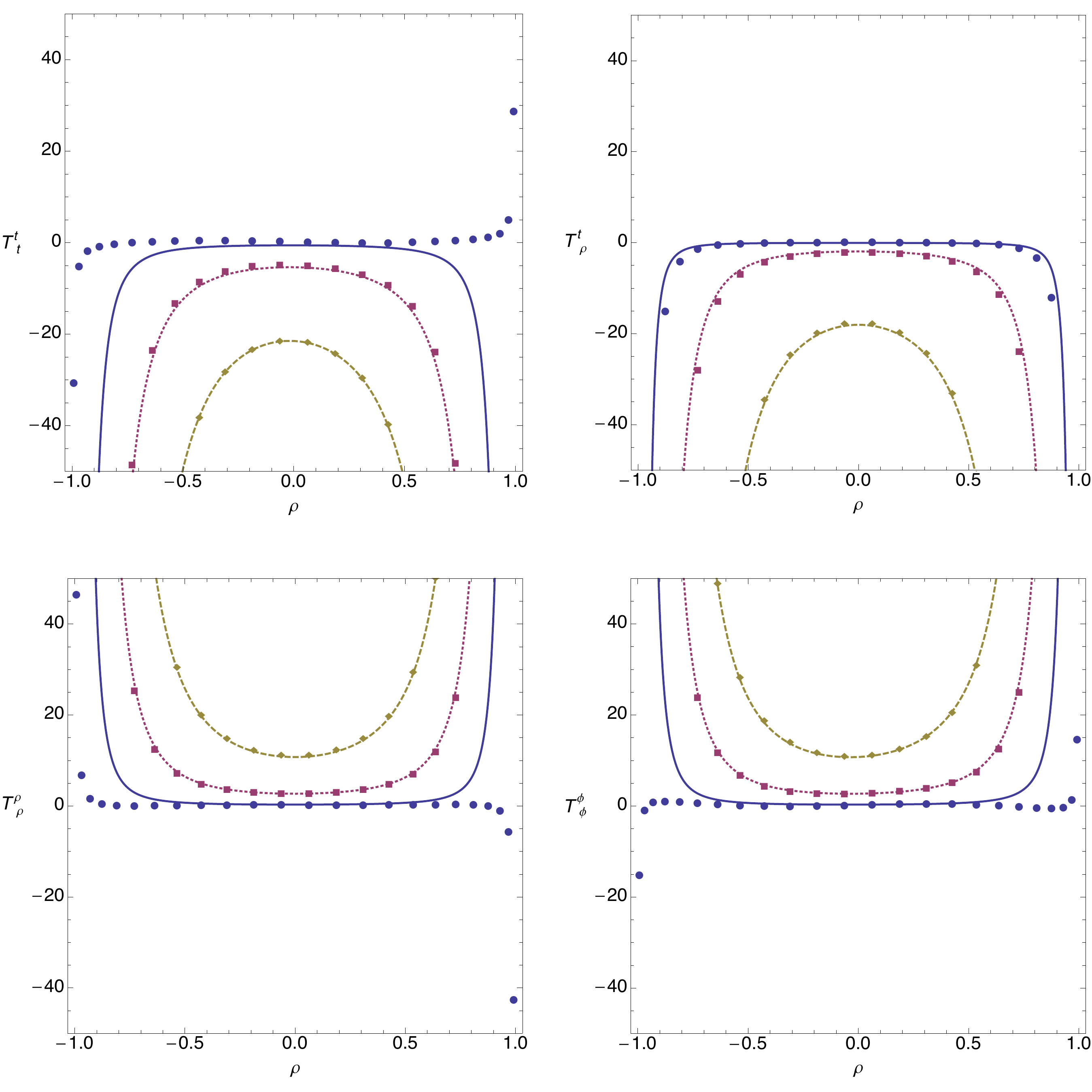}
}
\caption{Results from \cite{Fischetti:2012vt} for components of the CFT stress tensor as a function of a coordinate $\rho$ on ${\cal B}_d$ running from $\rho = -1$ at the colder boundary black hole horizon to $\rho = +1$ at the hotter boundary black hole horizon.  The panels compare linearized first order hydrodynamic predictions (lines) to numerical data (symbols).  The disks and solid line show $T_{CFT} = T_{\partial BH}$ for $\Delta T/T_\infty = .080$, the squares and dashed line show $T_{CFT} \approx 1.30 T_{\partial BH}$ for $\Delta T/T_\infty = .050$, and the diamonds and dotted lines show $T_{CFT} \approx 1.43 T_{\partial BH}$ for $\Delta T/T_\infty = .034$.  }
\label{fig:FFStress}
\end{figure}

It remains to discuss the stress tensor and the associated heat flux.  Here we content ourselves with displaying Figs. \ref{fig:FFFlux} and \ref{fig:FFStress} which are reproduced from \cite{Fischetti:2012vt}.  They indicate that the fluid approximation is indeed highly accurate for $T_{CFT}/T_{\partial BH} \gg 1$, and even perhaps when this parameter is as small as $1.4$.  However, notable differences appear for $T_{CFT}/T_{\partial BH}$ near 1.  Most of these, such as the discrepancy in the total heat flux $\Phi$ are merely quantitative.  However, an interesting qualitative feature concerns the sign of the CFT energy density ($-T^t_t$) near the horizon of the hotter (right) black hole.  While this quantity is necessarily positive in the fluid approximation, it clearly becomes negative for $T_{CFT}/T_{\partial BH}$ near 1.

What is particularly interesting is that this happens only for the hotter boundary black hole.  In free field theory, even static Hartle Hawking states have negative energy density at the horizon \cite{Candelas:1980zt,Page:1982fm}  (and see discussion in \cite{Fischetti:2013hja}), but an analogue of this effect would afflict both horizons equally.   On the other hand, one may note that the boundary stress tensor vanishes identically for $T_{CFT} = T_{\partial BH}$ and ${\cal B}_3$ built from two equal temperature $d=3$ BTZ black holes  (see \cite{Gregory:2008br} for an analogous result for $d=4$).  This may be interpreted as a cancelation between negative vacuum polarization and positive plasma energy analogous to that in \S\ref{sec:DFTstory}.  Turning on a small temperature difference $\Delta T$ then results  in a CFT stress tensor at linear order that is anti-symmetric under $\Delta T \rightarrow -\Delta T$.  This then implies anti-symmetry under exchange of the right- and left- black holes, and thus opposite signs for the energy density $-T^t_t$ at the two horizons.  One would like to verify perturbatively that this occurs at linear order with a sign matching Fig.~\ref{fig:FFStress}.  Perhaps it might suffice to study perturbations of pure AdS${}_d$ in Rindler coordinates, which as discussed in \S\ref{sec:DFTstory} provides a local model for any tuned funnel at the boundary black hole horizon as described in \S\ref{sec:DFTstory}.

%
\section{Discussion}
\label{sec:6}
%

As discussed above, the holographic AdS/CFT correspondence,
provides a powerful tool to explore the physics of strongly coupled physics in curved spacetimes. In addition to providing insight into some of the questions raised in \S\ref{sec:intro}, it also leads us to understand novel features of gravitational dynamics in AlAdS spacetimes.

In particular, we have seen that in certain phases strongly coupled quantum fields can be rather poor conductors of energy-momentum and heat. More generally one can study the out-of-equilibrium dynamics in the holographic models, obtaining universal results valid in the near-equilibrium hydrodynamic regime, and accessing the detailed process of equilibration following quenches. Furthermore, the holographic techniques also allowed us to discover CFT phases in which thermal plasmas can exchange heat only parametrically slowly with physically-small heat sources -- even when the heat source is a black hole, whose coupling to all fields is in a sense universal.

From a gravitational perspective, the presence of droplets, funnels and associated solutions adds to the rich class of black hole geometries which are present in the holographic theories. An interesting novelty is the presence of stationary black hole geometries with non-Killing (non-compactly generated) horizons.
Some of the solutions that we have discussed are known analytically, although many have so far only been constructed numerically. It is worth emphasising that there are very few calculations of dynamical stability for these solutions, even within the universal gravity sector.
Outside the hydrodynamic regime stability is not guaranteed, and this is clearly an important direction for future work.

Our prime focus had been to explain some of the key results in the context of black hole physics. It is however clear that there are perhaps other interesting physics lessons to draw from the correspondence and in the following we list some other applications and lay out some open problems for future investigation.

\paragraph{Cosmology and QFTs on time dependent backgrounds:} While a large part of our discussion centered on the physics in black hole backgrounds, it is clear that similar ideas provide a useful window into cosmological scenarios. A holographic discussion  of the behaviour of  quantum fields  in de Sitter spacetime (together with detailed references to earlier literature on the subject) can be found in \cite{Marolf:2010tg,Blackman:2011in}.  Once again one is able to independently tune the background temperature of the field theory away from the de Sitter temperature and study the resultant dynamics. Genuinely time dependent cosmologies have also attracted some attention in the holographic context: \cite{Hertog:2004rz,Hertog:2005hu,Barbon:2011ta} examine duals to big crunches in gravity in terms of quantum dynamics, while \cite{Chu:2006pa, Das:2006dz,Das:2006pw, Chu:2007um,  Awad:2007fj,Chu:2007zz} study quantum fields in time-dependent singular backgrounds.

\paragraph{Correlation functions:} For the most part we have given results for the simplest observable, the one-point function of the energy-momentum tensor of the quantum fields. While this captures the novel physics of vacuum polarization, there are other more detailed probes of the various phases that one could consider. For instance one could compute correlation functions of the gauge invariant operators in the field theory using holographic methods. In particular this computation would be interesting in contexts where we have novel physics such as in the droplet context, where one anticipates low correlation between the different regions of the plasma. While explicit evaluation of higher point correlation functions is likely to involve some technical challenges for the holographic constructions, it should be noted that an eikonal approximation to two-point function of heavy QFT operators \cite{Louko:2000tp} should prove instructive to extract some basic lessons about the correlation functions. This could already be used for example to learn about interesting features of two-point functions such as bulk-cone singularities \cite{Hubeny:2006yu}.

\paragraph{Entanglement measures:} Another interesting class of field theory probes is provided by a novel non-local probe: the entanglement entropy associated with a particular spatial region of the QFT. Holographic prescriptions \cite{Ryu:2006bv, Ryu:2006ef, Hubeny:2007xt} allow computation of this quantity in terms of the geometric data -- it is given by the area of an extremal surface anchored on the boundary of the region of interest. Over the past few years this quantity has been studied intensively as a potential probe of QFT dynamics and as a potential conduit for extracting some general lessons about the holographic dualities. It would be interesting to examine the behaviour of this quantity in some of the non-trivial examples of black hole spacetimes discussed herein. In the brane-world context \cite{Emparan:2006ni} made a compelling case for interpreting the entropy of the brane-world black hole (i.e., droplet) as entanglement entropy. The behaviour of entanglement has been recently investigated for field theories in cosmological spacetimes \cite{Maldacena:2012xp,Engelhardt:2013jda, Fischler:2013fba}. Likewise it would also be interesting to consider other holographically motivated measures of entropy and information \cite{Hubeny:2012wa, Kelly:2013aja} in the context of our discussion.

\paragraph{Equilibrium state of ${\cal N}=4$ SYM on black hole:} An important open problem is to complete the equilibrium phase diagram of ${\cal N}=4$ SYM on the Schwarzschild geometry.  Numerical constructions of funnel and droplet solutions now exist, though one needs to find the droplet solution with an equal temperature planar horizon. Once this has been found it would be interesting to examine the transition from the funnel to the droplet phase. Apart from intrinsic interest one also can enquire whether there is a preferred cone geometry that describes the transition \cite{Kol:2002xz}; see \cite{Emparan:2011ve} for a discussion in the context where unequal temperature horizons intersect (which would be relevant for some of the situations discussed herein).  Another interesting question is the phase structure on the Kerr background; the presence of rotation in the funnel solutions presents some novel issues which are yet to be analyzed.

\paragraph{Dynamics of CFTs on black holes:}  The instability of the simple Schwarzschild-AdS black string solution has been of significant historical interest.  While this solution is singular, the problem may be both regularized and generalized by considering instead the Schwarzschild-soliton black string as briefly described at the end of section \eqref{sec:dropsub}, and similarly unstable should arise for more general black holes.  One would like to understand the fate of this instability.  Some perturbations seem likely to produce droplets, but the lack of symmetry suggests that there should also be another end state associated with perturbations of the opposite sign.  Roughly speaking, one should be able to roll off of the effective-potential hill associated with the instability in at least two distinct directions.  In the soliton string context we may expect this second endpoint to have interesting structure at the tip of the soliton related to the physics of plasma balls.  Suppose, however, that the horizon continues to connect any any plasma ball-like feature to the boundary black hole. Then heat can flow in from infinity along this horizon and, depending on parameters, could drive continued expansion of the plasma ball.  In this way the solution may approximate a black funnel at late times (cf., Fig.~\ref{fig:solphases}).

\paragraph{Dynamics and mechanics of non-compact black holes:}   The laws of black hole mechanics provide strong constraints on dynamics when the horizons are compact, or with homogeneous horizons that may be compactified.  But our understanding of non-compact horizons is much less complete.  While this problem is quite difficult in general, one expects droplets and funnels to provide a context where progress can be made.  In particular, for bulk black holes which end on boundary black holes (say in the `tuned' case where $T_{bulk\ horizon} = T_{\partial BH}$), it is clear that time evolution can allow the dual CFT to exchange energy with the heat bath represented by the black hole -- and in particular that it may either gain or lose energy in this way.  As a result, an unstable solution may decay into one having either lesser or greater energy.  For the same reason, the change in entropy may have either sign.
Recall, however, that the second law constraints interactions with physical heat baths by requiring the non-decrease of the total entropy $S_{total} = S_{system} + S_{heat \ bath}$, including that of the heat bath.  And in the approximation that the bath remains at a truly constant temperature we have $\Delta S_{heat \ bath} = \Delta E_{heat \ bath}/T = - \Delta E_{system}/T$, so that in fact the free energy $F_{system} =  E_{system} - TS_{system}$ of our system should not increase.  In particular, this should be the case for the part of the dual CFT in the static region outside the boundary black hole.  It would be interesting to derive the corresponding result for the bulk gravity dual, and also to study natural analogues of the first law that involve simple changes of the boundary metric.  Such general results would inform discussions like that above about the dynamics of specific solutions.

\paragraph{Further Instabilities \& non-universal physics:} As we have emphasized at various points in the text, when one encounters geometries with structure on AdS scales, there is a potential for new physics arising due to the gravitational solutions wanting to localize on the transverse space. This typically is an issue whenever we are outside the hydrodynamic regime. Understanding the consequences of this phenomenon (apart from an obvious breaking of global symmetry) would be instructive in general. At the same time it is worth bearing in mind that there can be new phenomena associated with stringy degrees of freedom; whilst typically they require a breakdown of the supergravity approximation one may nevertheless have to contemplate these effects to be able to obtain a complete picture of the physics of quantum fields in curved spacetime via holographic methods.

\subsection*{Acknowledgements}

DM and MR would like to thank Veronika Hubeny  for collaboration and also for many interesting discussions on the physics of droplets and  funnels over several years.  We also thank Roberto Emparan, Pau Figueras, Sebastian Fischetti, Ted Jacobson, James Lucietti, Rob Myers, Harvey Reall, Simon Ross, Jorge Santos, and Benson Way for other related discussions.  We extend an additional special thanks to Rob Myers for permission to describe the previously unpublished work of \cite{AM} and to Jorge Santos for permission to use his Figs.~\ref{fig:Full} and \ref{fig:HP}.  Finally, we acknowledge the hospitality of Centro de Ciencias de Benasque Pedro Pascual and the participants of the Benasque workshops on Gravity: New Perspectives from Strings and Higher Dimensions in 2011 and 2013 for useful discussions.

DM  was supported in part by the National Science Foundation under Grant No PHY11-25915, by funds from the University of California, and as a Visiting Fellow College at Trinity College, University of Cambridge, UK.  He also thanks the Department of Applied Mathematics and Theoretical Physics at the University of Cambridge for their hospitality while writing this review. MR and TW are supported by STFC Consolidated Grants ST/J000426/1 and ST/J000353/1 respectively.

%
%

\providecommand{\href}[2]{#2}\begingroup\raggedright\endgroup


\end{document}